\documentclass[a4paper,11pt]{article}
\usepackage[dvips]{graphicx}
\usepackage{amsmath,amssymb}
\usepackage[T1]{fontenc}
\usepackage{xspace}

\usepackage{rotating}

\usepackage{mathrsfs}
\begin{document}


%
%
\renewcommand{\floatpagefraction}{0.7}
\renewcommand{\textfraction}{0.1}
\renewcommand{\bottomfraction}{0.6}

\newcommand{\Szero}{\ensuremath{\mathrm{H}^0}\xspace}
\newcommand{\mS}{\ensuremath{m_{\Szero}}\xspace}
\newcommand{\HSM}{\ensuremath{\mathrm{H}^0_{\mathrm{SM}}}\xspace}
\newcommand{\OPAL}{{\small OPAL}\xspace}
\newcommand{\LTHREE}{{\small L3}\xspace}
\newcommand{\LEP}{{\small LEP}\xspace}
\newcommand{\Zzero}{\ensuremath{\mathrm{Z}^{0}}\xspace}
\newcommand{\hzero}{\ensuremath{\mathrm{h}^{0}}\xspace}
\newcommand{\Hzero}{\ensuremath{\mathrm{H}^{0}}\xspace}
\newcommand{\dEdx}{\ensuremath{\mathrm{d}E/\mathrm{d}x}\xspace}
\newcommand{\gce}{{\small GCE}\xspace}
\newcommand{\sm}{{\small SM}\xspace}
\newcommand{\SM}{{Standard Model}\xspace}
\newcommand{\MSSM}{{\small MSSM}\xspace}
\newcommand{\mc}{Monte Carlo\xspace}
\newcommand{\MC}{Monte Carlo\xspace}
\newcommand{\klein}[1]{{\small #1}\xspace}
\newcommand{\hl}[1]{{\itshape #1}}
\newcommand{\degree}[1]{\ensuremath{\mathrm{#1}^\circ}}
\newcommand{\pa}{\ensuremath{\phi_a}\xspace}
\newcommand{\wa}{\ensuremath{\alpha}\xspace}
\newcommand{\aiso}{\ensuremath{\alpha_{\mathrm{iso}}}\xspace}
\newcommand{\minv}{\ensuremath{m_{\mathrm{inv}}}\xspace}
\newcommand{\tpmiss}{\ensuremath{\theta(\vec{p}_{\mathrm{miss}})}\xspace}
\newcommand{\pmiss}{\ensuremath{p_{\mathrm{miss}}}\xspace}
\newcommand{\nn}{\ensuremath{\nu\overline{\nu}}\xspace}
\newcommand{\mm}{\ensuremath{\mu^+\mu^-}\xspace}
\newcommand{\lplm}{\ensuremath{\mathrm{l}^+\mathrm{l}^-}\xspace}
\newcommand{\ee}{\ensuremath{\mathrm{e}^+\mathrm{e}^-}\xspace}
\newcommand{\bbar}{\ensuremath{\mathrm{b\overline{b}}}\xspace}
\newcommand{\keV}{\ensuremath{\mbox{keV}}\xspace}
\newcommand{\MeV}{\ensuremath{\mbox{MeV}}\xspace}
\newcommand{\GeV}{\ensuremath{\mbox{GeV}}\xspace}
\newcommand{\gev}{\ensuremath{\mbox{GeV}}\xspace}
\newcommand{\TeV}{\ensuremath{\mbox{TeV}}\xspace}
\newcommand{\Evec}{\ensuremath{\vec{E}}\xspace}
\newcommand{\tbc}{(\emph{\ldots to be completed \ldots})\xspace}
\newcommand{\mrec}{\ensuremath{m_{\mathrm{r}}}\xspace}
\newcommand{\sq}{\ensuremath{k}\xspace}
\newcommand{\Nnf}{\ensuremath{\mathrm{N^{95}}}\xspace}
\newcommand{\sqnf}{\ensuremath{\mathrm{k^{95}}}\xspace}
\newcommand{\Nsm}{\ensuremath{\mathrm{N_{SM}}}\xspace}
\newcommand{\SigmaZH}{\ensuremath{\sigma_{\mathrm{ZH^{SM}}}}\xspace}
\newcommand{\mH}{\ensuremath{m_{\mathrm{H}}}\xspace}
\newcommand{\MH}{\ensuremath{M_{\mathrm{H}}}\xspace}
\newcommand{\Mh}{\ensuremath{M_{\mathrm{H}}}\xspace}
\newcommand{\mHsm}{\ensuremath{m_{\mathrm{H}}^{\mathrm{SM}}}\xspace}
\newcommand{\mh}{\ensuremath{m_{\mathrm{h^0}}}\xspace}
\newcommand{\Gh}{\ensuremath{\Gamma_{\mathrm{H}}}\xspace}
\newcommand{\mhwo}{\ensuremath{m}\xspace}
\newcommand{\mhi}{\ensuremath{m_{\mathrm{h^0_i}}}\xspace}
\newcommand{\hi}{\ensuremath{\mathrm{h^0_i}}\xspace}
\newcommand{\SigmaZh}{\ensuremath{\sigma_{\mathrm{Zh}}}\xspace}
\newcommand{\err}[3]{\,\ensuremath{\mathrm{#1}\pm\mathrm{#2\,(stat.)}\pm\mathrm{#3\,(syst.)}}\xspace}
\newcommand{\erro}[2]{\,\ensuremath{\pm\mathrm{#1}\pm\mathrm{#2}}\xspace}
\newcommand{\mathd}{\ensuremath{\mathrm{d}}}
\newcommand{\deltam}{\ensuremath{\Delta m}}
\newcommand{\rb}[1]{\raisebox{-2ex}{#1}}
\newcommand{\pb}{\ensuremath{\mathrm{pb}^{-1}}}
\newcommand{\Ztoee}{\ensuremath{\Zzero\to\ee}}
\newcommand{\Ztomm}{\ensuremath{\Zzero\to\mm}}
\newcommand{\vnr}[1]{\vphantom{\rule{0mm}{#1}}\xspace}
\newcommand{\rr}{\raggedright\small}
\newcommand{\prelim}{\large\textbf{Preliminary}\xspace}
\newcommand{\G}{\mbox{$\mathrm{GeV}$}}
\newcommand{\sqrts}{\mbox{$\sqrt {s}$}}
\newcommand{\mZ}{$m_{\mathrm{Z}}$\xspace}

\newcommand{\Mz}{\ensuremath{M_{\mathrm{Z}}}\xspace}
\newcommand{\bb}{\mbox{$\mathrm{b}\bar{\mathrm{b}}$}}
\newcommand{\qq}{\mbox{$\mathrm{q}\bar{\mathrm{q}}$}}
\newcommand{\etal}{\mbox{\it et al.}}
\newcommand{\mysec}{\mbox{$ \sigma(\Mh,\Gamma_{\mathrm{H}})\times \mathrm{BR}(\mathrm{H}\to E_{\mathrm{MIS}})$}}
\newcommand{\nuenue}{\mbox{$\nu_{\mathrm{e}}\bar{\nu_{\mathrm{e}}}$}}
\newcommand{\mHrec} {\mbox{$m_{\mathrm{H}}^{\mathrm{rec}}$}}
\newcommand{\mHtest} {\mbox{$m_{\mathrm{H}}$}}
\newcommand{\gevcs} {\mbox{${\mathrm{GeV}}/c^2$}}
\newcommand{\gevcm} {\mbox{${\mathrm{GeV}}/c$}}
\newcommand{\czcz}  {\mbox{$\chi^0_1\chi^0_1$}}
\newcommand{\cz}    {\mbox{$\chi^{0}$}}
\newcommand{\co}    {\mbox{${\tilde{\chi}_1^0}$}}
\newcommand{\ct}    {\mbox{${\tilde{\chi}_2^0}$}}
\newcommand{\coct}  {\mbox{$\chi^0_1\chi^0_2$}}
\newcommand{\ctcoz} {\ct\ra\Zs}
\newcommand{\ctcog} {$\ct\ra\co\gamma$}
\newcommand{\hczcz} {\ho\ra\co\co}
\newcommand{\hcoct} {\ho\ra\co\ct}
\newcommand {\Ho}   {\mbox{$\mathrm{H}$}}
\newcommand {\ho}   {\mbox{$\mathrm{h}^{0}$}}
\newcommand {\Zo}   {\mbox{$\mathrm{Z}$}}
\newcommand{\Hpm}   {\mbox{$\mathrm{H}^{\pm}$}}
\newcommand{\Hp}    {\mbox{$\mathrm{H}^+$}}
\newcommand{\Hm}    {\mbox{$\mathrm{H}^-$}}
\newcommand{\qqp}   {\mbox{$\mathrm{q\overline{q}^\prime}$}}
\newcommand{\qpq}   {\mbox{$\mathrm{q^\prime\overline{q}}$}}
\newcommand{\qppqppp}{\mbox{$\mathrm{q^{\prime\prime}\overline{q}^{\prime\prime\prime}}$}}
\newcommand{\tnt}{\mbox{${\tau\nu_{\tau}}$}}
\newcommand{\tpnu}{\mbox{${\tau^+\nu_{\tau}}$}}
\newcommand{\tmnu}{\mbox{${\tau^-{\bar{\nu}}_{\tau}}$}}
\newcommand{\lnu}{\mbox{$\ell\nu$}}
\newcommand{\tnu}{\mbox{$\tau\nu$}}
\newcommand{\HH}{\Hp\Hm}
\newcommand{\mHpm}{\mbox{$m_{\mathrm{H}^{\pm}}$}}
\newcommand{\nbar}{\mbox{$\overline{\nu}$}}
\newcommand{\llnunu}{\mbox{\lpair$\nu$\nbar}}
\newcommand{\lpair}{\mbox{$\ell^+\ell^-$}}
\newcommand{\smc}{Standard Model Monte Carlo}
\newcommand{\smp}{Standard Model processes}
\newcommand{\dm}{\mbox{$\Delta m$}}
\newcommand{\Zs}         {\mbox{${\mathrm{Z}}^{*}$}}
\newcommand{\Zgs}        {\mbox{$\mathrm{(Z/\gamma)}^{*}$}}
\newcommand {\Wpm}       {\mbox{$\mathrm{W}^{\pm}$}}
\newcommand {\Wsp}       {\mbox{$\mathrm{W}^{*+}$}}
\newcommand {\Wsm}       {\mbox{$\mathrm{W}^{*-}$}}
\newcommand{\MZ}{M_{\mathrm Z}}
\newcommand{\qqbar}{\mbox{\mathrm{q} \bar{\mathrm{q}}}}
\newcommand{\gaga}       {\mbox{$\gamma\gamma$}}
\newcommand{\WW}         {\mbox{$\mathrm{W}^+\mathrm{W}^-$}}
\newcommand{\ZZ}         {\mbox{$\mathrm{Z}\mathrm{Z}$}}
\newcommand{\Wp}         {\mbox{$\mathrm{W}^+$}}
\newcommand{\Wm}         {\mbox{$\mathrm{W}^-$}}
\def\rm       {\mathrm}
\def\mrm       {\mathrm}
\newcommand{\gsim}{\;\raisebox{-0.9ex}
           {$\textstyle\stackrel{\textstyle >}{\sim}$}\;}
\newcommand{\lsim}{\;\raisebox{-0.9ex}{$\textstyle\stackrel{\textstyle<}
           {\sim}$}\;}
\newcommand{\ipb}         {\mbox{pb$^{-1}$}}
\newcommand{\Ecm}         {\mbox{$E_{\mathrm{cm}}$}}
\newcommand{\Evis}      {\mbox{$E_{\mathrm{vis}}$}}
\newcommand{\Rvis}      {\mbox{$R_{\mathrm{vis}}$}}
\newcommand{\Mvis}      {\mbox{$M_{\mathrm{vis}}$}}
\newcommand{\Ebeam}       {E_{\mathrm{beam}}} 
\newcommand{\Mrec}      {M_{\mrm{recoil}}}
\newcommand{\mgg} {$M_{\gamma \gamma}$}
\newcommand{\omax} {5.9}

%
%
\topsep0pt plus 1pt
\begin{titlepage}
\begin{center}
  {\large  EUROPEAN ORGANIZATION FOR NUCLEAR RESEARCH}
\end{center}
\bigskip
\begin{flushright}
  OPAL PR417 \\
  CERN-EP-2006-026   \\ August 3, 2006
\end{flushright}
\bigskip\bigskip\bigskip\bigskip\bigskip
\begin{center}
  {\huge\bfseries
    Search for Invisibly Decaying Higgs Bosons\\[1.2ex]
    with Large Decay Width Using the \\[1.2ex]{OPAL Detector} at LEP
  }
\end{center}
\bigskip\bigskip
\begin{center}
  {\LARGE The OPAL Collaboration}
\end{center}
\bigskip\bigskip\bigskip
\begin{center}
  {\large  Abstract}
\end{center}
{ \noindent This paper describes a topological search for an invisibly decaying Higgs boson, $\mathrm{H}$,
  produced via the Bjorken process ($\ee\to\mathrm{H}{}\mathrm{Z}$). The analysis is based on data recorded using the \OPAL detector at \LEP at centre-of-mass energies from 183 to 209\,\GeV corresponding to a total integrated luminosity of 629\,pb$^{-1}$.
In the analysis only hadronic decays of the Z boson are considered.
  A scan over Higgs boson masses from  1 to 120\,\GeV and decay widths from 1 to 3000\,\GeV revealed no indication for a signal in the data. From a likelihood ratio of expected signal and \SM background we determine upper limits on cross-section times branching ratio to an invisible final state. For moderate Higgs boson decay widths, these range from about 0.07\,pb (\Mh = 60\,\GeV) to 0.57\,pb (\Mh = 114\,\GeV).
For decay widths above 200\,\GeV the upper limits are of the order of 0.15\,pb.
 The results can be interpreted in general scenarios predicting a large
  invisible decay width of the Higgs boson. As an example we interpret the
  results in the so-called stealthy Higgs scenario. The limits from this analysis exclude a large part of the parameter range of this scenario experimentally accessible at \LEP\,2.}

\bigskip\bigskip\bigskip\bigskip\bigskip\bigskip
\begin{center}

  {\large (Submitted to Eur. Phys. J.)}

\end{center}

\end{titlepage}

\newpage

\begin{center}{\Large        The OPAL Collaboration
}\end{center}\bigskip
\begin{center}{
G.\thinspace Abbiendi$^{  2}$,
C.\thinspace Ainsley$^{  5}$,
P.F.\thinspace {\AA}kesson$^{  7}$,
G.\thinspace Alexander$^{ 21}$,
G.\thinspace Anagnostou$^{  1}$,
K.J.\thinspace Anderson$^{  8}$,
S.\thinspace Asai$^{ 22}$,
D.\thinspace Axen$^{ 26}$,
I.\thinspace Bailey$^{ 25}$,
E.\thinspace Barberio$^{  7,   p}$
% ...
}\end{center}\bigskip
\bigskip
$^{  1}$School of Physics and Astronomy, University of Birmingham,
Birmingham B15 2TT, UK
\newline
$^{  2}$Dipartimento di Fisica dell' Universit\`a di Bologna and INFN,
I-40126 Bologna, Italy
\newline
$^{  3}$Physikalisches Institut, Universit\"at Bonn,
D-53115 Bonn, Germany
\newline
% ...
$^{  5}$Cavendish Laboratory, Cambridge CB3 0HE, UK
\newline
% ...
$^{  7}$CERN, European Organisation for Nuclear Research,
CH-1211 Geneva 23, Switzerland
\newline
$^{  8}$Enrico Fermi Institute and Department of Physics,
University of Chicago, Chicago IL 60637, USA
\newline
$^{  9}$Fakult\"at f\"ur Physik, Albert-Ludwigs-Universit\"at 
Freiburg, D-79104 Freiburg, Germany
\newline
% ...
$^{ 21}$Department of Physics and Astronomy, Tel Aviv University,
Tel Aviv 69978, Israel
\newline
$^{ 22}$International Centre for Elementary Particle Physics and
Department of Physics, University of Tokyo, Tokyo 113-0033, and
Kobe University, Kobe 657-8501, Japan
\newline
% ...
$^{ 25}$University of Victoria, Department of Physics, P O Box 3055,
Victoria BC V8W 3P6, Canada
\newline
$^{ 26}$University of British Columbia, Department of Physics,
Vancouver BC V6T 1Z1, Canada
\newline
% ...
$^{  a}$ and at TRIUMF, Vancouver, Canada V6T 2A3
\newline
$^{  b}$ now at University of Alberta
\newline
% ...
$^{  d}$ now at Institute of Physics, Academy of Sciences of the Czech Republic
18221 Prague, Czech Republic
\newline 
$^{  e}$ and Department of Experimental Physics, University of Debrecen, 
Hungary
\newline
% ...
$^{  h}$ now at University of Liverpool, Dept of Physics,
Liverpool L69 3BX, U.K.
\newline
$^{  i}$ now at Dept. Physics, University of Illinois at Urbana-Champaign, 
U.S.A.
\newline
% ...
$^{  k}$ now at University of Kansas, Dept of Physics and Astronomy,
Lawrence, KS 66045, U.S.A.
\newline
$^{  l}$ now at University of Toronto, Dept of Physics, Toronto, Canada 
\newline
$^{  m}$ current address Bergische Universit\"at, Wuppertal, Germany
\newline
% ...
$^{  p}$ now at The University of Melbourne, Victoria, Australia
\newline
$^{  q}$ now at IPHE Universit\'e de Lausanne, CH-1015 Lausanne, Switzerland
\newline
$^{  r}$ now at IEKP Universit\"at Karlsruhe, Germany
\newline
$^{  s}$ now at University of Antwerpen, Physics Department,B-2610 Antwerpen, 
Belgium; supported by Interuniversity Attraction Poles Programme -- Belgian
Science Policy
\newline
% ...


\newpage
%
%
\section{Introduction}
An intense search for the Higgs boson was undertaken by all of the four \LEP experiments in various \SM and non--\SM search channels. Searches for the \SM Higgs boson, exploiting the prediction for its decay modes and also searches for invisible Higgs boson decays as predicted by various extensions of the \SM, excluded Higgs masses up to 114.4\,\GeV~\cite{Barate:2003sz,unknown:2001xz}.
These latter searches assumed a rather small invisible decay width comparable to the predicted \SM decay width for a light Higgs boson and well below the experimentally achievable mass resolution of about 3 to 5\,\GeV. \par  
Recent theories that postulate the existence of additional spatial dimensions offer a new possibility for invisible Higgs decays \cite{Giudice:2000av}. In such theories the Planck mass is lowered to the~\TeV range and a rich spectrum of new particles appears, like graviscalars in the case of flat extra dimensions. Hence the Higgs boson can mix with the graviscalars, which leads to a missing energy signature in the detector \cite{Giudice:2000av}. This mixing can result in a large invisible decay width of the Higgs boson, depending on the model parameters, and would therefore alter the \SM branching ratios. As a consequence of the broadening of the Higgs resonance in the recoil mass spectrum, the signal-to-background ratio can deteriorate significantly. In a worst case scenario, searches optimised under the assumption of a narrow Higgs resonance might have missed the detection of a kinematically accessible Higgs boson at \LEP.\par
This paper describes a search for the Higgs boson, H, which imposes no constraints on the total decay width. The Higgs boson is assumed to be produced in association with a Z boson via the Bjorken process, $\ee \to \mathrm{H}{}\mathrm{Z}$, where the Z is required to decay hadronically and the invisible Higgs boson is detected as missing energy $E_{\mathrm{MIS}}$ in the event.
The results are presented in a model-independent way in terms of limits on the Bjorken production cross-section times branching ratio, \mysec, at a centre-of-mass energy of 206\,\GeV, where $ \Gamma_{\mathrm{H}} $ is the Breit--Wigner width of the Higgs boson.
A simple model extending the \SM with additional SU(3)$_{\mathrm{C}}
\times$SU(2)$_{\mathrm{L}}\times$U(1)$_{\mathrm{Y}}$ singlet fields
which interact strongly with the Higgs boson (``stealthy Higgs scenario''~\cite{c:stealthy_higgs}) is chosen as an example for the interpretation of the result. This interaction gives rise to a large invisible decay width of the Higgs boson. This dedicated search expands on the previous decay-mode-independent search \cite{Abbiendi:2002qp} carried out by the \OPAL Collaboration which reported for the first time limits on the HZ production cross-section, interpreted in the stealthy Higgs model.\par
The paper is organised as follows. Section 2 introduces the stealthy Higgs scenario. Section 3 gives details about the modelling of signal and background. Section 4 describes the search and the results are interpreted in Section 5. We summarise the results in Section 6.    
%
%
\section{The stealthy Higgs scenario} 
In general renormalisable theories there might be other fundamental scalars, in addition to the \SM Higgs boson, that do not interact with normal matter. 
To investigate the influence of a hidden scalar sector on the Higgs observables 
the stealthy Higgs scenario conjectures the existence of additional SU(3)$_{\mathrm{C}}\times$SU(2)$_{\mathrm{L}}\times$U(1)$_{\mathrm{Y}}$ singlet fields called phions. Radiative corrections to weak processes are not sensitive to the presence of singlets in the theory because no Feynman graphs containing singlets appear at the one-loop level. Since effects at the two-loop level are below the experimental precision, the presence of a singlet sector is not ruled out by any \LEP\,1 precision data~\cite{c:stealthy_higgs}.~These phions would not interact via the strong or electro-weak forces, but couple only to the Higgs boson~\cite{c:stealthy_higgs}, thus offering invisible decay modes to the Higgs. The width of the Higgs resonance can become large if either the number of such singlets, $N$, or the coupling, $\omega$, is large, thus leading to a broad mass spectrum recoiling against the reconstructed Z boson. \par
The Lagrangian of the scalar sector in this model contains only four additional parameters compared to the \SM. Not listing the unchanged couplings of the Higgs boson to the fermions, the scalar part of the Lagrangian is given by
\begin{eqnarray} \label{model}
{\cal L}_{\mathrm{scalar}} &=& {\cal L}_{\mathrm{Higgs}} + {\cal L}_{\mathrm{phion}} + {\cal L}_{\mathrm{interaction}}  \\
{\cal L}_{\mathrm{Higgs}}  &=&
 - \partial_{\mu}\phi^\dagger \partial^{\mu}\phi -\lambda \,
 (\phi^\dagger \phi - \frac{v^2}{2})^2 \\ 
{\cal L}_{\mathrm{phion}}  &=& - \frac{1}{2}\,\partial_{\mu} \vec\varphi \, 
\partial^{\mu}\vec\varphi
     -\frac{1}{2} m_{\mathrm{phion}}^2 \,\vec\varphi^2 - \frac{\kappa}{8N} \, 
     (\vec\varphi^2 )^2  \\ 
{\cal L}_{\mathrm{interaction}} &=& -\frac{\omega}{2\sqrt{N}}\, \, \vec\varphi^2 \,\phi^\dagger \phi.  
\end{eqnarray} 
The term ${\cal L}_{\mathrm{Higgs}}$ describes the usual \SM Higgs doublet $\phi$ acquiring the \SM vacuum expectation value, $v$, and having its self-coupling $\lambda$. In the free Lagrangian of scalar singlets, ${\cal L}_{\mathrm{phion}}$, the singlets with mass $m_{\mathrm{phion}}$ are denoted as the O($N$)-symmetric multiplet $\vec{\varphi}$. The phions also have a self-coupling $\kappa$, which is fixed at $\kappa(2M_{\mathrm{Z}}) = 0$, to allow for the widest parameter range of the model. The self coupling term entering loop calculations is suppressed like 1/$N$. The interaction term between the Higgs and the additional phions, ${\cal L}_{\mathrm{interaction}}$, leads to the phenomenological consequence of invisible Higgs decays because the Higgs boson couples to the phions independently of their mass. The strength of the coupling is instead proportional to the coupling constant $\omega$, which is a free parameter of the model. Even though the vacuum-induced mass term of the phions after the symmetry-breaking is suppressed like $ 1/\sqrt{N}$ \cite{c:stealthy_higgs}, the phions occur in loop corrections to the Higgs boson propagator and therefore affect the resonance width of the Higgs boson.
 An analytic expression \cite{vanderBij:2006ne} for the change in the Higgs width compared to the \SM decay width, $\Gamma_{\mathrm{SM}}$, can be found in the limit $N \to \infty$, when neglecting the self-coupling of the phions as a
small effect:
\begin{equation}\label{eq:higgs_width}
  \Gamma_{\mathrm{H}}(\Mh) = \Gamma_{\mathrm{SM}}(\Mh) + \frac{\omega^2 v^2}
                  {32\, \pi\, \Mh}\times \sqrt{1-4m_{\mathrm{phion}}^2/\Mh^2}.
\end{equation}
The cross-section for the Bjorken process can be calculated from
Equations~9 and 10 of reference~\cite{c:stealthy_higgs}.
Using the parametrisation for the invisible decay width (Equation 1 and 2 in ~\cite{Binoth:1999ay}) one can express the total cross-section
for the production and invisible decay by 
\begin{eqnarray}\label{eq:totalhiggs}
\sigma_{(\mathrm{e^+e^-}\rightarrow \mathrm{Z}+E_{\mathrm{MIS}})} = \int ds_I \, \sigma_{(\mathrm{e}^+\mathrm{e}^- \rightarrow \mathrm{ZH})}(s_I) \,
\frac{\sqrt{s_I} \quad \Gamma_{\mathrm{H}}^{\mathrm{inv}}}
{\pi ((M_{\mathrm{H}}^2-s_I)^2+s_I\,\Gamma_{\mathrm{H}}^2)}.
\end{eqnarray}
Here $s_I$ denotes the invariant mass squared of the invisible decay products of the Higgs boson. The production rate of these invisible masses is given by the \SM cross-section\footnotemark[1]\footnotetext[1]{By choosing $\omega >0$ one can prevent the phions from acquiring a non-zero vacuum expectation value and avoid a Higgs-phion mixing due to a non-diagonal mass matrix. In case of non-zero mixing, the couplings of the lightest scalar to the gauge boson would decrease proportional to the cosine of the mixing angle. As a consequence the cross-section of the Bjorken process would be lowered.} $\sigma_{(\mathrm{e}^+\mathrm{e}^-\rightarrow \mathrm{ZH})}(s_I)$ for a Higgs boson of mass $\sqrt{s_I}$. Hence the \SM cross-section completely determines the dependence of the total cross-section on the centre-of-mass energy (see e.g. in \cite{c:hunter}). Therefore the total cross-section goes rapidly to zero for Higgs boson masses above the kinematic limit.
The effect of the convolution with the Breit-Wigner-like function is a broadening of the resonance in the recoil mass spectrum and hence a dilution of the signal-to-background ratio. In extreme cases of large invisible decay width one could expect the Higgs recoil mass spectrum to mimic the background.
In such extreme cases even a light and kinematically accessible Higgs boson might have escaped detection at \LEP. \par   
In Section~\ref{s:stealthy} we derive limits on the stealthy Higgs
model under the assumption of $m_{\mathrm{phion}} = 0$. By simulating signal spectra for different Higgs boson masses \Mh and widths $\Gamma_{\mathrm{H}}$, we set
limits in the $\omega$-$M_\mathrm{H}$ plane in the large $N$ limit.
%
%
\section{Data sets and Monte Carlo samples}

\subsection{The \OPAL detector and event reconstruction}

The \OPAL detector~\cite{c:detector}, operated between 1989 and 2000 at \LEP, had
nearly complete solid angle\footnotemark[3]\footnotetext[3]{ \OPAL used a right-handed coordinate
  system. The $z$ axis pointed along the direction of the electron beam
  and the $x$ axis was horizontal pointing towards the centre of the
  LEP ring. The polar angle $\theta$ was measured with respect to the
  $z$ axis, the azimuthal angle $\phi$ with respect to the $x$ axis.} coverage and excellent hermeticity.
The innermost detector of the central tracking was a high-resolution
silicon microstrip vertex detector~\cite{simvtx} which lay immediately
outside the beam pipe.  
The silicon microvertex detector
was surrounded by a high precision 
vertex drift chamber,
a large volume jet chamber, and $z$--chambers which measured the $z$ coordinates of tracks, all in a uniform 0.435\,T axial magnetic field. A lead-glass electromagnetic calorimeter with presampler was located outside the magnet coil. In combination with the forward calorimeters, a forward ring of lead-scintillator
modules (the ``gamma catcher''), a forward scintillating tile 
counter~\cite{c:detector,mip}, and the silicon-tungsten
luminometer~\cite{sw}, the calorimeters provided 
a geometrical acceptance down to 25~mrad from the beam direction. The silicon-tungsten luminometer served to measure the integrated luminosity using small angle Bhabha
scattering events~\cite{lumino}.
The magnet return yoke was instrumented with streamer tubes and thin gap
chambers for hadron calorimetry and is surrounded by several layers 
of muon chambers.\par
The analysis is based on data collected with the \OPAL detector at
\LEP\,2 from 1997 to 2000 at centre-of-mass
energies between 183 and 209\,\GeV.  The integrated luminosity
analysed is 629.1\,pb$^{-1}$. To compare with the \smc ~the data are binned in five
nominal centre-of-mass-energy points, corresponding to the energies at which the \MC is produced, as detailed in Table~\ref{t:lumicor}. \par
A fast online filtering algorithm classifies the events as multi-hadronic.
~Events are reconstructed from tracks and energy
deposits (``clusters'') in the electromagnetic and hadronic calorimeters.
~All tracks and energy clusters satisfying quality
requests similar to those described in \cite{Abbiendi:2003} are associated to
form ``energy flow objects''.
The measured energies are corrected for double counting of energy in the tracking chambers and calorimeters by the algorithm described in \cite{Abbiendi:2003}. Global event variables, such as transverse momentum and visible mass, are then reconstructed from these objects and all events are forced into a two-jet topology using the Durham algorithm \cite{Brown:1990nm}. \par

\subsection{Signal and background modelling}
To determine the detection efficiency for a signal from an invisibly decaying
Higgs boson and the amount of expected background from \SM processes, several
\MC samples are used. 
Signal events for a hypothetical Higgs boson mass \Mh decaying with arbitrary broad width \Gh are simulated by reweighting invisibly decaying events of type $\mathrm{H} \to \chi^0_1\chi^0_1$.
The mass of neutralinos $\chi^0_1$ is chosen such that the Higgs boson with mass $m_i$ can decay into a pair of neutralinos, which leave the detector without being detected.  
 These Higgs bosons with decays into `invisible' particles are generated with masses $m_i$ from 1\,\GeV to 120\,\GeV  with the
\klein{HZHA}~\cite{c:MC_HZHA} generator. The HZHA events are generated assuming the \SM production
cross-section $\sigma_{(\mathrm{e}^+\mathrm{e}^-\rightarrow \mathrm{ZH})}$~for
 the Higgs boson. The test masses $m_i$ are spaced in steps of 1\,\GeV.
The spacing of the test masses is chosen such that they are not resolved by the detector in the signal yielded after a reweighting procedure described in the following.  
From Equation~\ref{eq:totalhiggs} one extracts the event weights $w_i(m_i;M_{\mathrm{H}},\Gamma_{\mathrm{H}})$ for a mass point $m_i$ contributing to the search for a Higgs boson of mass $M_{\mathrm{H}} $ and total decay width
  $\Gamma_{\mathrm{H}}$. The total decay width \Gh is defined as the sum of \SM width and invisible width $\Gamma_{\mathrm{H}}^{\mathrm{inv}}$.
\begin{eqnarray}
\label{eq:EWG}
w_i(m_i;M_{\mathrm{H}},\Gamma_{\mathrm{H}})&=& 
\frac{ \frac{d\sigma}{d m_i}(m_i) }{ \sum_{m_j=1 \,\mathrm{GeV}}^{120\,\mathrm{GeV}}{\frac{d \sigma}{dm_j}( m_j)}}
\end{eqnarray}
\begin{eqnarray}
\label{eq:EWG1}
\frac{d \sigma}{dm_i}(m_i)&=& \frac{\sigma_{(\mathrm{e}^+\mathrm{e}^-\rightarrow \mathrm{ZH})}(m_i)~2 m_i^2\Gamma_{\mathrm{H}}^{\mathrm{inv}}}{\pi ((M_{\mathrm{H}}^2-m^{2}_{i})^2+m^{2}_{i}\Gamma_{\mathrm{H}}^2)} .
\end{eqnarray}
The \SM cross-section $\sigma_{(\mathrm{e}^+\mathrm{e}^-\rightarrow \mathrm{ZH})}$ for the Bjorken production process in Equation~\ref{eq:EWG1} propagates the centre-of-mass energy dependence of the total cross-section into the
weights. The unweighted signal Monte Carlo samples contain 2000 events per mass point
$m_i$. In the reweighted signal \MC sample all test masses contribute according to their weight.
The reweighted masses \Mh range from 1 to 120\,\GeV spaced in steps of 1\,\GeV. The smallest width simulated by this procedure is a $\Gamma_{\mathrm{H}}$~of 1\,\GeV and the largest a $\Gamma_{\mathrm{H}}$~of 3\,\TeV. The detection efficiency for a Higgs boson with \Mh and \Gh is estimated by
the sum of selected event weights assuming binomial errors.\par

The classes of \SM background processes considered are two-photon\footnotemark[2]\footnotetext[2]{Two-photon interactions occur when an electron and a positron at high energies and in close proximity emit a pair of photons which interact via the electromagnetic force to generate a fermion pair.}, two- and four-fermion processes. 
For simulation of background processes the following generators are used:
 \klein{KK2F}~\cite{c:MC_KK2F} and \klein{PYTHIA}~\cite{c:MC_PYTHIA}
(q$\bar{\mathrm{q}}(\gamma)$), \klein{GRC4F}~\cite{c:MC_GRC4F}
~(four-fermion processes), \klein{PHOJET}~\cite{c:MC_PHOJET},
\klein{HERWIG}~\cite{c:MC_HERWIG}, Vermaseren~\cite{c:MC_VERMASEREN}
(hadronic and leptonic two-photon processes).
For Monte Carlo generators other than \klein{HERWIG}, the hadronisation is done
using \klein{JETSET} 7.4~\cite{c:MC_PYTHIA}. The integrated luminosity of the main background Monte
Carlo samples is at least 15 times the statistics of the data for the
two-fermion background, 24 times for the four-fermion background and 30
times for the two-photon background. 
The \MC events are passed through a detailed simulation of the \OPAL
detector~\cite{c:GOPAL} and are reconstructed using the same
algorithms as for the real data.
\section{Search for \boldmath $\mathrm{e}^+\mathrm{e}^- \to \mathrm{H}{}\mathrm{Z} $ \unboldmath with \boldmath
  $ \mathrm{Z} \to \qq$ \unboldmath and \boldmath$ \mathrm{H} \to E_{\mathrm{MIS}}$ \boldmath final state}
\label{s:decay width_independent_search}
The event selection is intended to be efficient for the complete
range of possible Higgs masses \Mh and corresponding decay widths \Gh studied in this search.
The preselection cuts remain relatively loose and are intended to accumulate signal like event topologies in the data. The final discrimination between signal and background is done by a likelihood-based selection.
The optimised likelihood selection has to account for the fact that the kinematical properties of the signal change considerably over the range of masses and width hypotheses considered.
\subsection{Event topologies}
The signal signature is generally characterised by an acoplanar two-jet system from the Z boson decay. We use the term `acoplanar' for jet pairs if the two jet axes and the beam axis are not consistent with lying in a single plane. The decay products of the Z boson are preferentially emitted into the central part of the detector, recoiling against the
invisibly decaying Higgs boson. This is because, in contrast
  to the irreducible background of ZZ $ \to \qq ~\nn$ which is produced with an
  angular dependence of the differential cross-section proportional to $ \cos^2 \theta $, the Bjorken process is proportional to $ \sin^2 \theta$. The Higgs boson decay leads to a large
missing momentum and a significant amount of missing energy.   
 In two-photon processes, where the incoming electron and positron
 are scattered at low angles, usually one or both of the
  electrons remain undetected. Events of this type have large
  missing momentum with the missing momentum vector, $\vec{p}_{\mathrm{MIS}}$, pointing at low angles to the beam axis. The two-photon events have a small visible invariant
 mass $M_{\mathrm{VIS}}$ and a tiny transverse momentum  $p^{\mathrm{T}}_{\mathrm{MIS}}$ but a
 considerable longitudinal momentum along the $z$-axis in the common case that
 the two photons do not have equal energy. Due to these special
 characteristics this background can be easily reduced to a negligible level.\par
The two-fermion background important for this search consists of $\mathrm{Z} / \gamma^* \to \qq(\gamma)$ events. These events tend to have a big cross-section
if one or more initial state radiation photons (abbreviated as ISR photons)
 are emitted so that the effective centre-of-mass energy $
\sqrt{s^{'}}$ is reduced to a value near the Z-resonance (so-called radiative return events). The emission of ISR photons happens predominantly at small polar angles. In case of a mismeasurement or escape of the ISR photons through the beam pipe these events have a sizeable missing momentum preferentially oriented at small polar angles, close to the beam pipe. In such events the two jets are almost coplanar.\par
 The most difficult background to separate is four-fermion processes with neutrinos in the final state, such as
\WW$\to \ell^{\pm} \nu$  \qq ~and $W^{\pm} e^{\mp} \nu \to $ \qq  $e^{\mp}\nu$
with the charged lepton escaping detection. The irreducible background to this
search stems from ZZ $ \to \nn \qq $~(about 28\,\% of all ZZ decays) leading
to a signature indistinguishable from a signal event with a Higgs mass close to the Z boson mass. The vector bosons are usually not produced at rest, leading to a transverse momentum of the two-jet system and therefore to a large
acoplanarity of the jets, as in the signal case. Furthermore, the missing
momentum vector points into the central detector more often than for the two-fermion case. To discriminate between this background and the signal one can exploit the difference in the angular distribution of the differential production cross-section.
\subsection{Preselection} 
\begin{figure}
\centering
  \vspace*{-10mm}

\includegraphics[width=.41\textwidth]{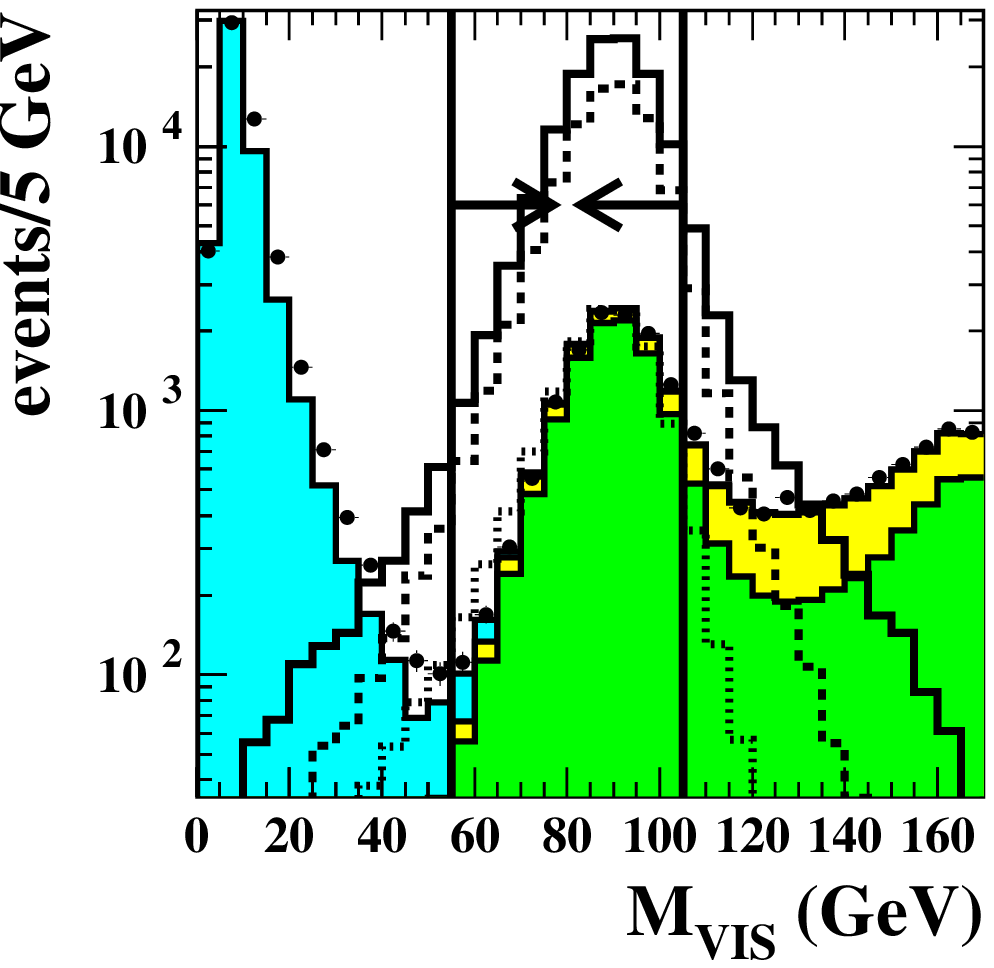}
\includegraphics[width=.41\textwidth]{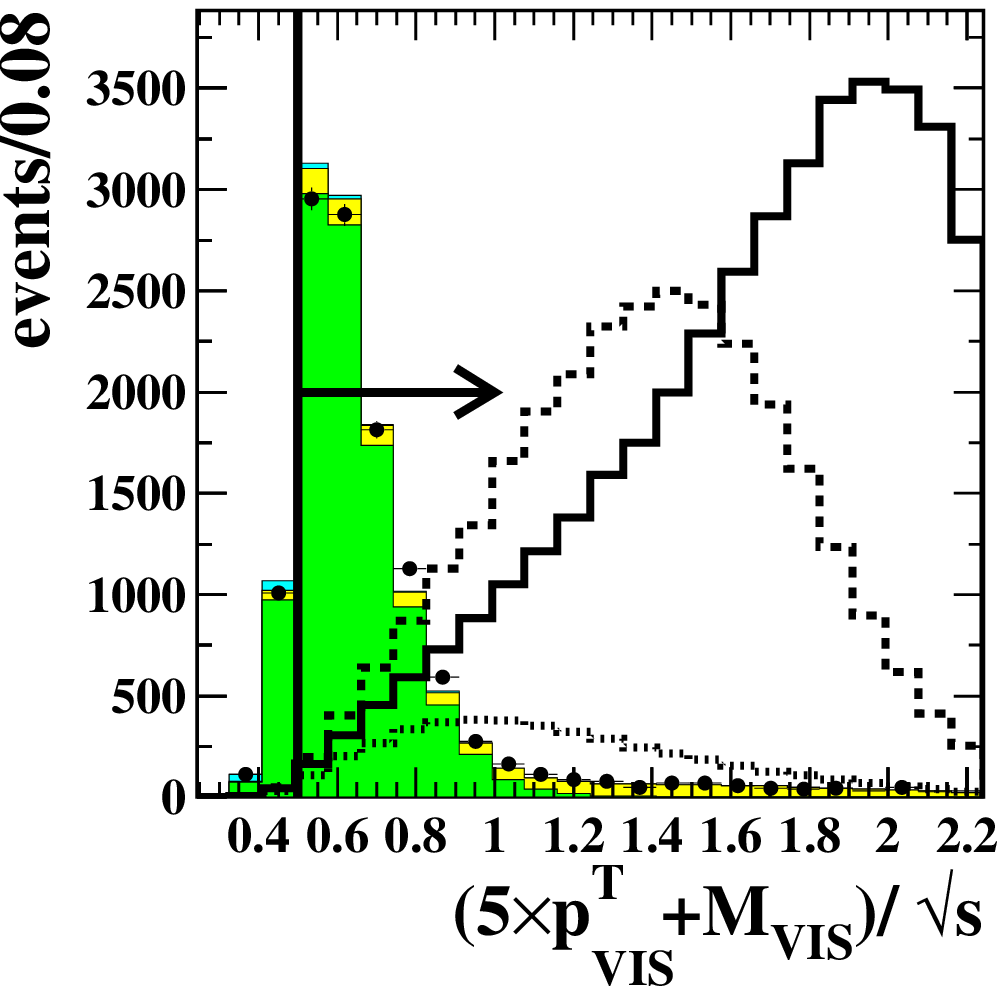}\\  \vspace*{-10mm}
\includegraphics[width=.41\textwidth]{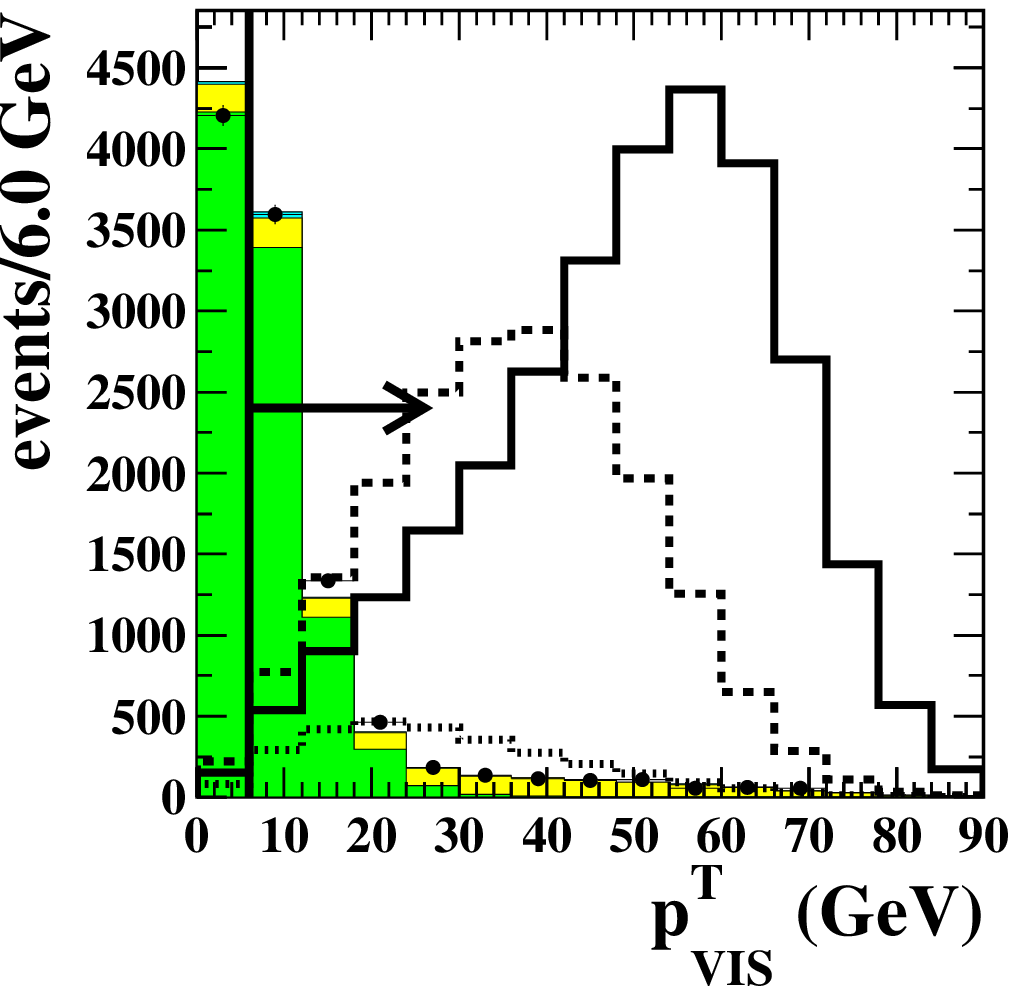}
\includegraphics[width=.41\textwidth]{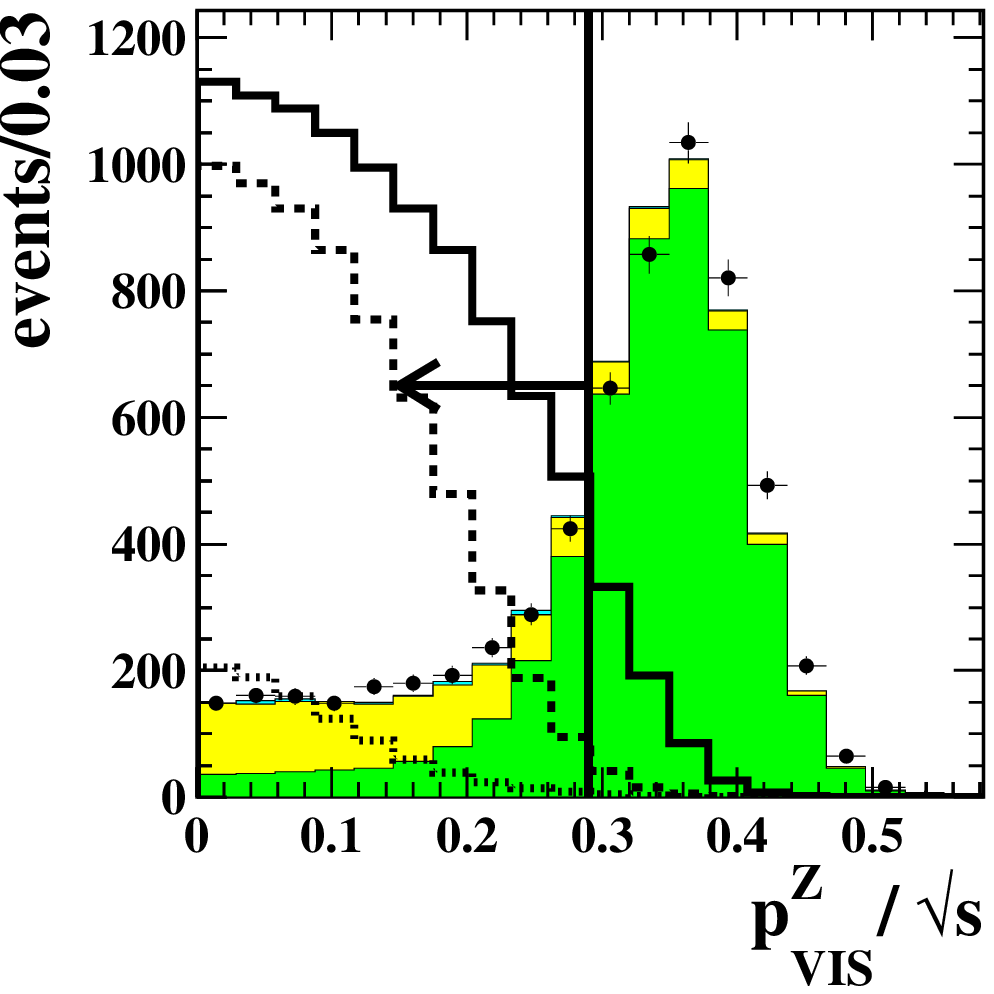}\\  \vspace*{-10mm}
\includegraphics[width=.41\textwidth]{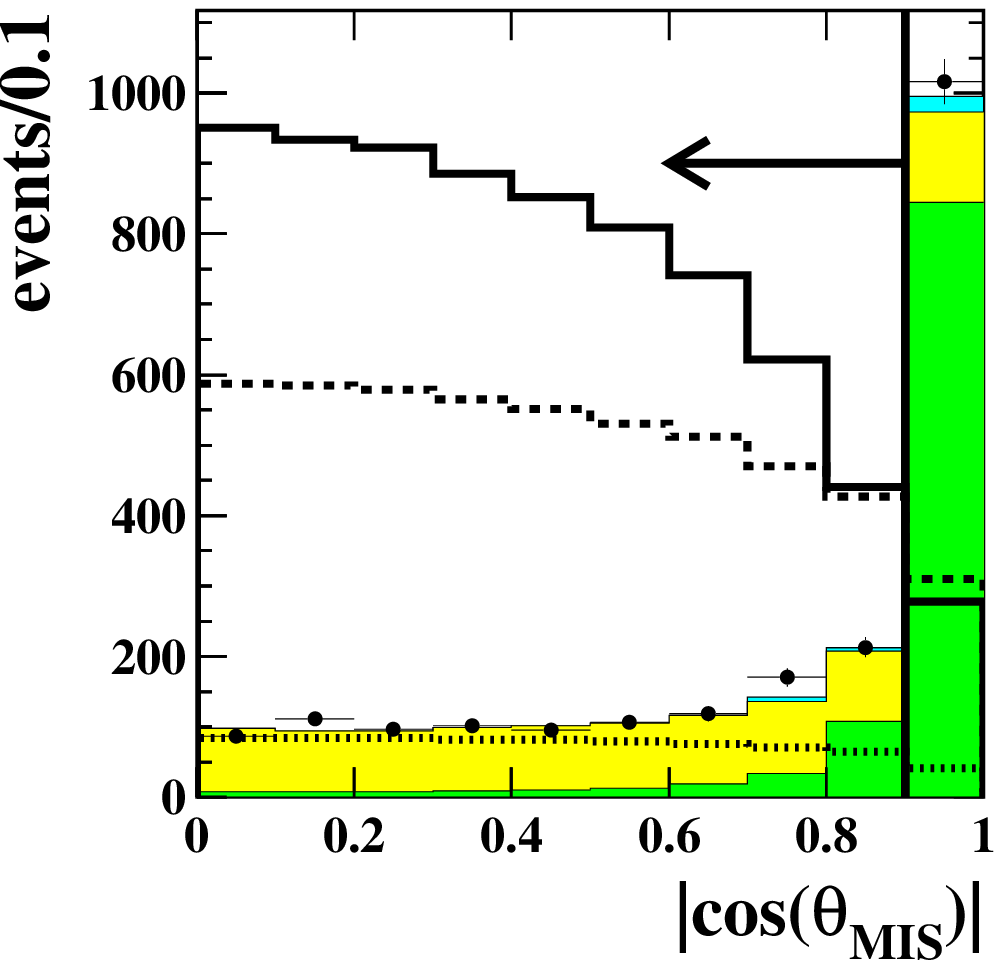}
\includegraphics[width=.41\textwidth]{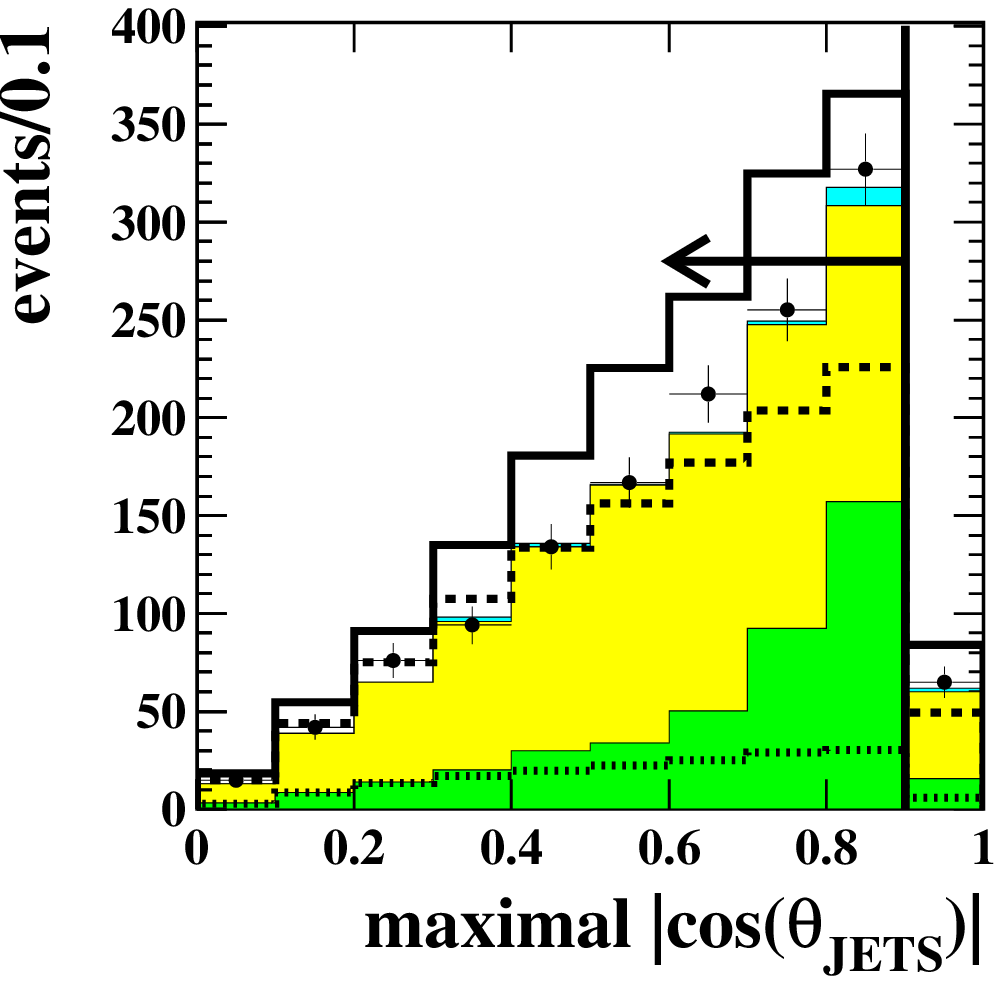}\\  \vspace*{-10mm}
\includegraphics[width=.41\textwidth]{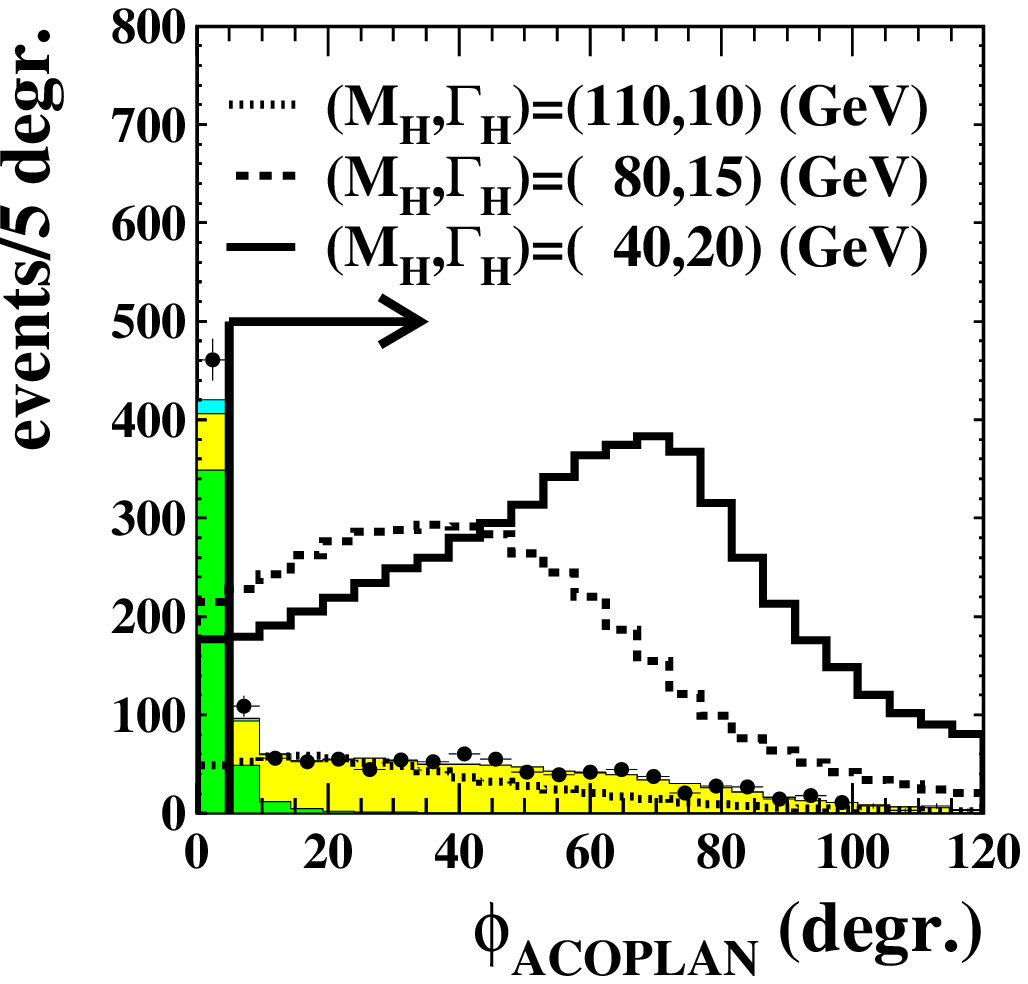}
\includegraphics[width=.41\textwidth]{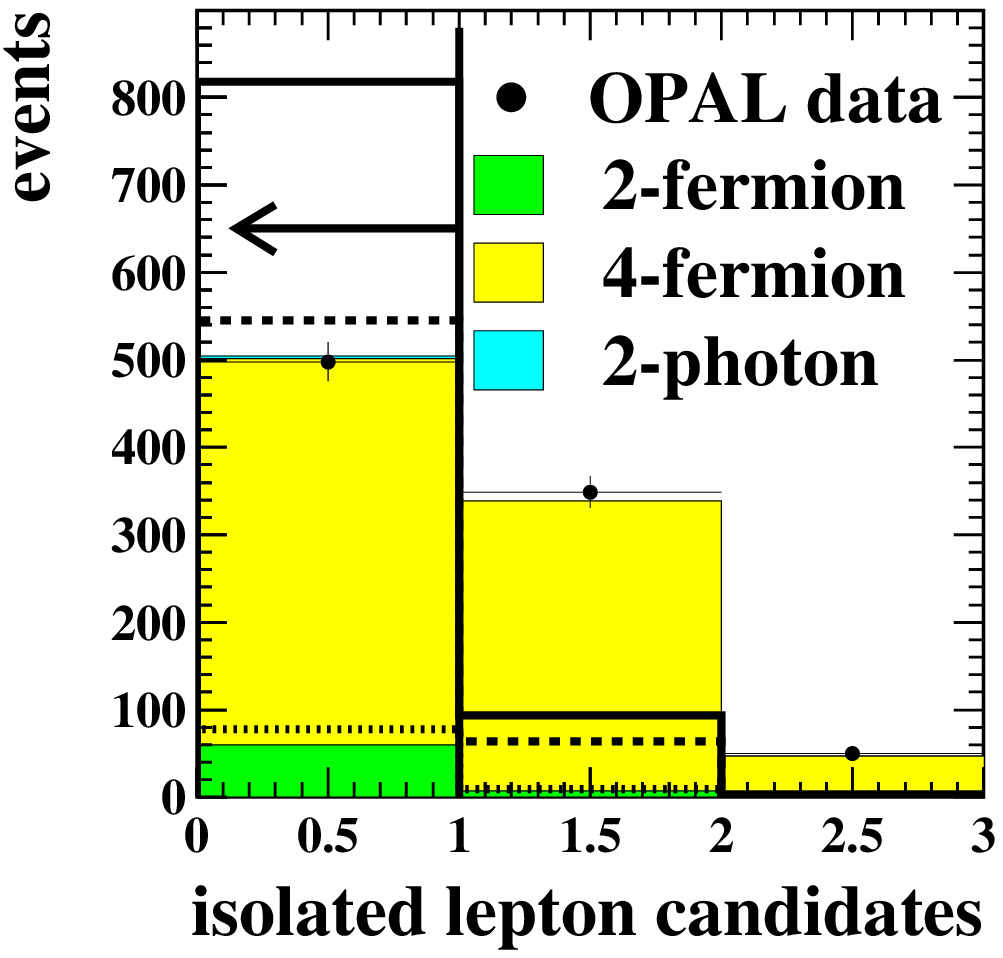}\\ 
\caption{\label{f:cutpix} \sl Distribution of the preselection variables after the preselection cuts (1)-(5).  All classes of \SM background and data are added for all analysed centre-of-mass energies. The distributions of three arbitrarily scaled signal hypotheses at $\sqrt{s} = 206$\,\GeV are displayed as open histograms.}
\end{figure}

In order to reduce the amount of background data only a events fulfilling the following quality criteria are analysed. From cut No.(5) onwards, the cut values were defined using as a guide a simple figure of merit based on the efficiency and expected background.
The following cuts remove almost all the two-photon background:
\begin{itemize}

\item[(1)] To reduce two-photon and accelerator induced background,
 track criteria are applied demanding that more than 20\,\% of all tracks be qualified as good measured tracks \cite{Akrawy:1990bt} and that at least 6 of them be found.
\item[(2)] A forward energy veto rejects events with more than 5\,\GeV in either the left or right compartment of the gamma catcher  calorimeters or the silicon tungsten luminometers. Events with more than 2\,\GeV in the forward calorimeters are also removed.
\item[(3)] The missing transverse momentum $p^T_{\mathrm{MIS}}$ should exceed 1\,\GeV and $M_{\mathrm{VIS}}$ has to be larger than 4\,\GeV.
\item[(4)] Less than 20\,\% of the measured visible energy $E_{\mathrm{VIS}}$ should be located close to the beam pipe in the region $|\cos\theta| > 0.9$.
\item[(5)] The visible energy $E_{\mathrm{VIS}}$ must be less than 90\,\% of $\sqrt{s}$.
\item[(6)] It is required that the visible mass of the event should be of order \Mz, i.e.
$55\,\GeV <  M_{\mathrm{VIS}}  < 105\,\GeV$. An asymmetric cut around the Z mass is chosen, since with increasing Higgs mass \Mh the Z bosons will be more and more off-shell. 
\end{itemize}
\noindent The remaining backgrounds at this stage, which are more difficult to remove, are mismeasured $ \mathrm{Z} / \gamma^* \to \qq $ events, four-fermion processes with neutrinos in the final state, such as
$\WW \to \ell^{\pm} \nu \qq $ ~and $W^{\pm} e^{\mp} \nu \to \qq e^{\mp}\nu$
with the charged lepton escaping detection (see Table \ref{t:preselcuts}).
\begin{itemize}
\item[(7)] To select events that are well measured in the detector with a 
visible mass $ M_{\mathrm{VIS}}$ close to \Mz and a sizeable transverse momentum $p^T_{ \mathrm{VIS}}$ the following criterion is applied: $ M_{\mathrm{VIS}}+ 5 \times p^T_{\mathrm{VIS}} > \sqrt{s}/2 $
\item[(8)]
A large part of the $\qq$ events and the remaining two-photon background is eliminated by requiring the visible transverse
momentum $p^T_{\mathrm{VIS}} > 6$\,\GeV. 
\item[(9)]
To remove backgrounds in which particles go undetected down the
beam pipe, the projection of the visible momentum along the beam
axis, $p^{z}_{\mathrm{VIS}}$, is required to be less than $0.294 \sqrt{s}$.
\item[(10)] To reduce the radiative \qq ($\gamma$) 
  background, the polar angle of the missing momentum vector must lie within
  the region $|\cos \theta_{\mathrm{MIS}}|<0.9$.
\item[(11)] 
  The axes of both jets, reconstructed with the Durham algorithm,
  are required to have a polar angle satisfying $|\cos\theta| < 0.9$
  to ensure good containment. Furthermore this cut exploits the fact that
  events of the WW and ZZ background are produced according to an angular
  distribution proportional to $\cos^2 \theta$.
\item[(12)]  The remaining background from
$\mathrm{Z} / \gamma^* \to \qq $~is characterised 
 by two jets that tend to be back-to-back with small acoplanarity angles, in contrast to signal events in which the jets are expected to have some acoplanarity angle due to the recoiling Higgs boson. Here the acoplanarity 
angle $\phi_{\mathrm{ACOPLAN}}$ is defined as 180$^{\circ} - \phi_{jj}$ 
where $\phi_{jj}$ is the angle between the two jets in the plane perpendicular to the beam axis.
 This background is suppressed by requiring that the jet-jet acoplanarity angle be larger than $5^{\circ}$. 
\item[(13)]
\WW~events with one of the W bosons decaying leptonically and the other decaying into hadronic jets are rejected by requiring that the events have no isolated leptons.
In this context, leptons are low-multiplicity jets with one, two or three tracks, associated to electromagnetic or hadronic energy clusters, having an invariant mass of less than 2.5\,\GeV and momentum in excess of 5\,\GeV. In the case of a single-track candidate, the lepton is considered isolated if there are no additional tracks within an isolation cone of 25$^{\circ}$ half-angle, and if the electromagnetic energy contained between cones of 5$^{\circ}$ and 25$^{\circ}$ half-angle around the track does not exceed 5\,\% of the sum of the track energy and the electromagnetic energy within the 5$^{\circ}$ half angle cone. In the case of a two- or three-track candidate, consisting of the tracks and electromagnetic or hadronic energy clusters confined to a cone of 7$^{\circ}$ half-angle, the lepton is considered isolated if the sum of track and electromagnetic energy between the 7$^{\circ}$ half-angle cone and a 25$^{\circ}$ half-angle isolation cone does not exceed 15\,\% of the lepton energy.
\end{itemize}
For each individual centre-of-mass energy there is good agreement between the numbers of 
expected background events and observed candidates after the preselection. 
Table \ref{t:preselcuts} gives the number of preselected events summed over all centre-of-mass energies. Figure \ref{f:cutpix} shows the distributions for background classes summed over all centre-of-mass energies and three arbitrarily scaled signal distributions (at a centre-of-mass energy of 206\,\GeV). The efficiencies of the preselection vary on average between 39\,\% and 55\,\% for small decay widths and between 45\,\% and 53\,\% for larger decay widths above \Gh = 100\,\GeV.
\subsection{Likelihood analysis}
To consider the changing kinematic properties of the signal hypotheses in an optimal way, five different likelihood-based analyses for the signal and background discrimination were applied after the preselection.
By a likelihood analysis we denote the combination of a set of likelihood input variables, a so-called likelihood, and the corresponding reference distributions of these variables. The reference distributions are filled with events of the specific classes for which the likelihood is calculated. The classes considered in this search are the two- and four-fermion backgrounds and the signal events. The two-photon events are negligible after the preselection. The search uses combinations of two likelihoods and three fixed signal mass ranges for unweighted reference histograms.\par
To compare the kinematic properties of a selected data event to the hypothesis (\Mh,\Gh) when evaluating the likelihood, one in principle has to fill weighted signal reference distributions for each hypothesis (\Mh,\Gh). This will soon lead to an unmanageable technical effort, given the number of hypotheses scanned. Therefore a compromise was sought in which certain kinematic properties of the signal were emphasised and simultaneously the number of reference histograms kept small. This was achieved by filling unweighted signal reference histograms. 
For most of the (\Mh,\Gh) hypotheses all signal masses were used for filling the reference histograms. This reflects the fact that for a very large decay width  of the Higgs boson the possible values of kinematical variables are also smeared out over a large range. It was, however, found that the sensitivity of the likelihood selection (i.e. the median expected upper limits on \mysec) could be increased further for small widths below 50\,\GeV by filling reference histograms with signal masses from 50-80\,\GeV and from 80-120\,\GeV for intermediate and heavy Higgs boson masses respectively.
~A first likelihood was designed for a signal consisting of small masses (\Mh < 80\,\GeV) or large masses and a very large width (\Gh $\geq 110$\,\GeV). In this likelihood input variables are used exploiting the characteristics of the dominant fraction of light masses in the signal mass distribution. However for signal masses above \Mh = 80\,\GeV and small or moderate (i.e. below 110\,\GeV) decay widths, the contribution of large masses dominates the signal mass distribution. In this case the kinematics and topology of the signal events are determined by the higher masses close to the kinematic limit. A second likelihood is therefore built with input variables optimised for such signal characteristics.  
 In the following the choice of the inputs for the two optimised likelihoods are presented.\par
\begin{figure}
  \centering
\includegraphics[width=.45\textwidth]{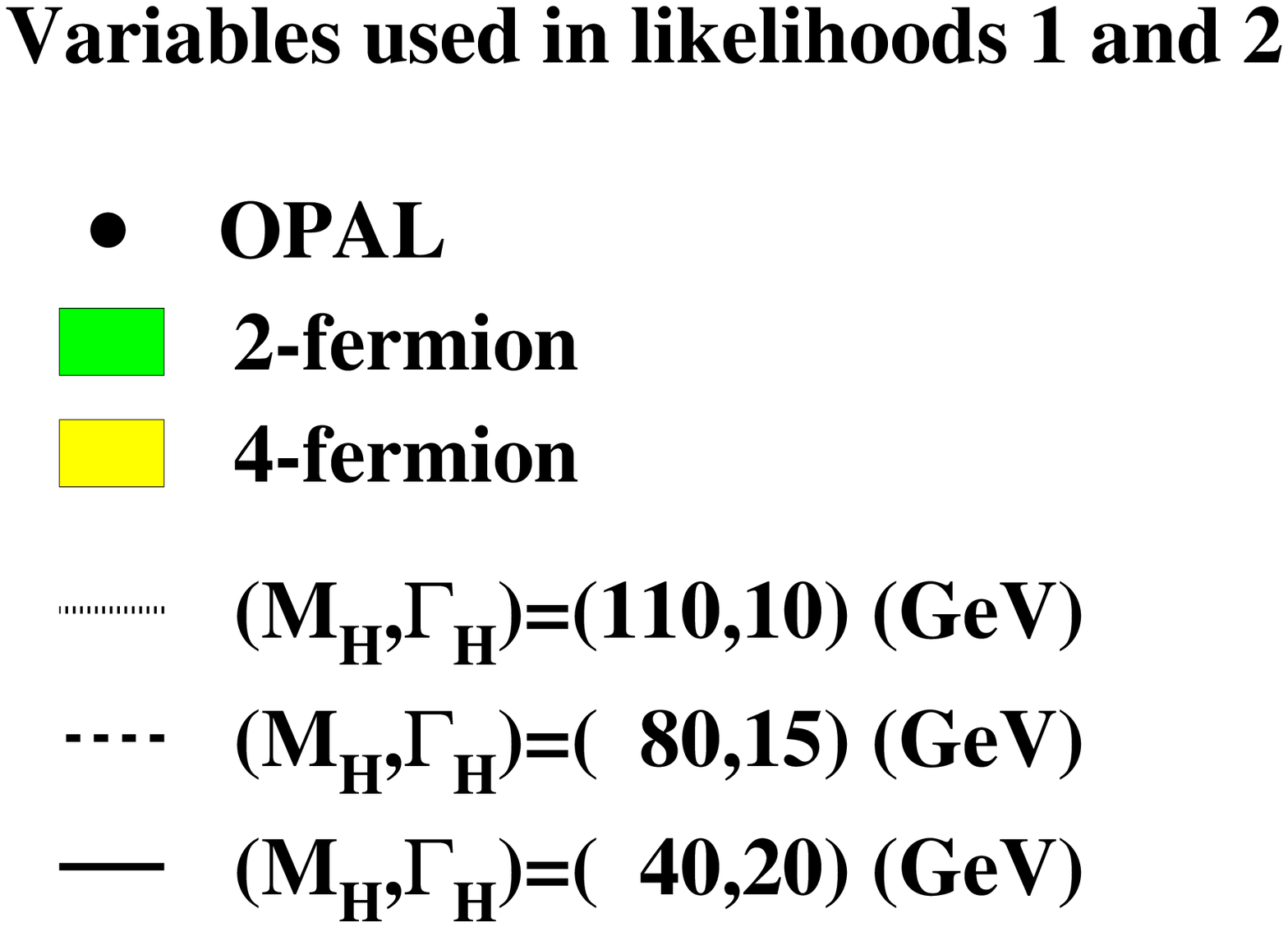}
\includegraphics[width=.5\textwidth]{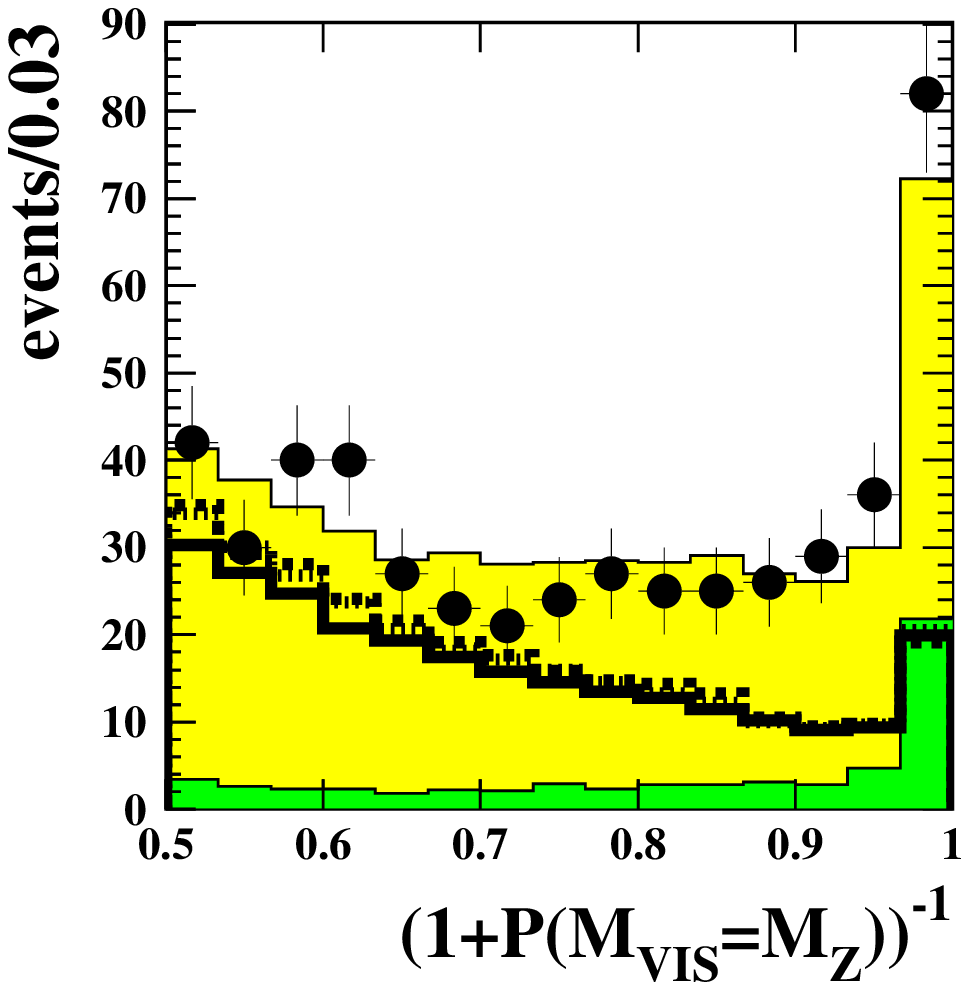}\\  \vspace*{-3mm} \hspace{-9mm}
\includegraphics[width=.5\textwidth]{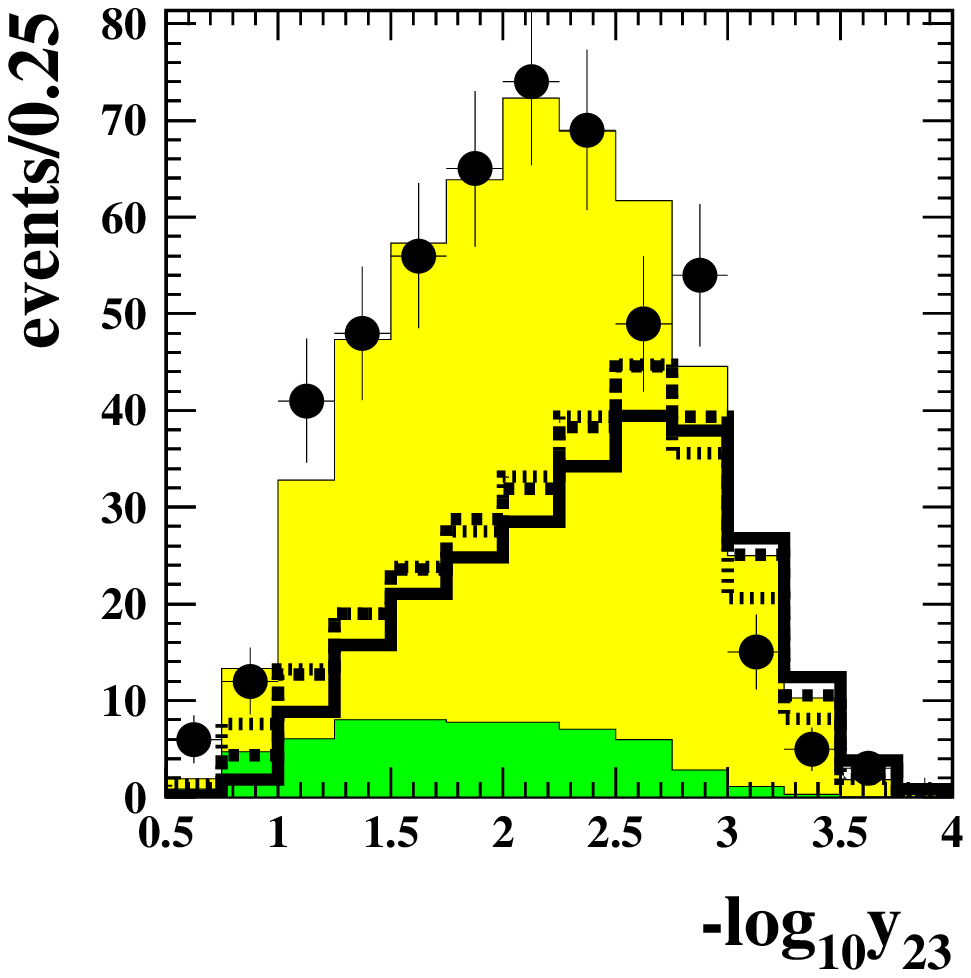}
\includegraphics[width=.5\textwidth]{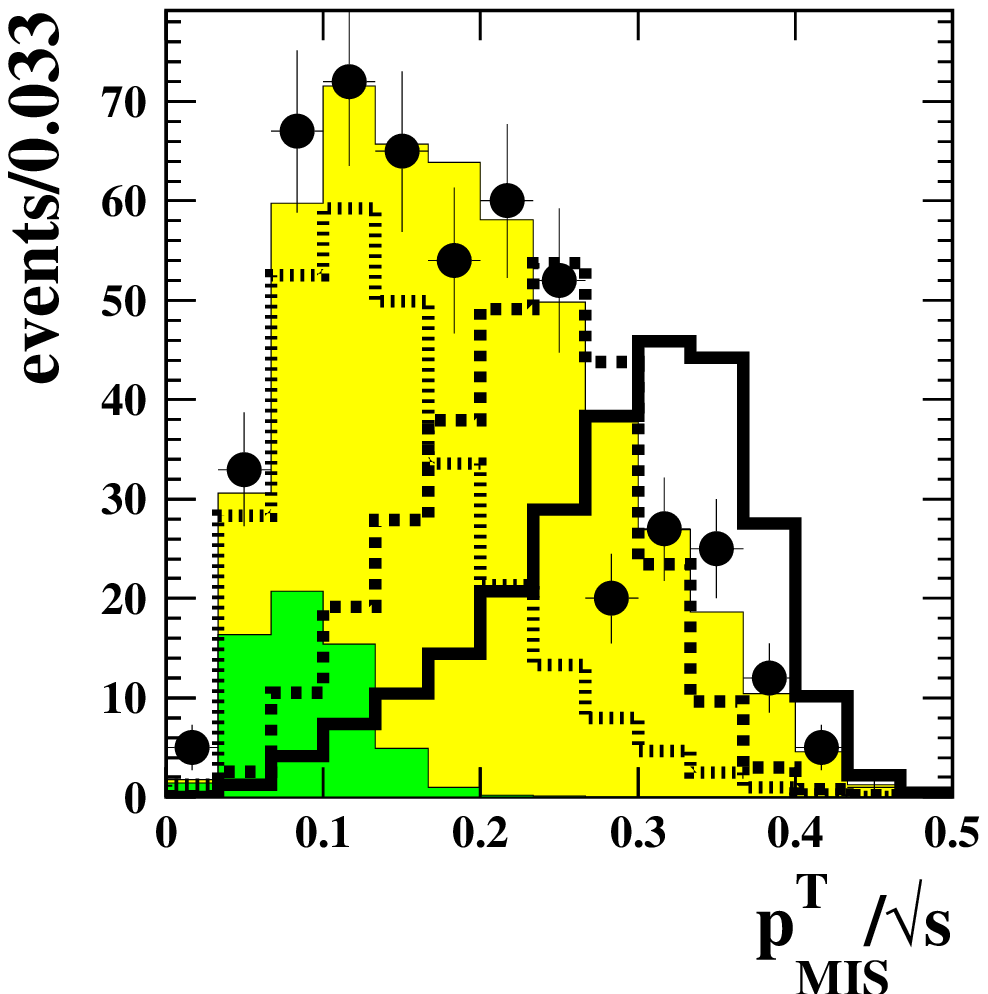}
   \caption{\label{f:inputsLH12} \sl Distributions of the likelihood variables. All classes of \SM background and data are added for all centre-of-mass energies analysed. The distributions of three arbitrary scaled signal examples at $\sqrt{s} = 206\,\GeV$ are displayed as open histograms. The variables shown contribute to likelihood 1 and 2 as they exploit general properties of the signal signature.}
\end{figure}
\begin{figure}
  \centering
\includegraphics[width=.45\textwidth]{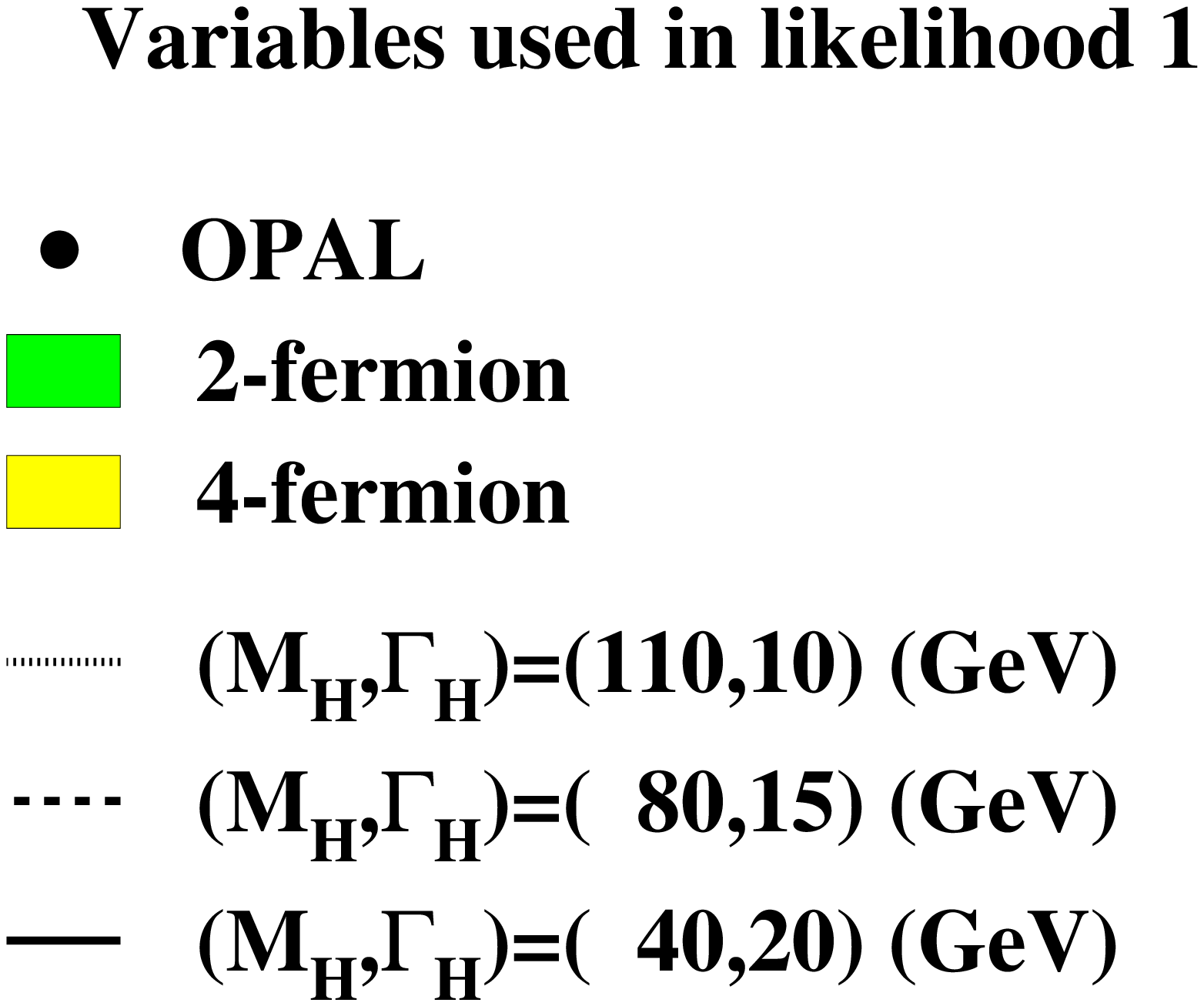}
\includegraphics[width=.5\textwidth]{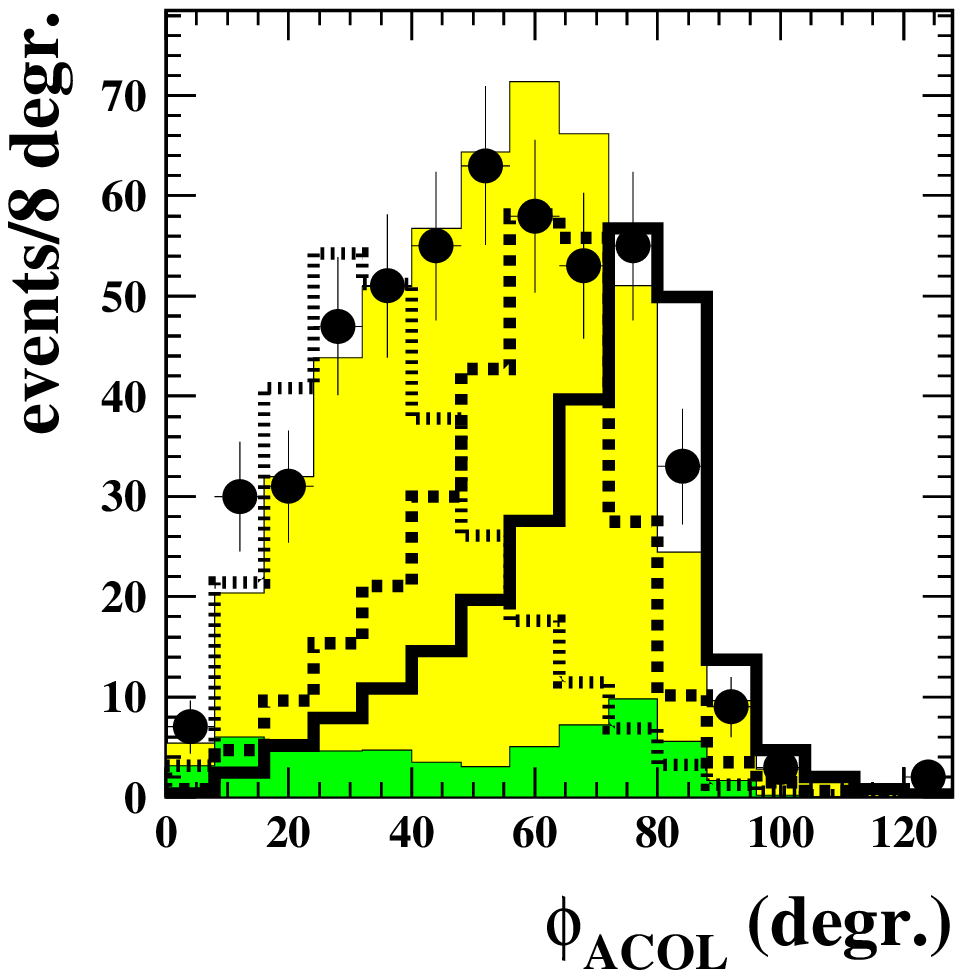}\\  \vspace*{-3mm} \hspace{-9mm}
\includegraphics[width=.5\textwidth]{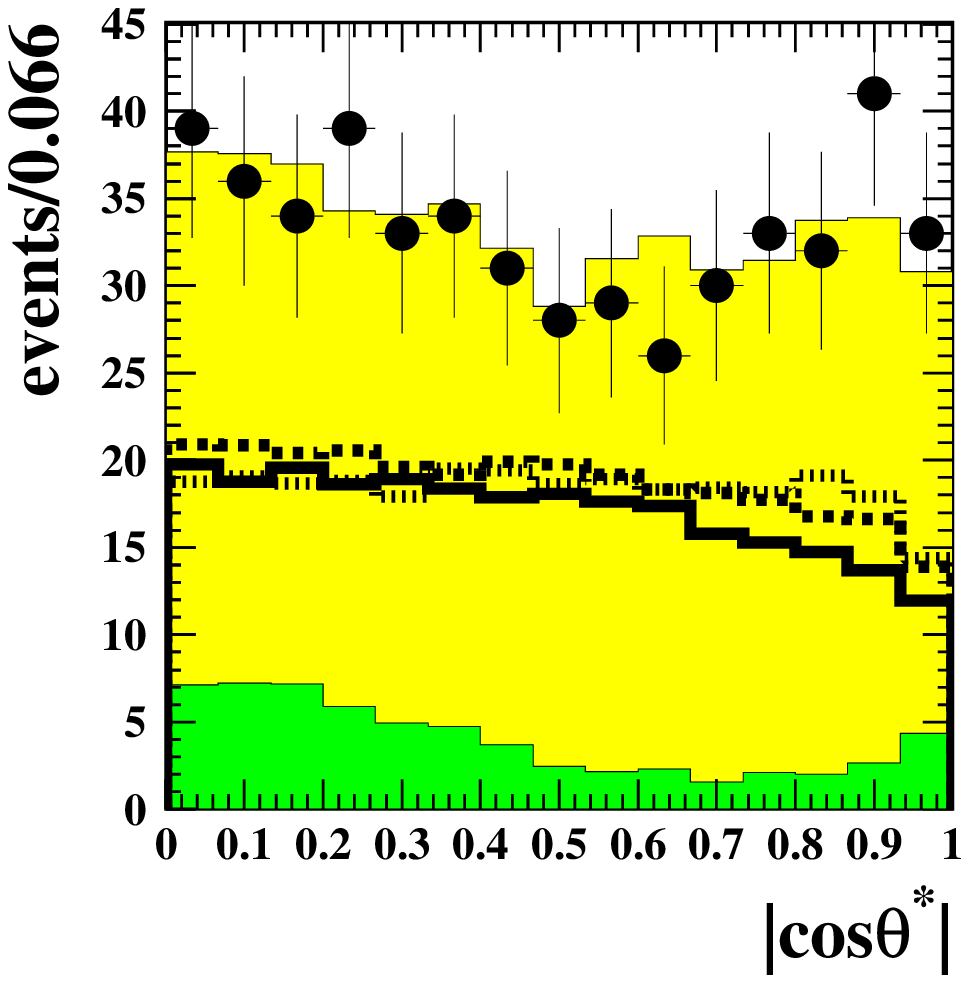}
\includegraphics[width=.5\textwidth]{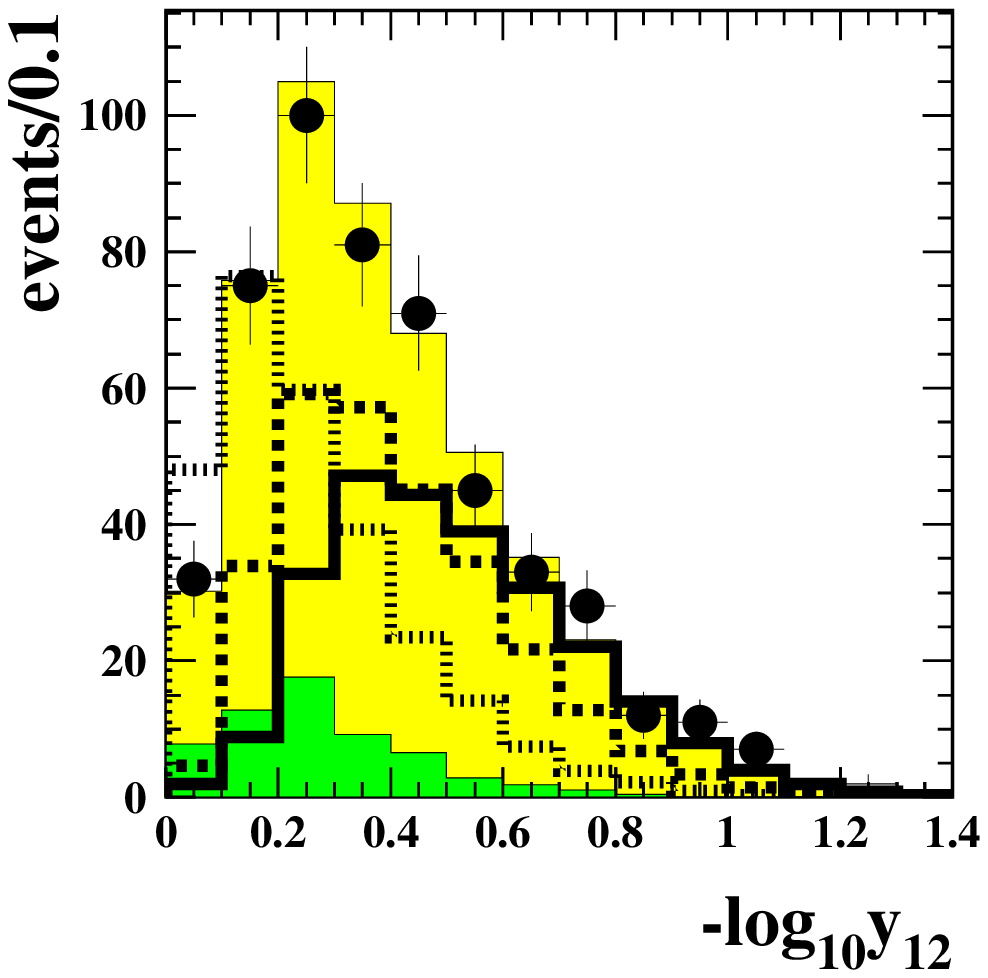}
   \caption{\label{f:inputsLH1} \sl Distributions of the likelihood variables. All classes of \SM background and data are added for all centre-of-mass energies analysed. The distributions of three arbitrary scaled signal examples at $\sqrt{s} = 206\,\GeV$ are displayed as open histograms. The variables shown are combined with the ones of Figure \ref {f:inputsLH12} to construct the likelihood 1 used in a general search at different \Mh and \Gh.}  
\end{figure}
\begin{figure}
  \centering
\includegraphics[width=.45\textwidth]{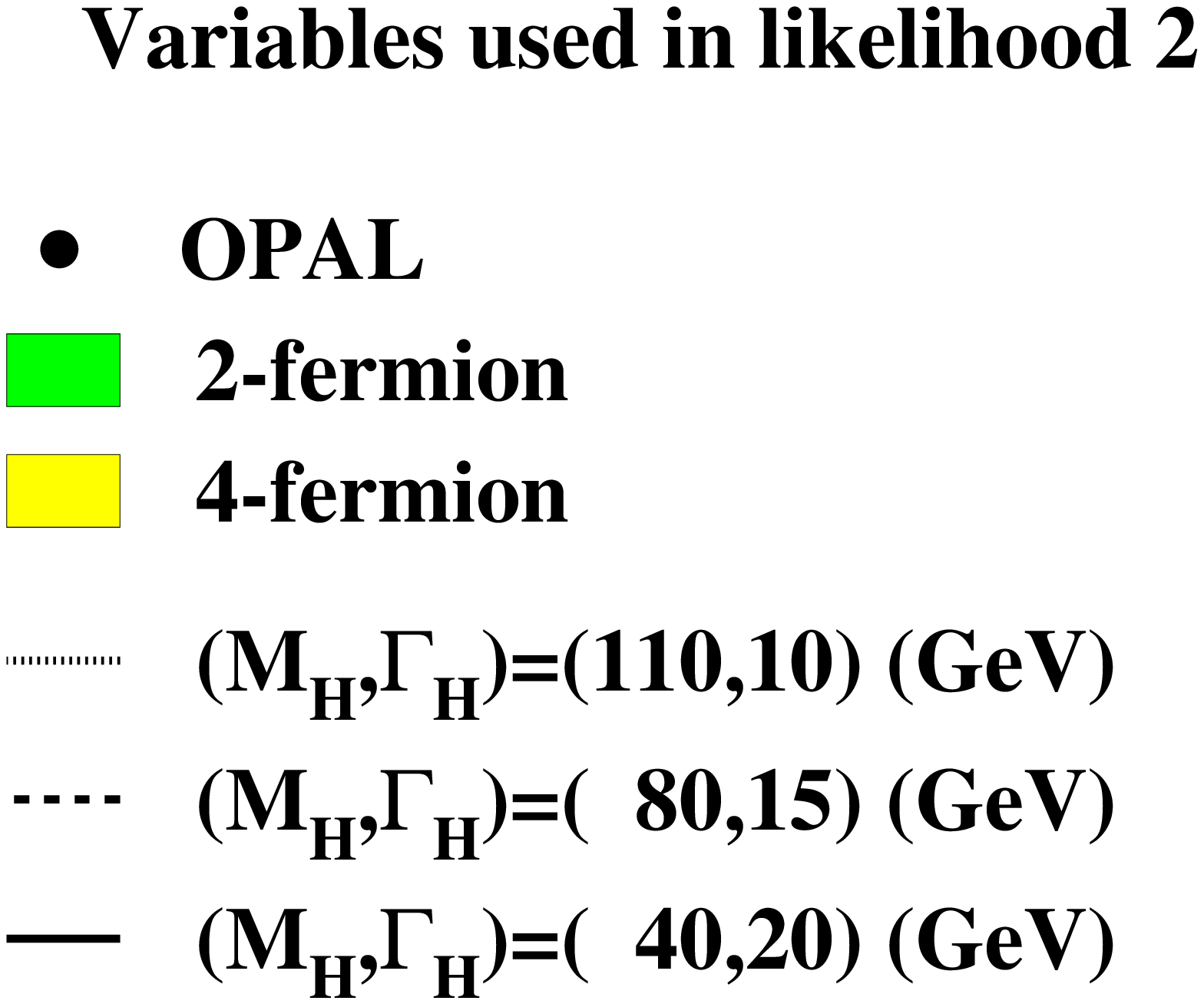}
\includegraphics[width=.5\textwidth]{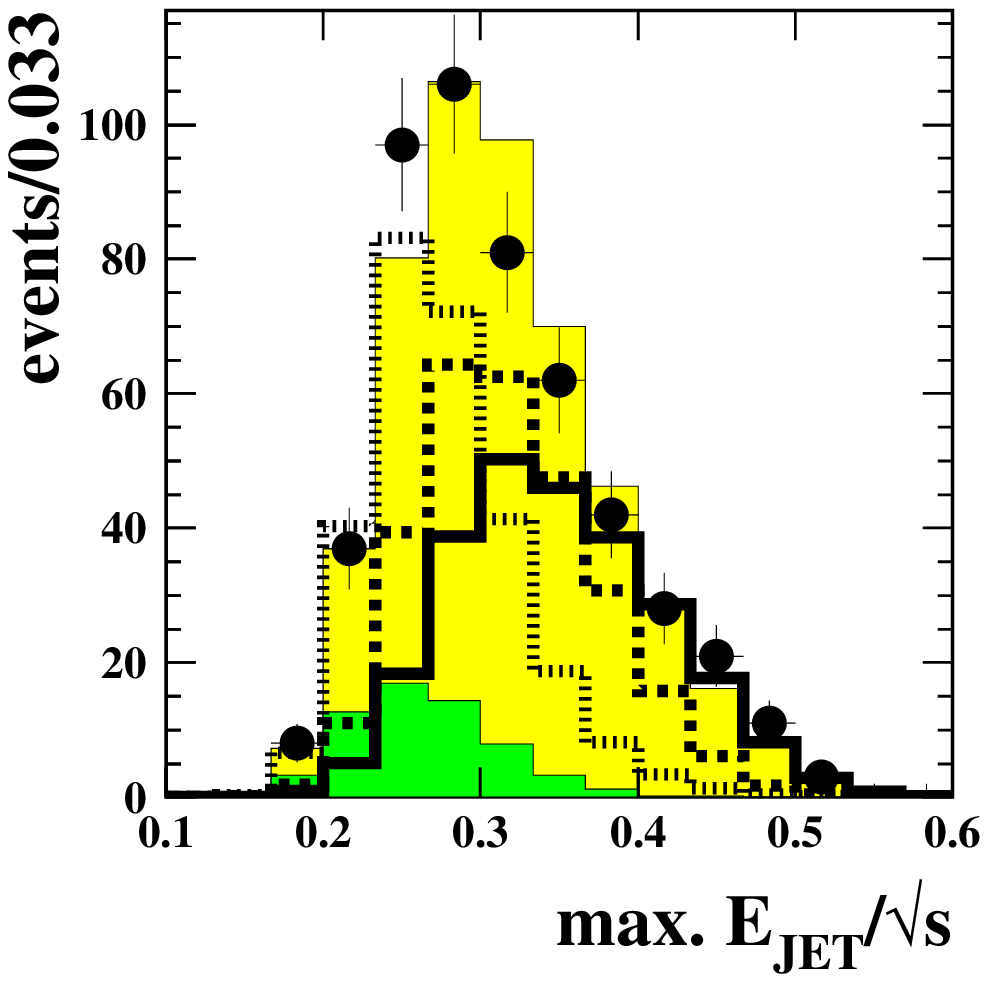}\\  \vspace*{-3mm} \hspace{-9mm}
\includegraphics[width=.5\textwidth]{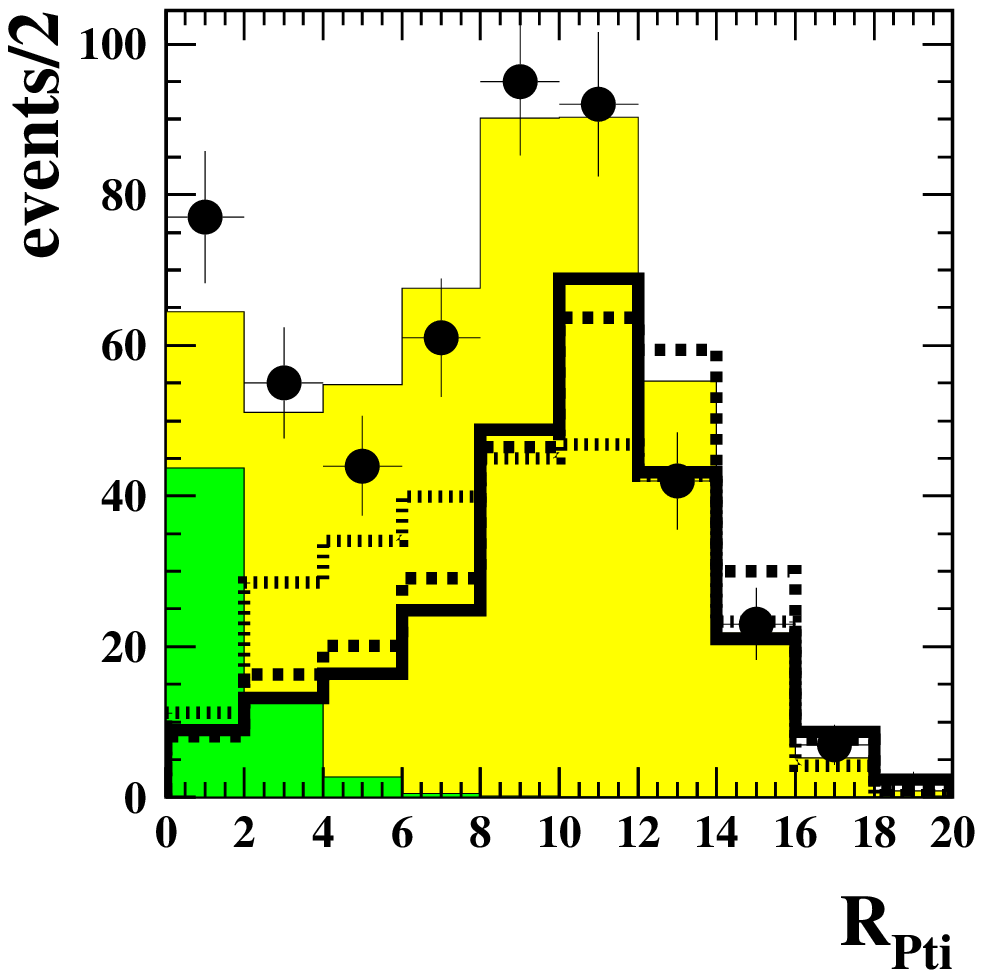}
\includegraphics[width=.5\textwidth]{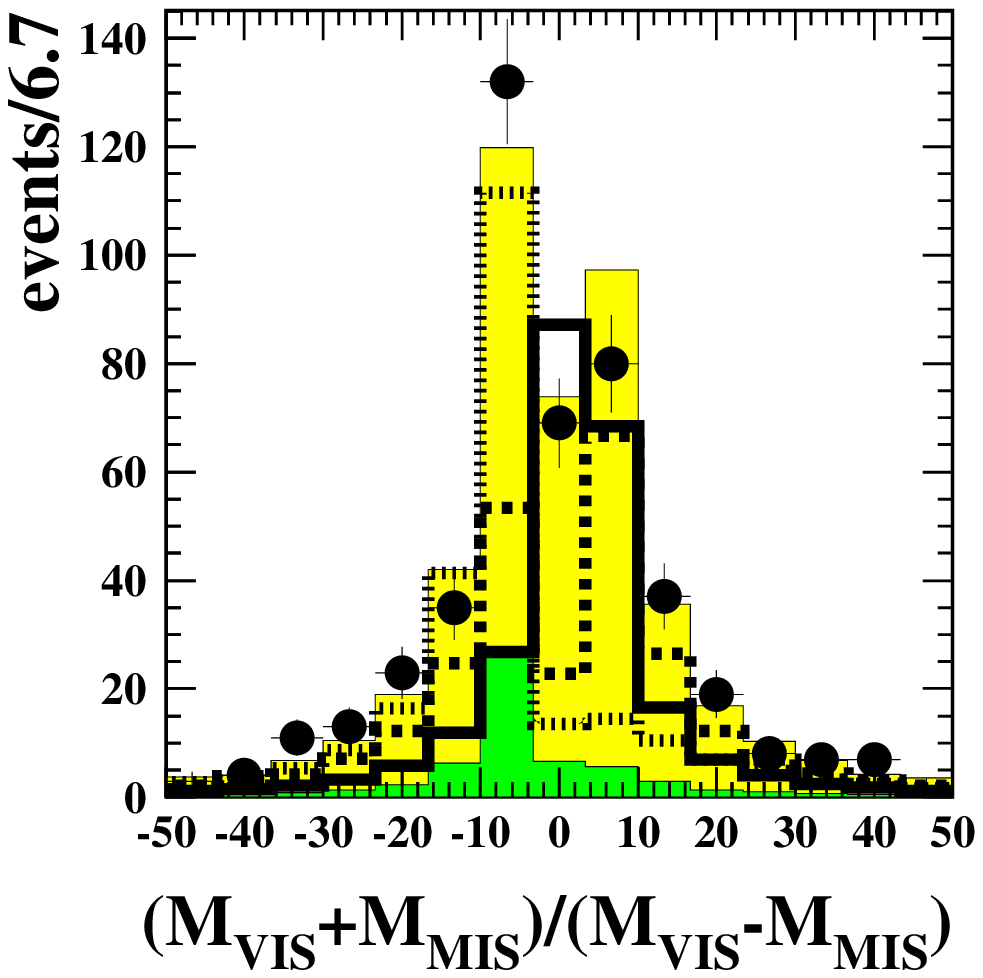}
   \caption{\label{f:inputsLH2} \sl Distributions of the likelihood variables. All classes of \SM background and data are added for all centre-of-mass energies analysed. The distributions of three arbitrary scaled signal examples at $\sqrt{s} = 206\,\GeV$ are displayed as open histograms. The variables shown have a larger discrimination power for a heavier Higgs boson and contribute with the variables of Figure \ref {f:inputsLH12} to the second likelihood.}  
\end{figure}

The first three input variables are used in both likelihoods (see Figure \ref{f:inputsLH12}) .\newline   
\begin{itemize}
\item[(1)] {\bf $(1+ {\cal P}(M_{\mathrm{VIS}}\equiv \mathrm{M_{Z}}))^{-1}$} \newline
  ${\cal P}(M_{\mathrm{VIS}}\equiv \mathrm{M_{Z}})$ is the probability of a kinematical $\chi^{\mathrm{2}}$ fit of the jet four-vectors under the assumption that the invariant mass of the two jets is compatible with the Z boson mass. The uncertainties on the measured jet energies are of the order of 5-10\,GeV,
 while the jet directions are measured to approximately 1-2$^{\circ}$\cite{Abbiendi:1998qs}. This variable depends only weakly on the Higgs mass. For events with non-converging fit the probability is set to zero. They therefore accumulate at a value of 1. 
\item[(2)]  {\bf$-\log y_{32}$}\label{jetaufl}\newline
The Durham algorithm groups two energy flow objects $i$ and $j$ into a jet as long as their separation in phase space $y_{ij} = 2\times min(E^2_i,E^2_j) \times (1-\cos(\theta_{ij})/E^2_{\mathrm{VIS}}$ is smaller then the cut value $y_{\mathrm{cut}}$. The number of jets in a event is predefined to be 2, $y_{32}$ is the value of $y_{\mathrm{cut}}$ where the two-jet topology of the event changes to a three-jet topology. Hence the negative logarithm of the so-called jet resolution parameter $y_{32}$ is a measure for the jet topology being more two-jet like (large value of $-\log y_{32}$) or three-jet like (small value of $-\log y_{32}$).
\item[(3)]  {\bf$p^{T}_{\mathrm{MIS}}/\sqrt{s}$}\newline
The transverse missing momentum $p^{T}_{\mathrm{MIS}}$ is one of the most
prominent characteristics of signal-like events, but depends very much on the
Higgs boson mass. For a heavy Higgs boson produced close to the kinematic threshold almost at rest, the Z boson has almost no boost and decays into two more or less back-to-back jets. In this case the discriminating power of the variable is lost.
\end{itemize}
The next three variables (see Figure \ref{f:inputsLH1}) complete the first likelihood, which is used for all Higgs masses in the domain of very large width > 110\,\GeV or low Higgs masses < 80\,\GeV. 
\begin{itemize}
\item[(4)]  {\bf$\phi_{\mathrm{ACOL}}$} \newline
The acolinearity angle $\phi_{\mathrm{ACOL}}$  of the two-jet system is obtained by  subtracting the three-dimensional angle between the reconstructed jet-axes from $180^{\circ}$.
Events containing a low-mass Higgs boson exhibit on average a larger acolinearity than the background.
\item[(5)]  {\bf$|\cos\theta^{\ast}|$}\newline
The Gottfried-Jackson angle $\theta^{\ast}$, is defined as the angle between the flight direction of the Z boson in the laboratory frame and the direction of the decay products of the Z boson boosted into the Z boson rest-frame. The variable tends to have smaller values for the signal. 
\item[(6)]  {\bf$-\log y_{21}$}\newline
The variable $-\log y_{21}$ is analogous to  $-\log y_{32}$ and measures the compatibility of the event with a two-jet topology. Two-jet events tend to accumulate at small values of $-\log y_{21}$. \newline
\end{itemize}
The last three variables (see Figure \ref{f:inputsLH2}) tune the second likelihood to become more sensitive for large Higgs boson masses and small to moderate widths.\par
\begin{figure}
  \centering
  \includegraphics[width=1.0\textwidth]{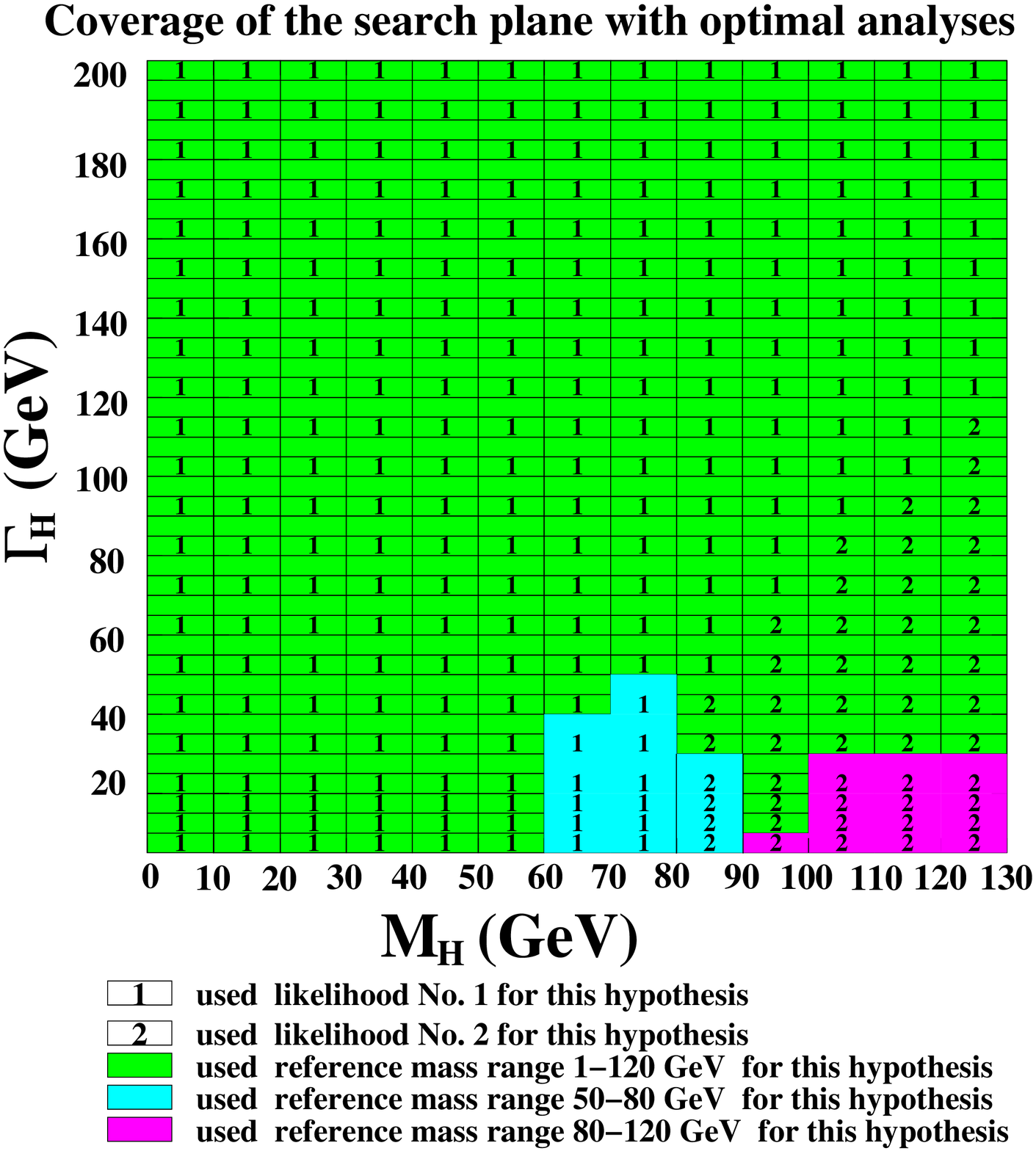}
 \caption{\label{f:pattern}\sl In total five analyses were used to cover the plane of hypothetical Higgs mass and decay width pairings. The analyses differ in whether the first or second likelihood was used (denoted by the number in the cell) and what signal masses where used in filling the reference histograms (depicted by the shading of the cell). The pattern resulted from an optimisation starting with \Gh = 5 \GeV up to 50 \GeV. Below \Gh = 5 \GeV the pattern was simply continued and not optimised anymore. Above 50 \GeV a simple continuation of the pattern was found and proved to be sufficiently sensitive.}
\end{figure}
\begin{figure}
  \centering
  \vspace*{-10mm}

\includegraphics[width=.4\textwidth]{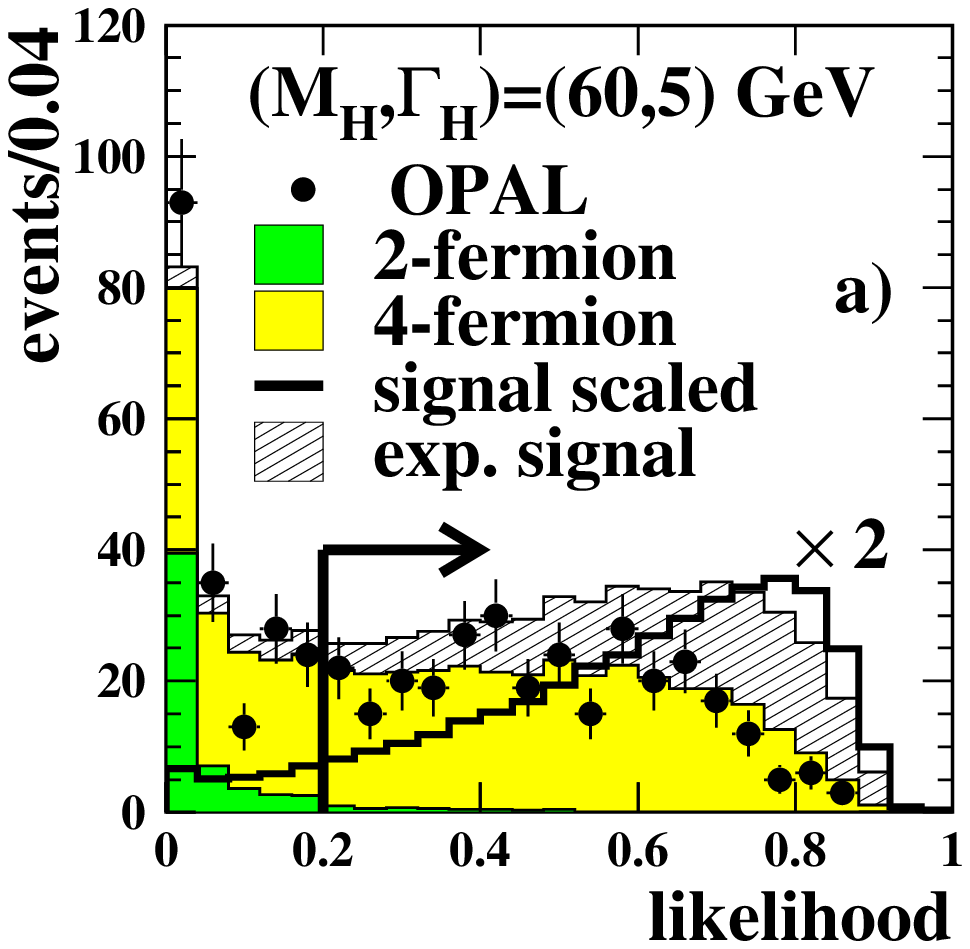}
\includegraphics[width=.4\textwidth]{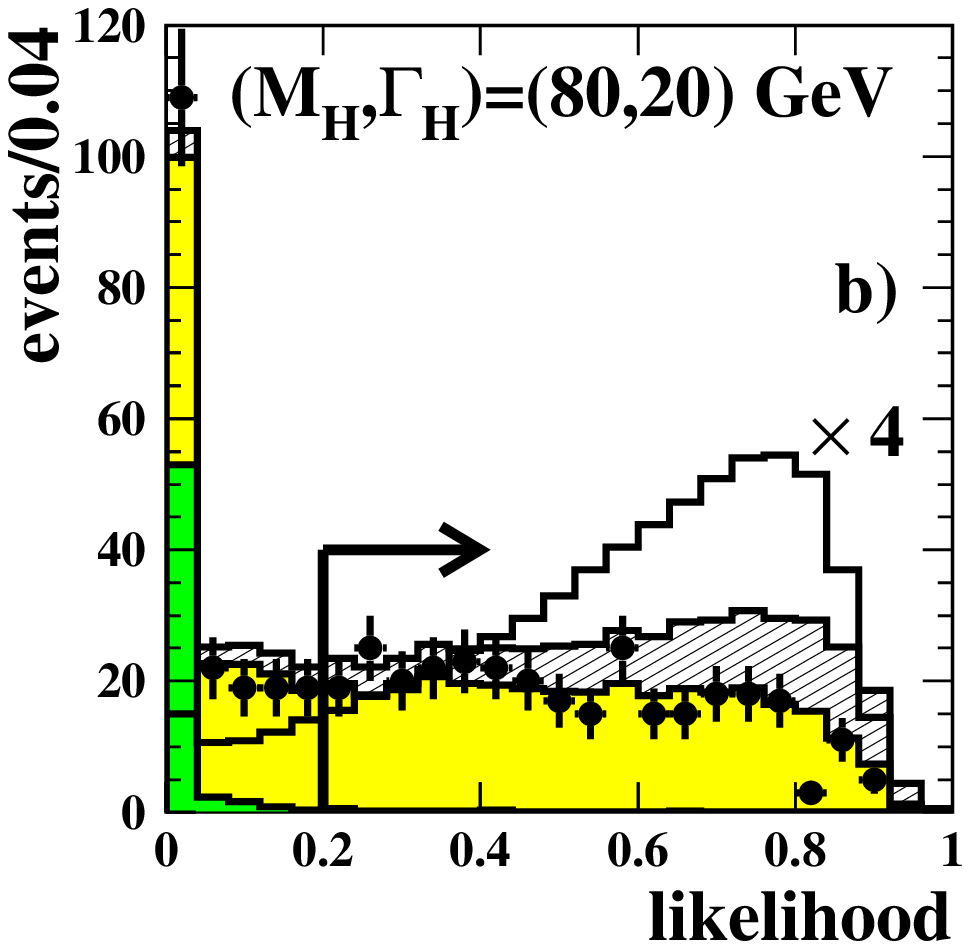}\\  \vspace*{-12mm}
\includegraphics[width=.4\textwidth]{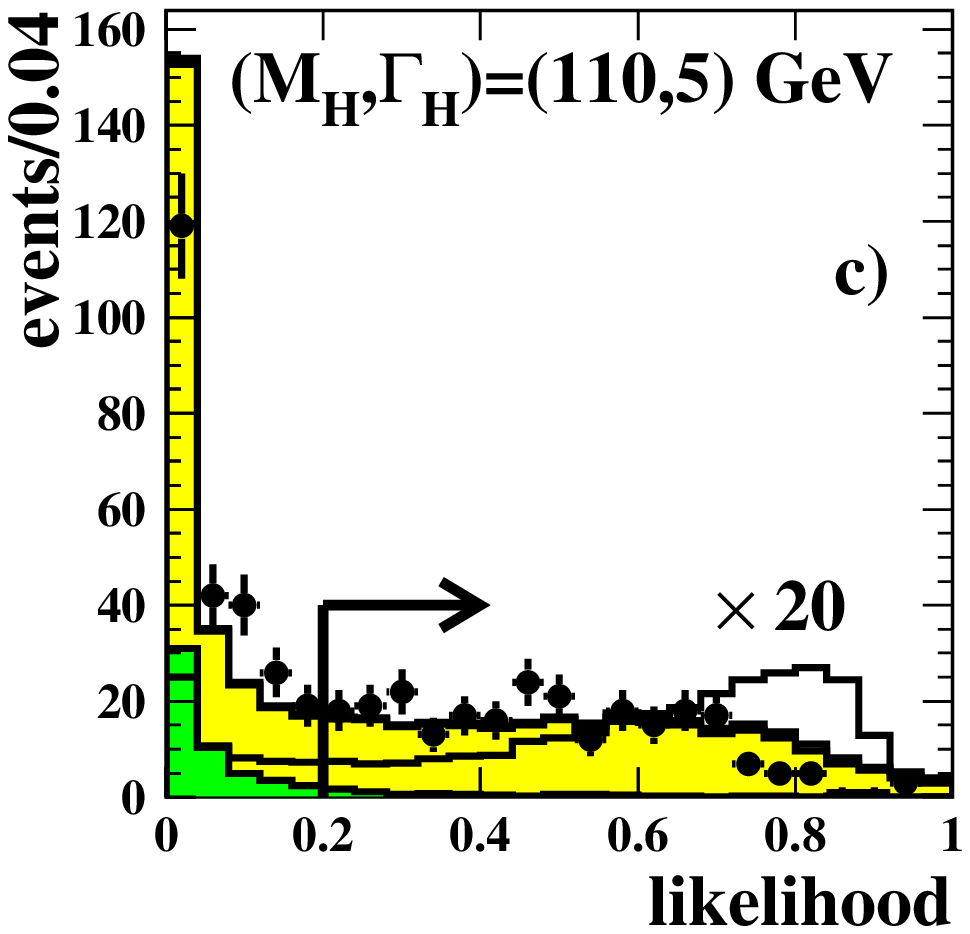}
\includegraphics[width=.4\textwidth]{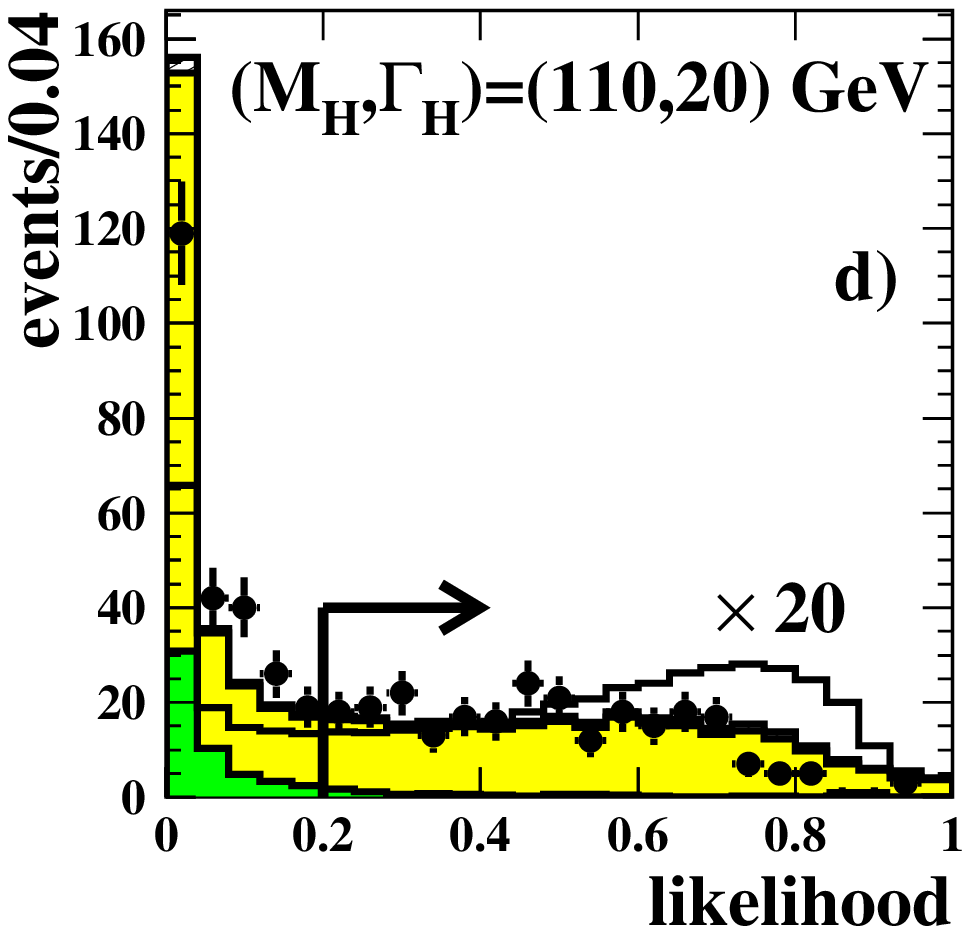}\\  \vspace*{-12mm}
\includegraphics[width=.4\textwidth]{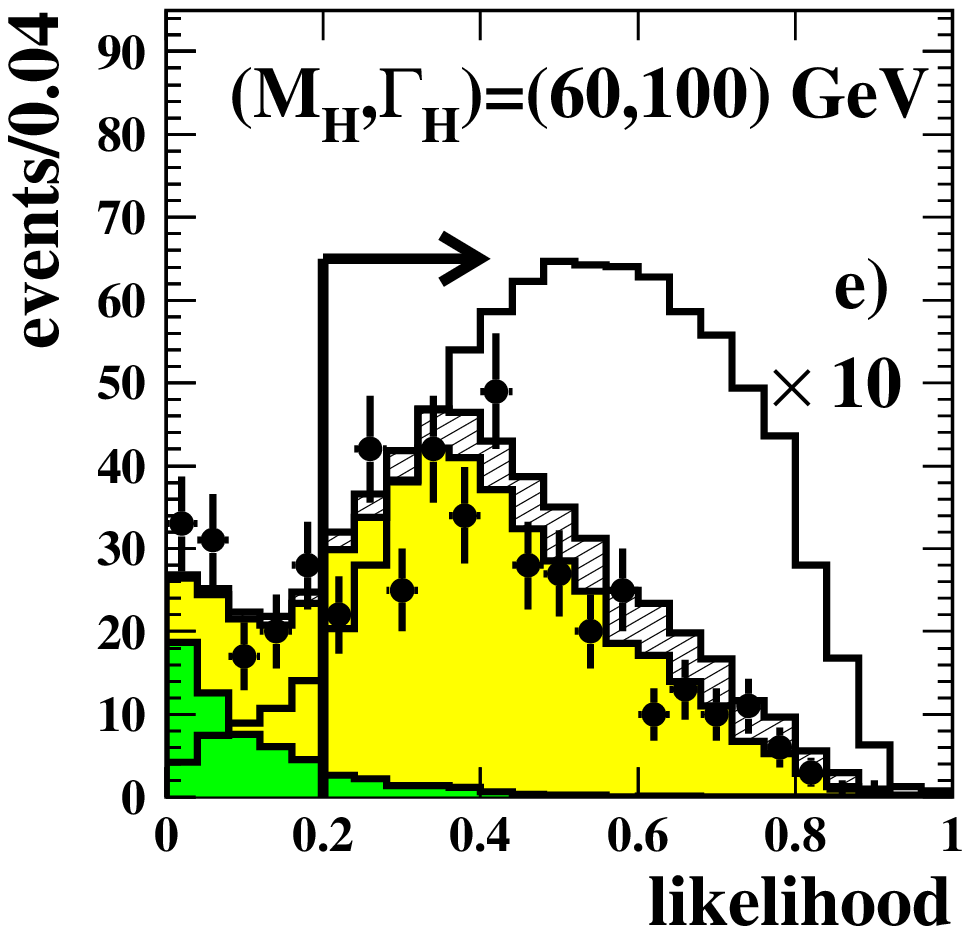}
\includegraphics[width=.4\textwidth]{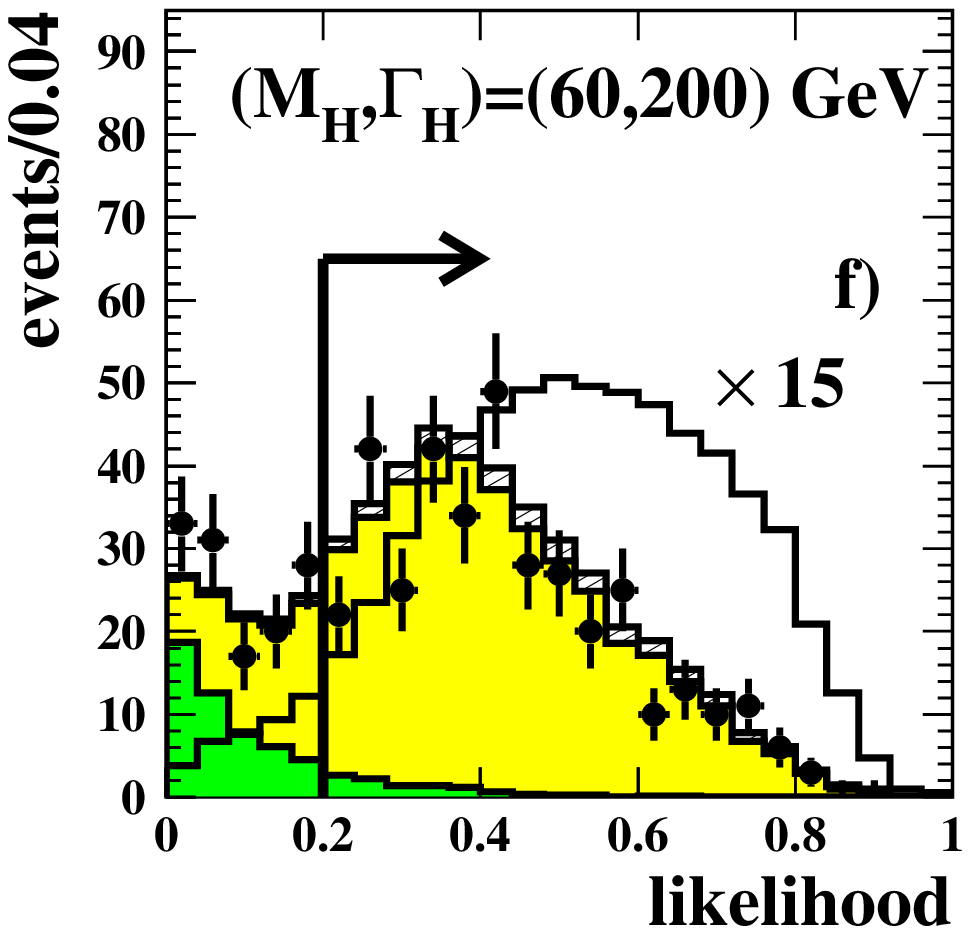}\\  \vspace*{-12mm}
\includegraphics[width=.4\textwidth]{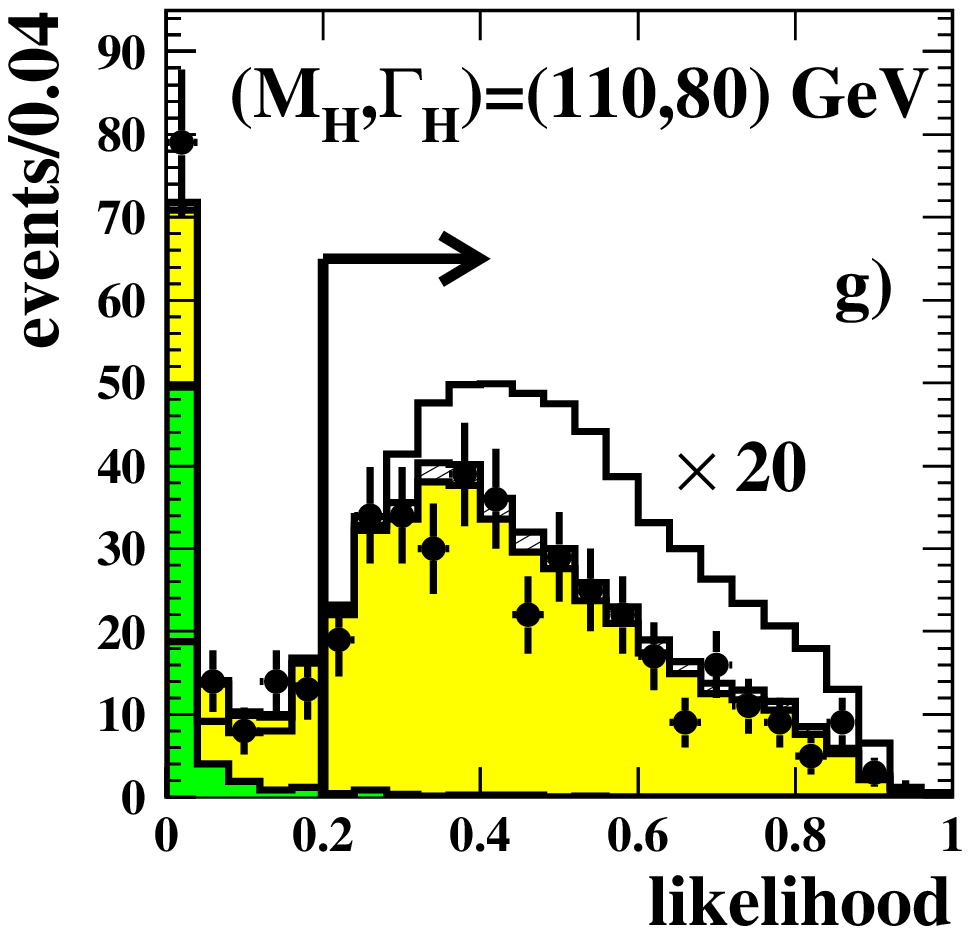}
\includegraphics[width=.4\textwidth]{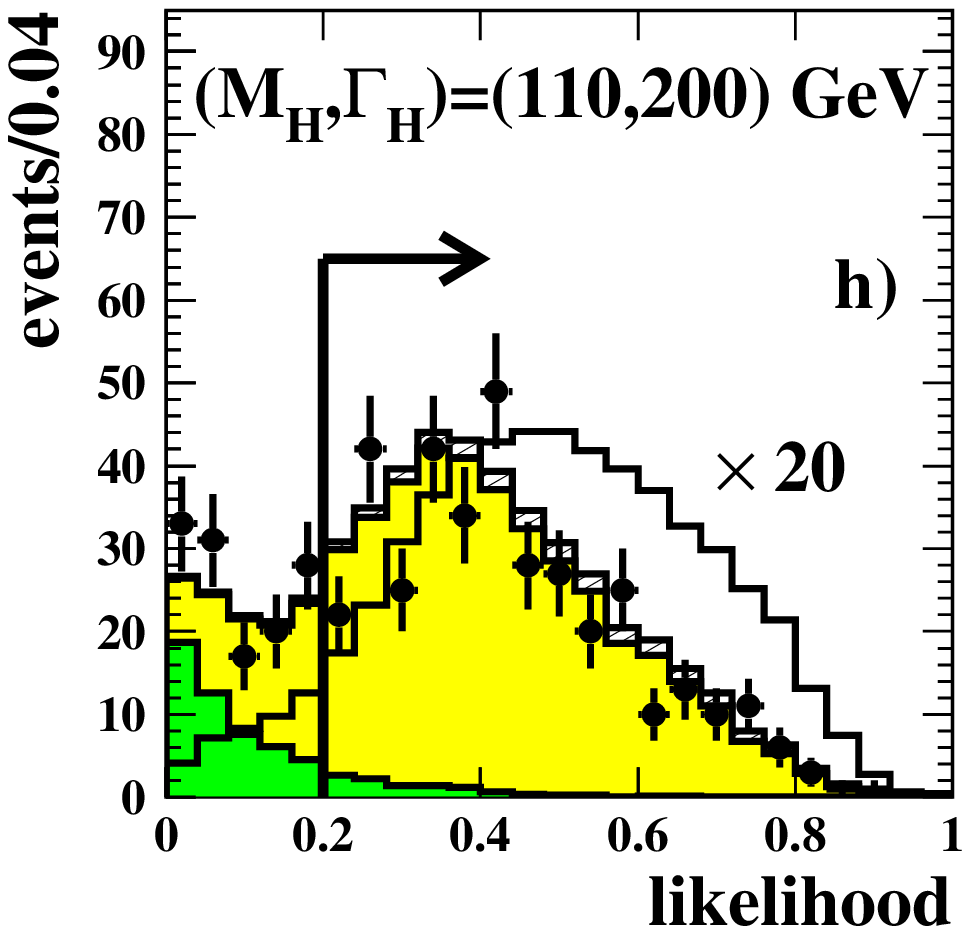}\\  \vspace*{-3mm}
\caption{\label{f:lhexamples} \sl Examples of some of the likelihood selections. Figure \ref{f:lhexamples} e), f), h) corresponds to analysis A1 (as labelled in Table \ref{t:sysres0} and \ref{t:lhoodsel}), a) to A3, b) to A4 and c), d) to A5. The \OPAL data and the expected 2-fermion and 4-fermion background are added for all analysed centre-of-mass energies. The signal hypothesis in the hatched histograms is normalised to the number of expected signal events and added to the background. The open histograms display the shapes of scaled signal distributions.}
\end{figure}

\begin{itemize}
\item[(7)]  {\bf $E^{\mathrm{Max}}_{\mathrm{JET}}/\sqrt{s}$ }\newline
The variable $E^{\mathrm{Max}}_{\mathrm{JET}}$ measures the energy of the most energetic of the two jets. 
This is on average higher for the four-fermion background, due to the boost of the W and Z pairs, whereas heavy Higgs bosons and a Z boson are produced at rest.
\item[(8)]  {\bf$R_{P_{ti}}$}\newline
 This variable is the significance of the acoplanarity between the
two jets, taking into account detector resolution and acceptance.
The discrimination power is enhanced by weighting the acoplanarity
with the average jet polar angle, since transverse jet directions are more
precisely measured at large polar angles.
Signal events tend to have a more significant acoplanarity
and thus larger values of $R_{P_{ti}}$ than background.
The precise definition of $R_{P_{ti}}$ can be found in the OPAL analyses of $\mathrm{ZZ\rightarrow q{\overline q }\nu {\overline \nu }} $ events \cite{strom}.

\item[(9)]  {\bf$(M_{\mathrm{VIS}}+M_{\mathrm{MIS}})/(M_{\mathrm{VIS}}-M_{\mathrm{MIS}})$}\newline
This variable, described in \cite{Acciarri:2000ec}, uses two strongly correlated quantities, the invariant missing mass $M_{\mathrm{MIS}}$ and the visible  mass of the event $M_{\mathrm{VIS}}$.
Depending on the mass reconstruction accuracy it can have positive or negative values. The signal distribution of this variable is broader and accumulates at higher values than for the two- and four-fermion events, which are distributed more narrowly around the origin.
\end{itemize}
From the two likelihoods and three ranges of signal masses filled in the reference histograms one has six analyses to search for the different hypotheses in \Mh and \Gh. The study of the median expected \mysec~shows that five of these six are sufficient to have an optimally efficient analysis for each signal hypothesis characterised by \Mh and \Gh (see Figure \ref{f:pattern}) in the range studied. Likelihood 1 was not used with the reference distribution filled for the signal mass range of 80 to 120\,\GeV.
Figure~\ref{f:lhexamples} a) to c) and g) to h) display examples for the likelihood distributions of all five analyses used. In the histograms the events selected at all five centre-of-mass energies are added up, although each centre-of-mass energy was evaluated separately in the limit setting as explained in Section \ref{s:xesclimits}. The appropriate likelihood was calculated for each background, data and signal event. In case of a signal event it was added to the histogram with the weight defined in Equation~\ref{eq:EWG}. 
The number of expected signal events is normalised according to Equation~\ref{eq:totalhiggs}. The use of different analyses gives rise to varying shapes of the likelihood distributions of the background. Also the various shapes of the signal likelihood distributions are visible for different \Mh and \Gh. Since the form of the likelihood distributions for signal and background can yield additional information in the limit calculation, only a loose cut is applied in the likelihood selection, requiring a signal likelihood larger than 0.2.

\subsection{Correction of background and signal efficiencies}
\label{s:RBC}
A correction is applied to the number of expected background events and the signal efficiencies due to noise in the detectors in the forward region which is not modelled by the Monte Carlo. The forward energy veto used in the preselection can accidentally be triggered by machine backgrounds.
The correction factor is derived from the study of
random beam crossings, and applied individually for each year of data
taking.  Random beam crossing events were recorded when no physics
trigger was active. The fraction of events that fail the veto on activity in the forward region is below 3.4\,\% for all runs analysed. The detailed breakdown of the fraction of accidentally vetoed events is given in the last column of Table~\ref{t:lumicor}. 

\subsection{Systematic uncertainties}\label{s:systematics}
A possible signal in the data would reveal itself by altering the shapes of the distributions of the discriminating variables.
 Thus a systematic deviation in the description of a reconstructed observable between \SM \MC  and  a data sample in which the signal is absent, could wrongly be attributed to the presence of a signal. \par
The systematic uncertainties in the Monte Carlo description of the kinematic
event variables are studied in two control samples at a
centre-of-mass energy of 206\,GeV. In the first control sample, called two-fermion control-sample in the following, radiative returns contributing to the $\qq(\gamma)$ processes with photons detected at large angles are selected and the tagged ISR photon is removed from the event in \MC and data. This creates a $\qq$-like topology with missing momentum at large angles. The second control sample, called four-fermion control-sample, is obtained by selecting $\WW \to\qq l \nu$ events and removing the identified isolated lepton from the events in \MC and data. After this procedure these two control samples possess a topology very similar to signal events.
For all kinematic variables $x$ of the preselection and the likelihood
selection the mean $\overline{x}$ and the width of the distribution 
($RMS$) are compared between \MC and data, for the two-fermion and four-fermion control-samples.
The observables in the two-fermion and four-fermion \MC are then modified separately according to $ x^{\mathrm{NEW}}_{\mathrm{MC}} = ( x^{\mathrm{OLD}}_{\mathrm{MC}} - {\overline x}_{\mathrm{MC}} ) \times \frac{RMS_{\mathrm{DATA}}}{RMS_{\mathrm{MC}}} + {\overline x}_{\mathrm{DATA}}$.
Then all five likelihood selections are repeated separately and the relative change in the number of selected events compared to the unmodified case is taken as the systematic uncertainty.\par 
Since the analysis labelled A1 in Table \ref{t:sysres0} is used over a large range of the search plane (see Figure \ref{f:pattern}), the systematic uncertainties on the background determined in this analysis A1 are taken as an estimate for the background for all analyses. 
To determine the effect of the systematic uncertainty on the signal efficiencies, one has to take into account the fact that the kinematic properties of the signal depend on the assumed Higgs mass and decay width. Twelve representative hypotheses are studied with \Mh chosen to be 20, 60 or 110\,\GeV and \Gh taking values of 5, 20, 70 and 200\,\GeV. For these hypotheses the signal \MC is modified according to the four-fermion correction factors, representing the dominant remaining background after the cut on the signal likelihood. The relative change in selected event weights compared to the unmodified case is then taken as an estimate of the systematic uncertainty on the signal efficiency for a given hypothesis. The root-mean-square of all twelve hypotheses is applied as an (\Mh,\Gh)-independent estimate for the whole search area and for all centre-of-mass energies (see Table \ref{t:sysres0a}).\par 
The W pairs are very effectively reduced in the preselection by the isolated lepton veto. Due to the importance of this veto the uncertainty from the lepton isolation angle and the vetoed cone energy is studied in the following way. The half-cone angle of the outer cone is increased and decreased by two degrees, following the studies in \cite{Abbiendi:1998rd}, and the relative effect on the selection determined. Furthermore the cone energy is varied by 7.4\,\% and the analyses are repeated. The value of the cone energy rescaling is determined by the relative deviation of the mean of the measured energy of the lepton candidates in the inner cone between data and \MC in the  $\WW \to\qq l \nu$ sample. For signal efficiencies an analogous study was performed at the twelve points described above.
~Both results for the relative change of the selection for the cone opening half-angle and cone energy variation are added in quadrature and the root-mean-square of the 12 (\Mh,\Gh) hypotheses was taken to yield the total uncertainty associated with the isolated lepton veto (see Table \ref{t:sysres0a}). \par 
The theoretical prediction on the cross-section for the two- and four-fermion processes adds an uncertainty of 2\,\% to the background uncertainty \cite{Grunewald:2000ju}. Finally, the uncertainty due to the limited \MC statistics is evaluated.\par
Table \ref{t:sysres} summarises the results of the studies.
All uncertainties are assumed to be uncorrelated and the individual
contributions are added in quadrature to obtain the total systematic uncertainties on
the background expectation and signal efficiency. The dominant systematic
uncertainties on the signal efficiency arise from the description
of the kinematic variables. The background expectation is more affected by the uncertainty in the isolated lepton veto, as the main contribution of the background stems from four-fermion processes. But the uncertainty associated with the description of the kinematic variables is of similar magnitude.  
The limits quoted in Section \ref{s:xesclimits} were calculated including the uncertainties of Table \ref{t:sysres}. To estimate the extent to which the limits depend on the size of the systematic uncertainties, the limit calculation was repeated doubling the systematic uncertainties.
A comparison of the limits with single and double systematic uncertainties, done at similar representative points as used for the systematic studies, showed that the excluded cross-sections typically decrease between a half and one and a half percent. A maximal reduction of 2.1\,\% was found.

%
%
\section{Results}\label{s:results}
\begin{figure}
  \centering
\includegraphics[width=.47\textwidth]{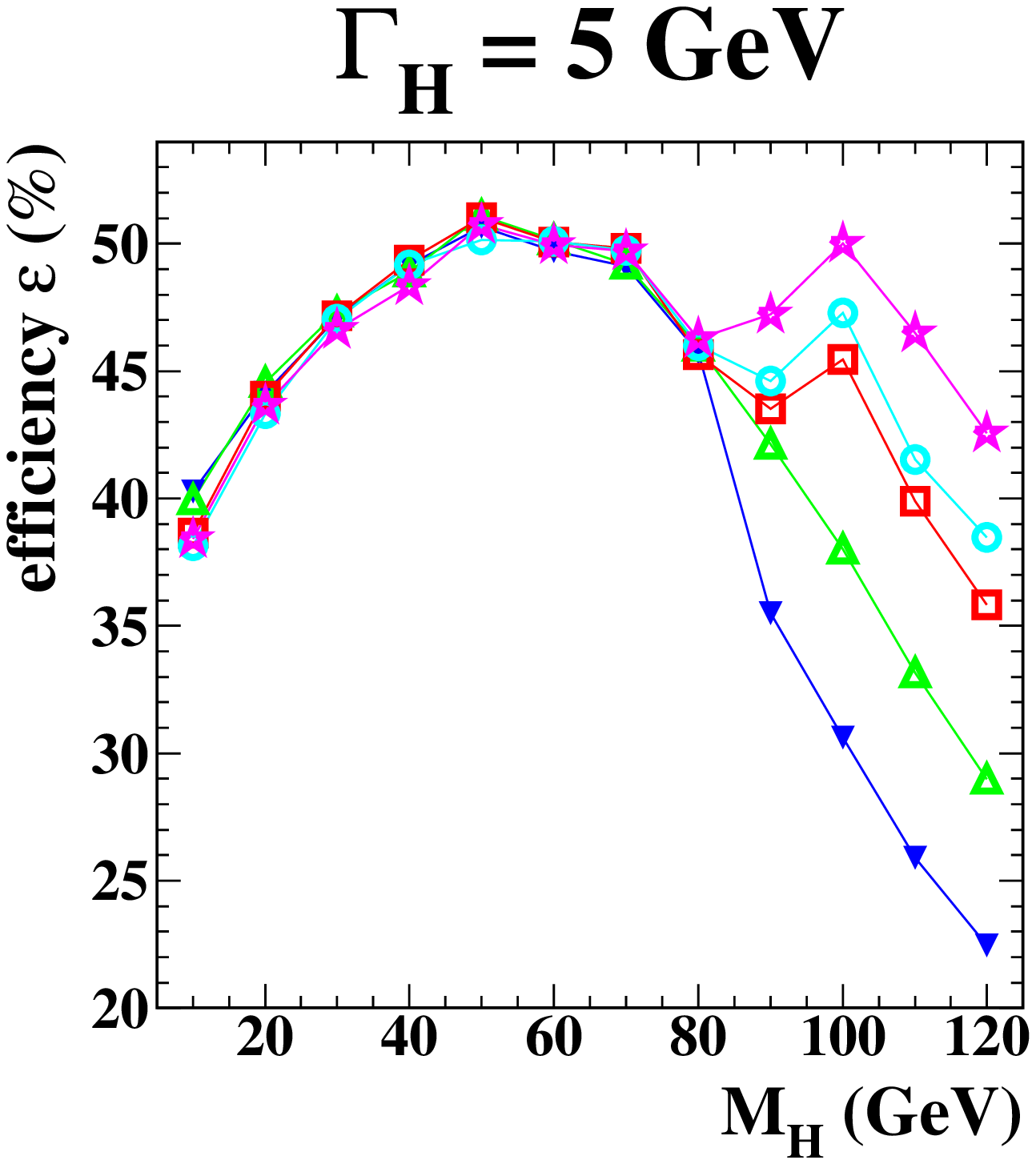} 
\includegraphics[width=.47\textwidth]{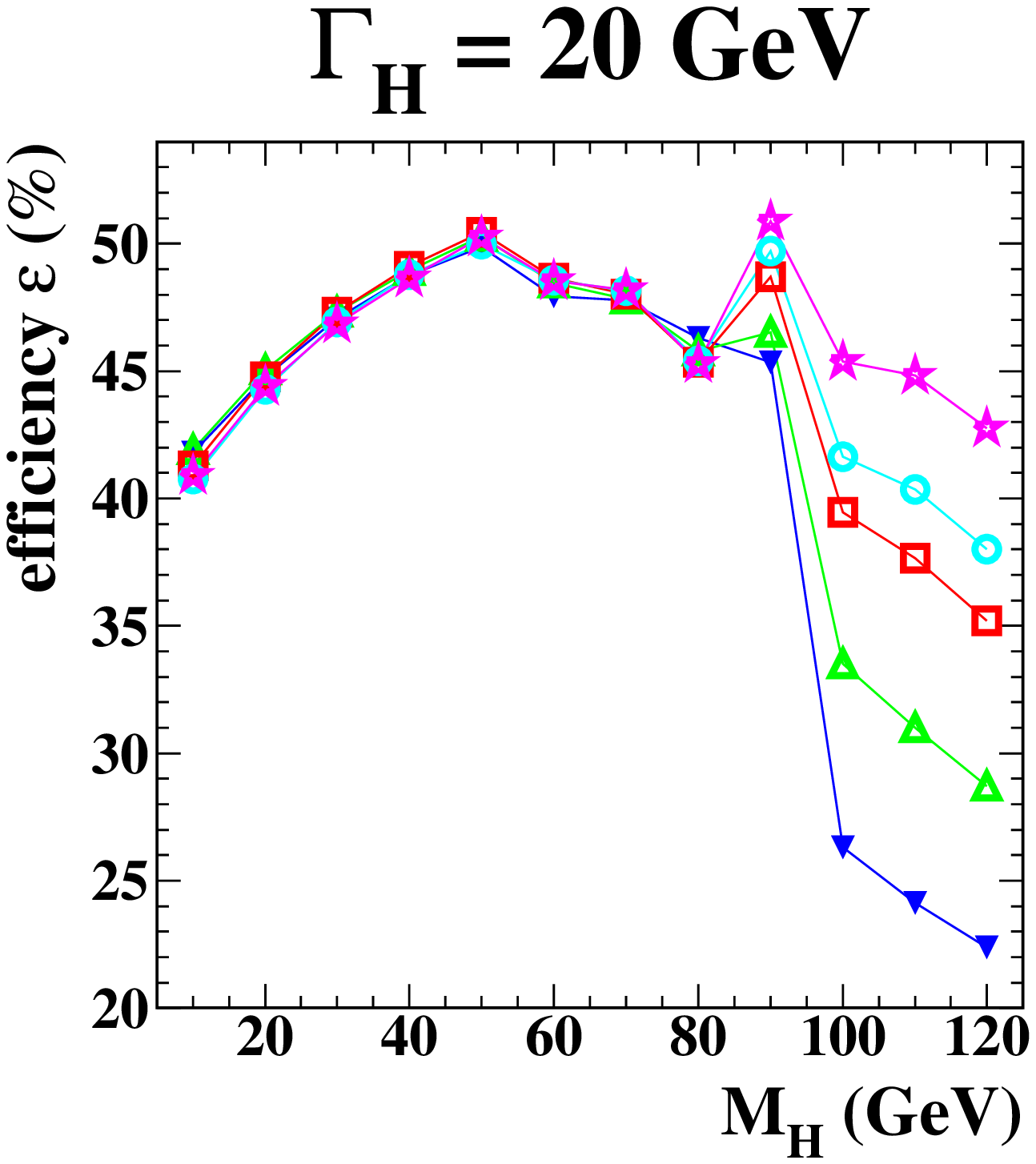}\\  \vspace*{1mm}
\includegraphics[width=.47\textwidth]{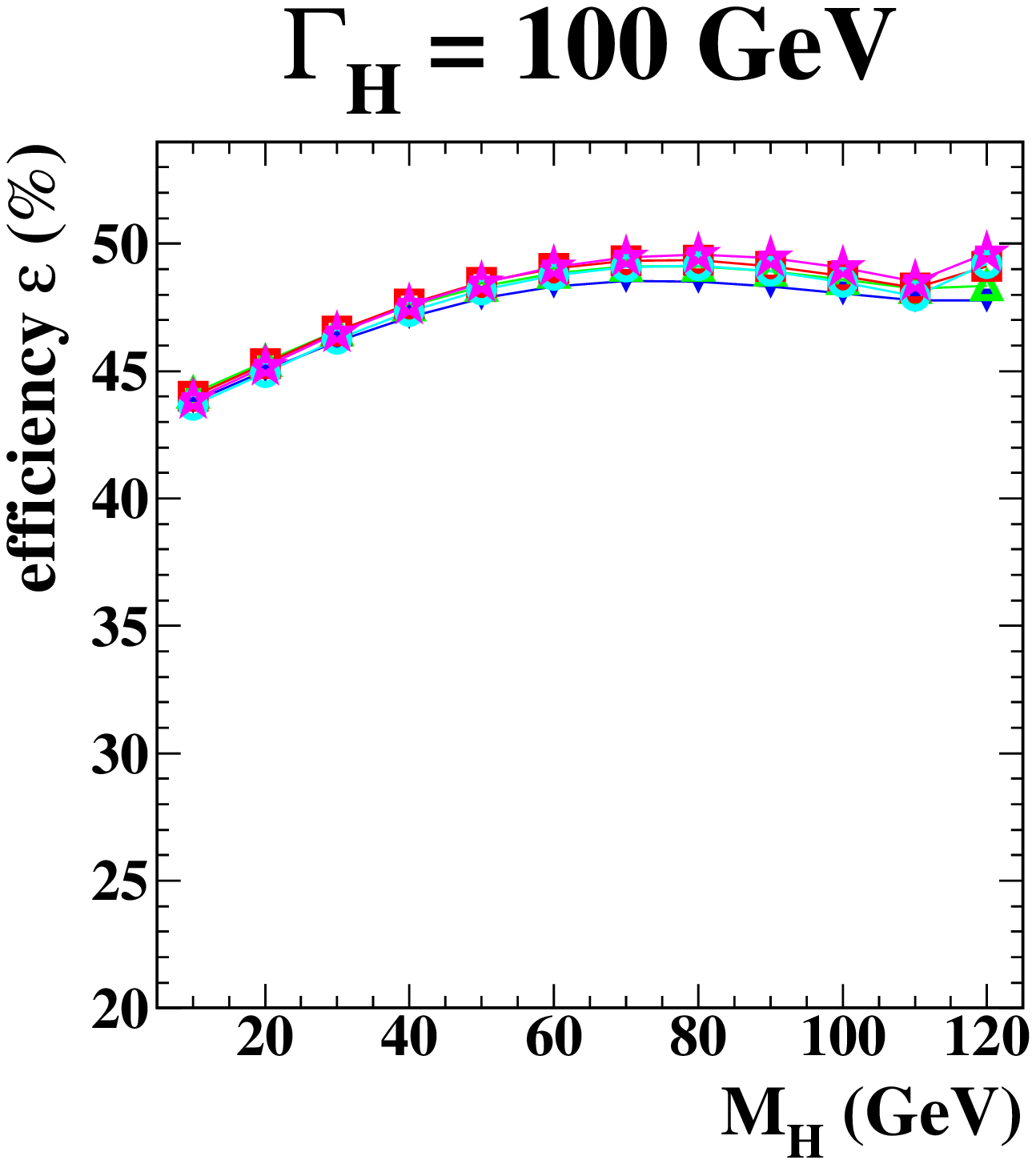}
\includegraphics[width=.47\textwidth]{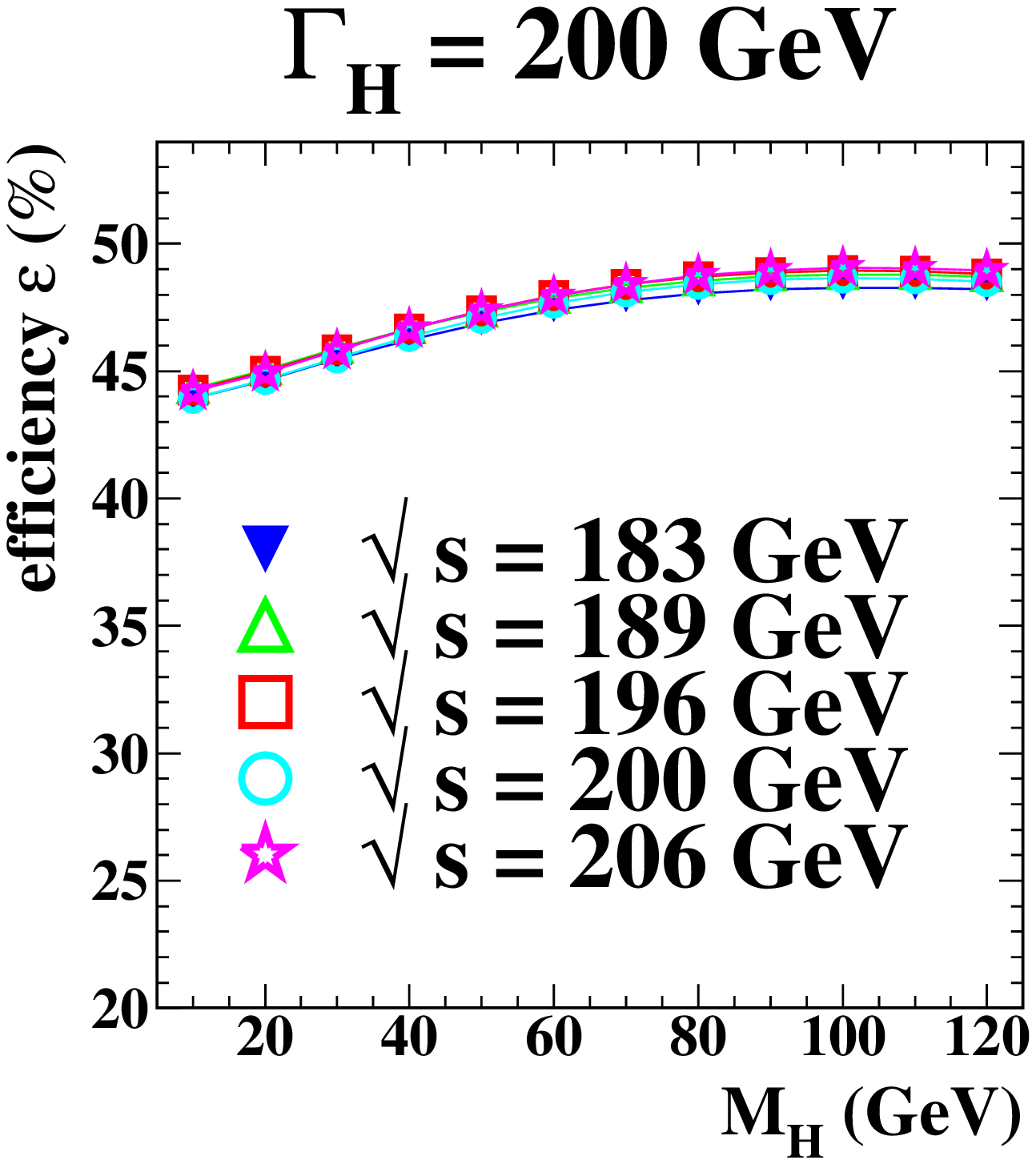}
  \caption{\label{f:eff5ecm}
 \sl  Examples for the selection efficiency after a cut on the signal likelihood greater than 0.2 versus the Higgs mass \Mh as function of the assumed decay width \Gh at the different $\sqrt{s}$. The error is the binomial error on the selected event weights and smaller than the markers. Lines are added to guide the eye. A signal in the range of 80 to 90 \,\GeV suffers from a drop in the efficiency due to the relatively large remaining W- and Z-pair backgrounds. For a smaller widths \Gh the efficiency to detect a relatively heavy (above 100\,\GeV) and more \SM like Higgs boson is more restricted by the available $\sqrt{s}$. For a large \Gh signal hypothesis, the kinematic distributions of events and the distribution of weights assigned to these events are broader. Therefore it is more likely to select a larger fraction of the event weights leading to more uniform efficiency, that does not depend very much on the centre-of-mass energy.}
\end{figure}

The results of the search using each of the five different likelihood selections, labelled A1-A5, after a cut on the likelihood larger than 0.2 are summarised in Table~\ref{t:lhoodsel}, which compares the numbers of observed candidates with the \SM background expectations. The data are compatible with the \SM background expectations. The remaining four-fermion background consists predominantly of W pairs, representing roughly three quarters of the background at energies above the Z pair threshold.
 Figure \ref{f:eff5ecm} shows examples for signal efficiencies. For small decay widths the dependence on the centre-of-mass energy is weak up to \Mh $\approx$ 80\,\GeV, and for large widths it is weak up to the kinematic limit.
Because of the centre-of-mass energy dependence of the Bjorken cross-section of a Higgs boson with mass \Mh, the lower centre-of-mass energies contribute more significantly to the sensitivity for lighter Higgs bosons (see e.g. in \cite{c:hunter}). For very light Higgs bosons the efficiency is moderately reduced by the preselection cuts demanding a sizeable amount of missing energy. In the case of broader Higgs resonances with high mass, one observes a generally enhanced efficiency since the chance of selecting events from the low mass tail compensates the suppression due to the falling production cross-section of a heavy Higgs boson.\par   

\subsection{The upper limits on the production cross-section times branching ratio}
\label{s:xesclimits} 
Upper limits are calculated on the model-independent cross-section \mysec\ scaled to $\sqrt{s} = 206$\,\GeV. 
~As the likelihood distributions are only loosely cut, one can use not only the information from the integral number of selected events (Table~\ref{t:lhoodsel}) but also from the shape in a likelihood ratio \cite{Read:2002hq} to set more sensitive upper limits. For each centre-of-mass energy separately, each bin with a signal likelihood larger than 0.2 in the distributions of expected signal, background and selected data is treated as a search channel. 
For each centre-of-mass energy the number of expected signal events is scaled to the total cross-section (Equation~\ref{eq:totalhiggs}). As with the analysis described in \cite{Abbiendi:2004gn}, the likelihood distributions are given as discriminating input to a limit program \cite{Junk:1999kv}. A likelihood ratio is used to determine the signal confidence level, $\mathrm{CL_S}$, defined in \cite{Read:2002hq,Junk:1999kv}, which excludes the presence of a possible signal according to the modified frequentist approach \cite{Junk:1999kv}. Additionally the program calculates the median upper number of signal events that could be excluded at 95\,\% confidence level (CL). This number is then scaled to the total cross-section at the centre-of-mass energy of 206\,\GeV for each (\Mh,\Gh) hypothesis.
The systematic uncertainties on the background expectations and signal selection efficiencies are included according a generalisation of the method described in~\cite{c:cousins}.\par 
A very fine scan of the (\Mh,\Gh)-plane was performed by simulating the spectra of Higgs bosons with a mass \Mh from 1 to 120\,\GeV and widths \Gh starting at 1\,\GeV  up to 3 \,\TeV. The Higgs boson mass was simulated in  steps of $\delta \Mh = 1$\,\GeV. Simulated values of \Gh are spaced in steps of 1\,\GeV up to 5\,\GeV. A spacing of  $\delta \Gh = 5$\,\GeV is chosen from \Gh = 5\,\GeV to \Gh = 750\,\GeV. Above this value steps of $\delta \Gh = 50$\,\GeV are adopted up to the maximal \Gh of 3\,\TeV.\par
\begin{figure}
  \centering
\includegraphics[width=0.38\textwidth]{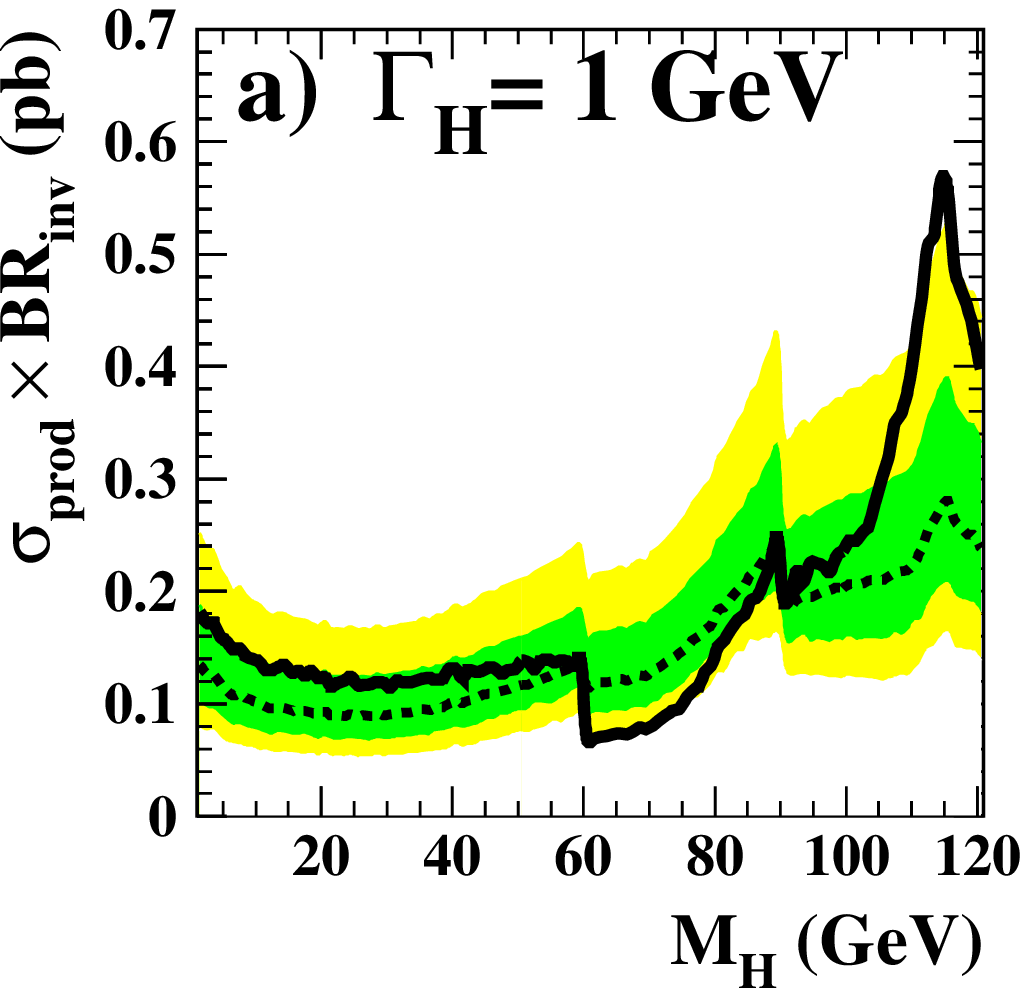}\hspace*{-8mm}
\includegraphics[width=0.38\textwidth]{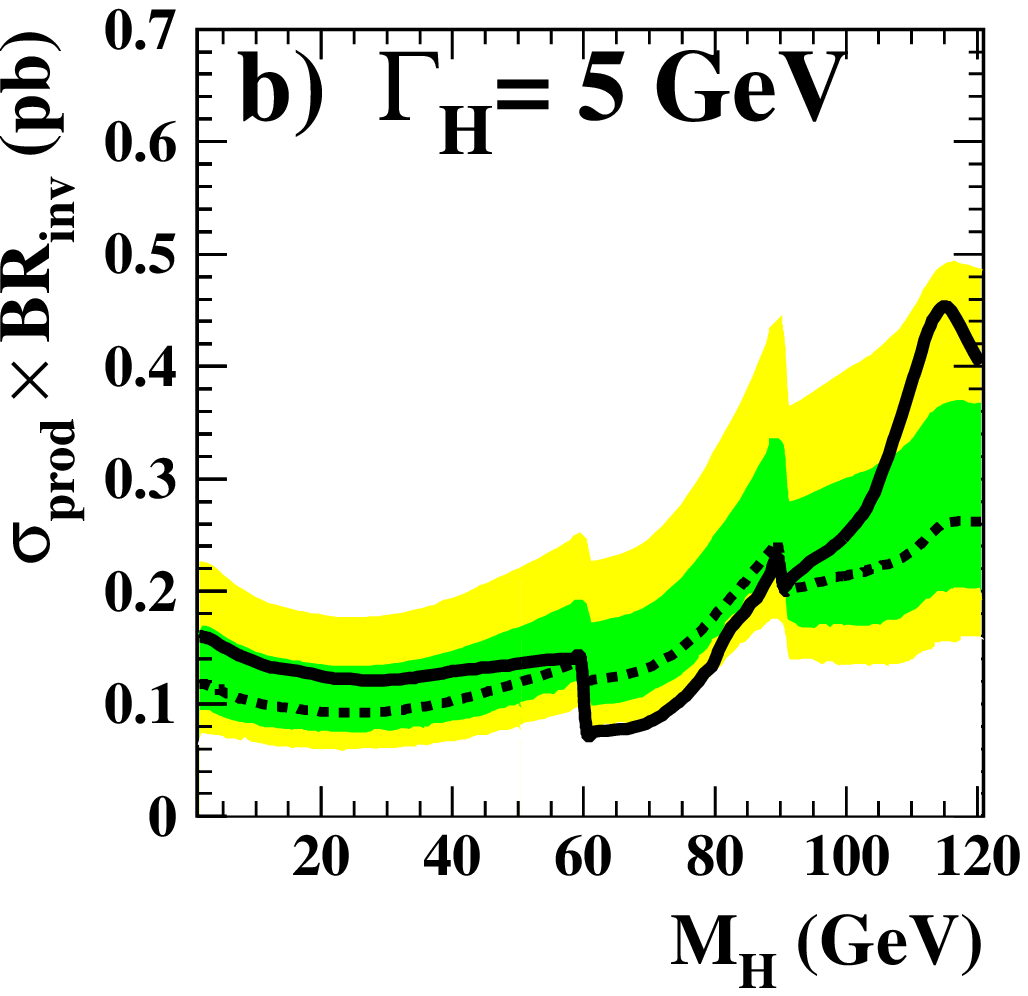}\hspace*{-8mm} 
\includegraphics[width=0.38\textwidth]{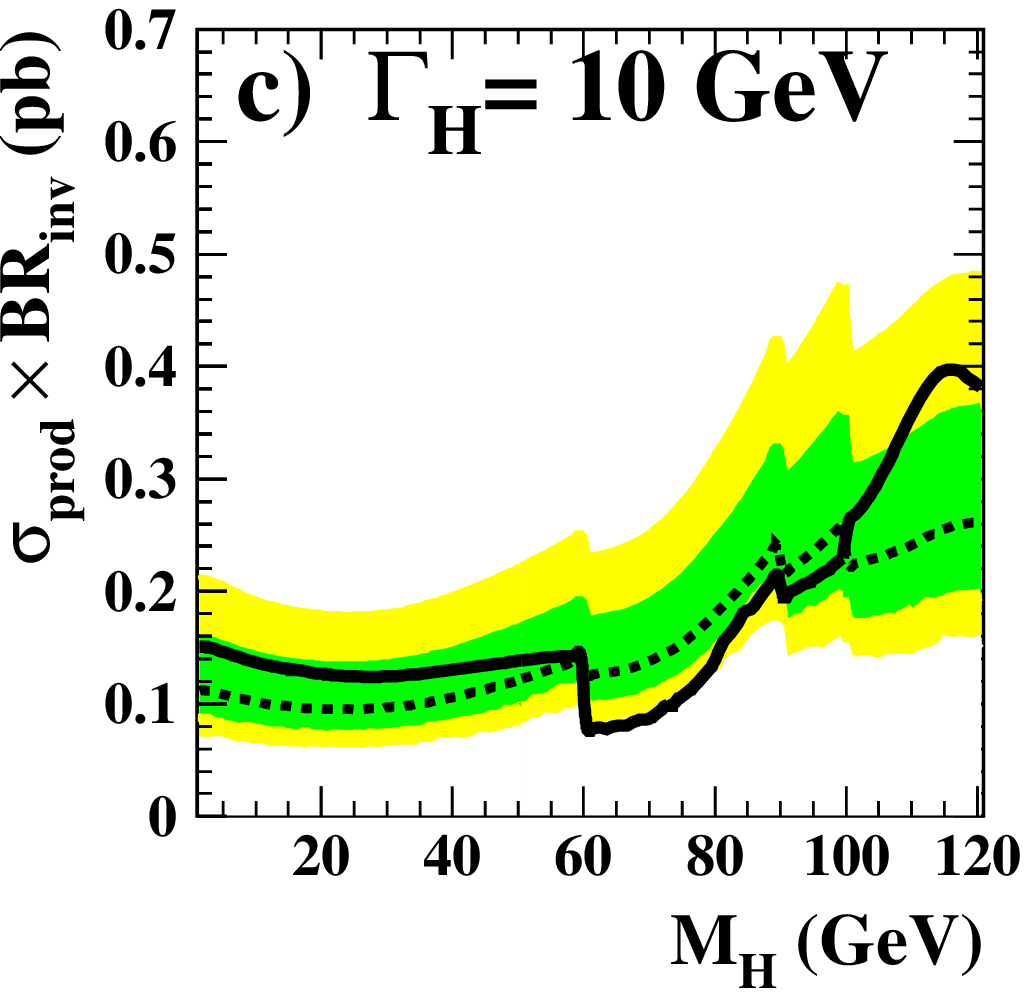}\\ \vspace*{-9mm}

\includegraphics[width=0.38\textwidth]{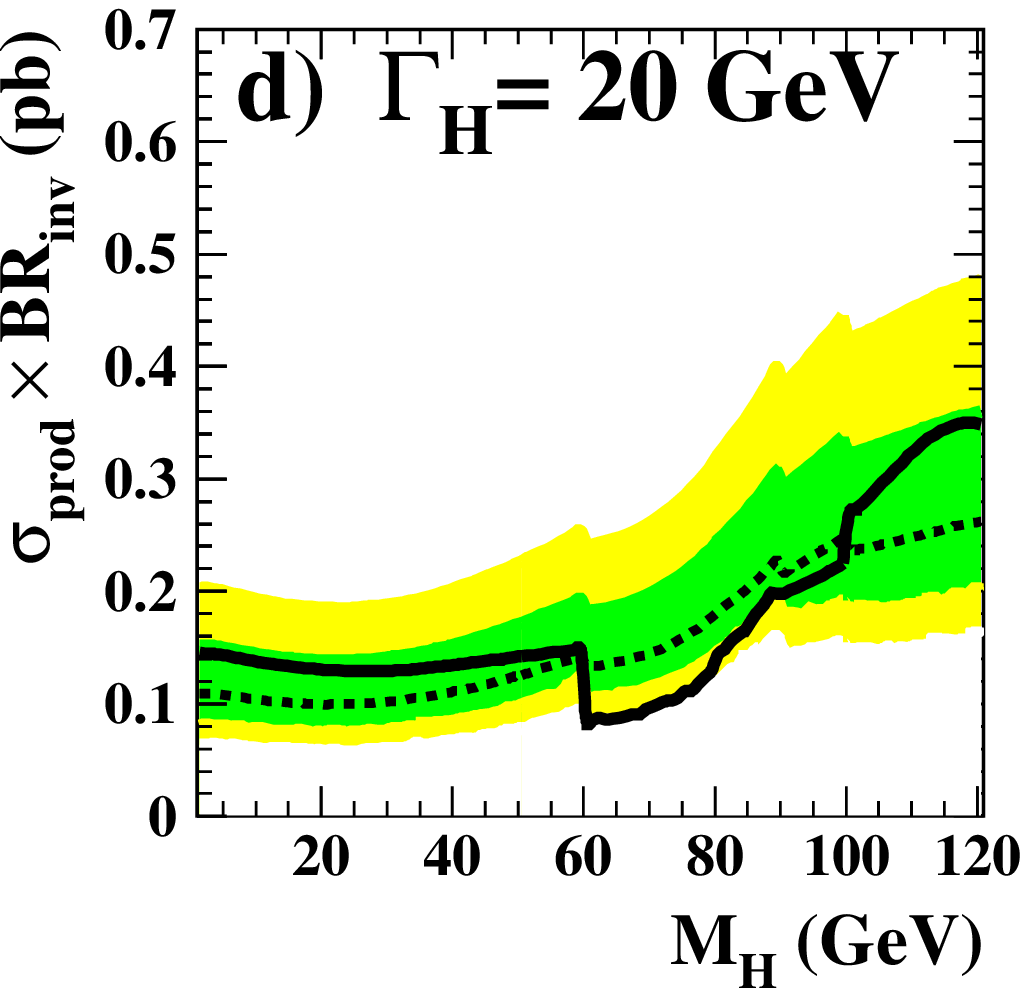}\hspace*{-8mm}
\includegraphics[width=0.38\textwidth]{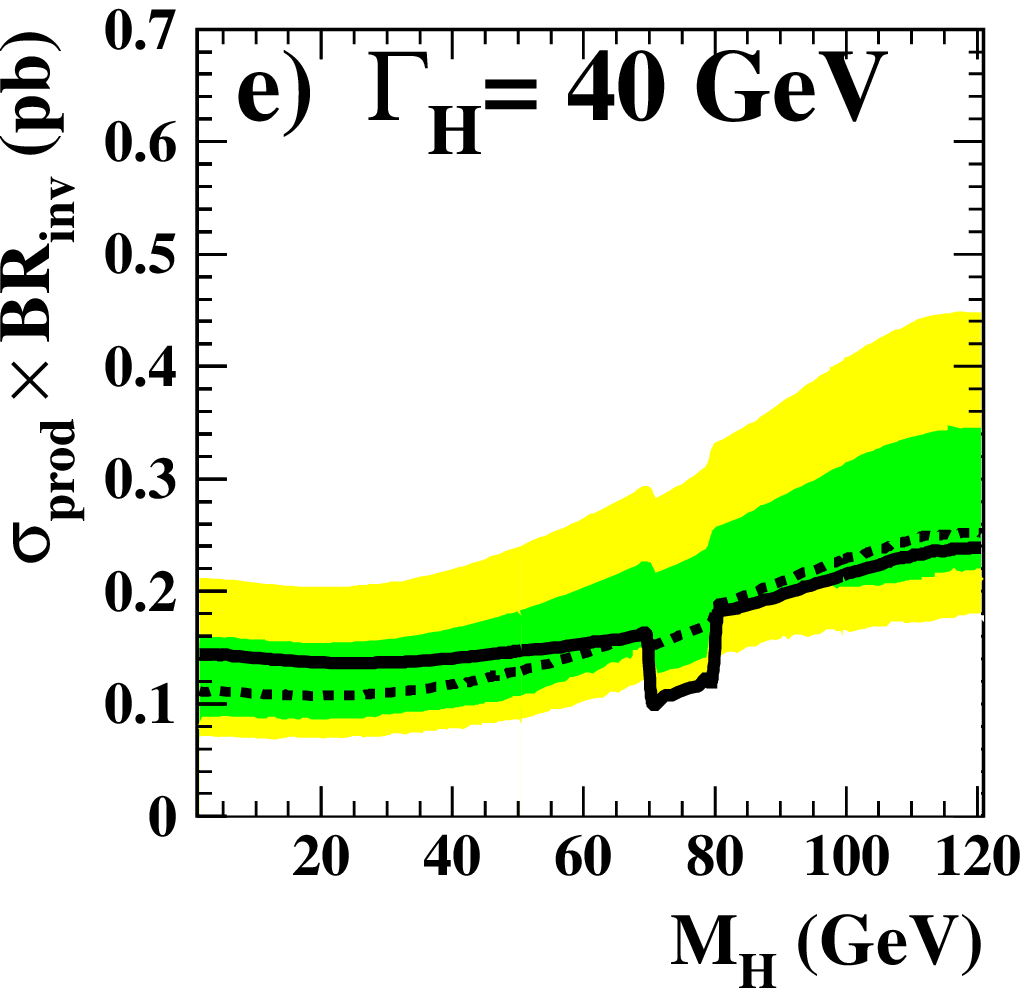}\hspace*{-8mm}
\includegraphics[width=0.38\textwidth]{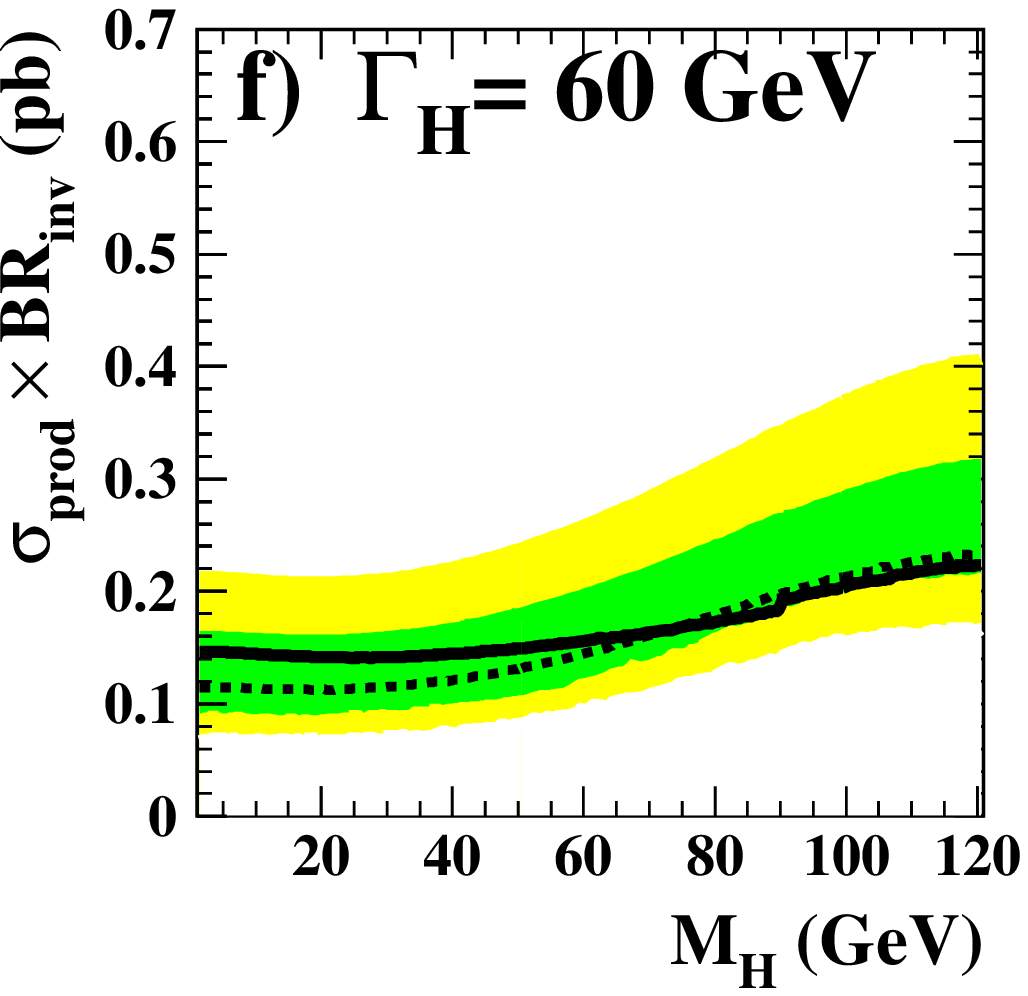}\\ \vspace*{-9mm}

\includegraphics[width=0.38\textwidth]{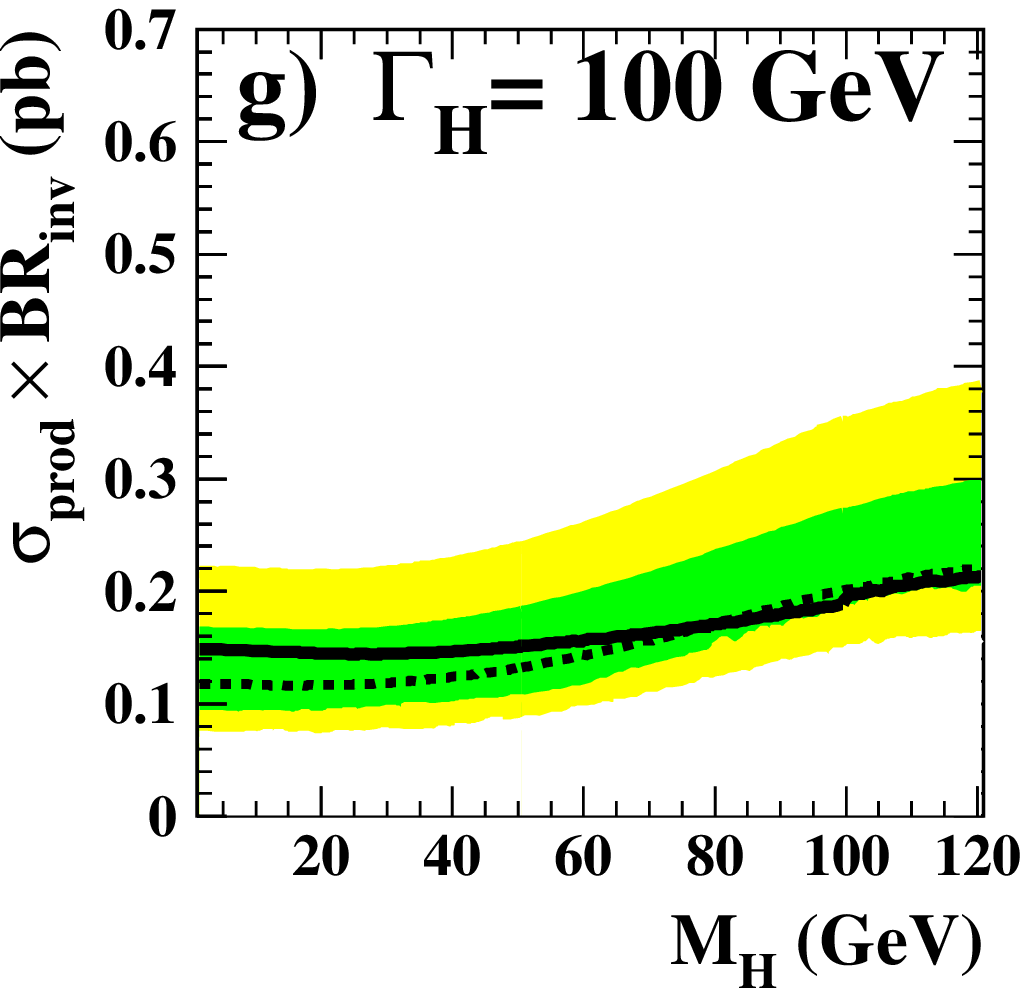}\hspace*{-8mm}
\includegraphics[width=0.38\textwidth]{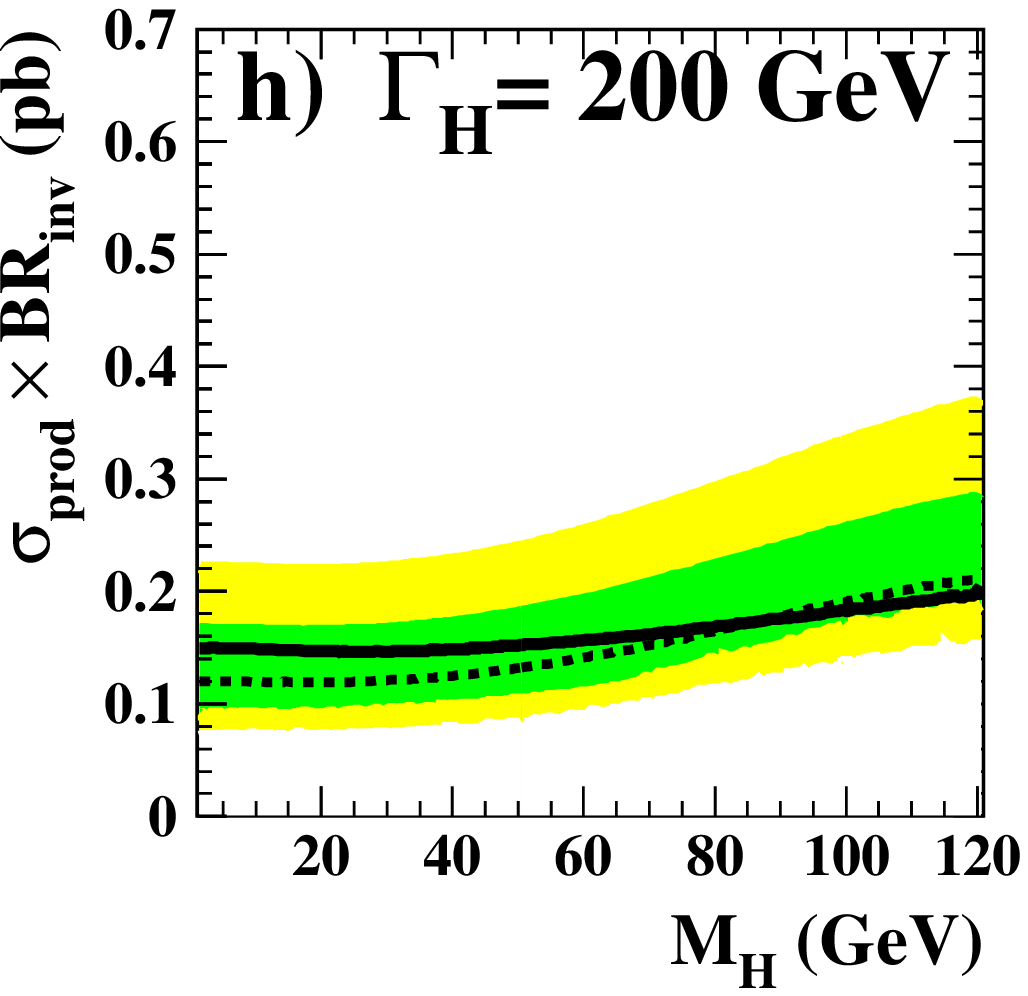}\hspace*{-8mm}
\includegraphics[width=0.38\textwidth]{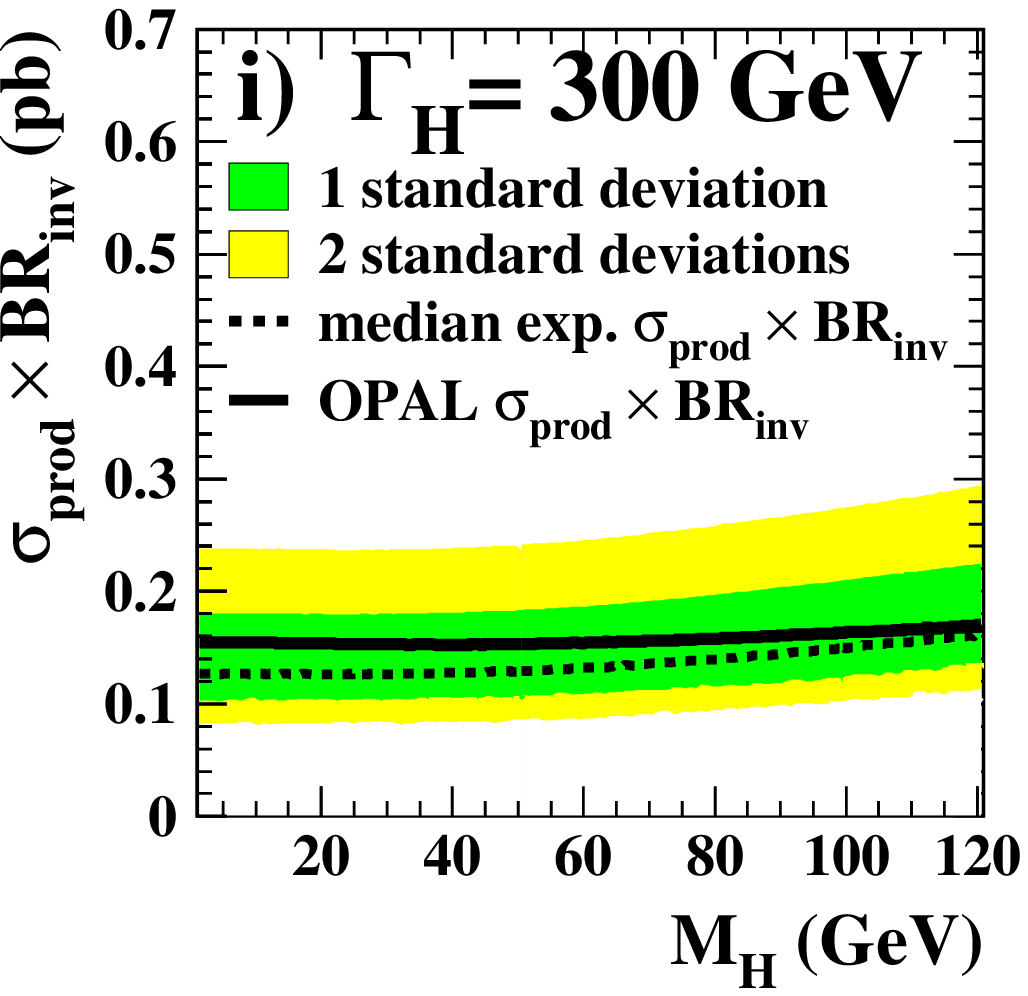}\\

  \caption{\label{f:xsbands} \sl The model independent upper limits at 95\,\% CL on the production cross-section times branching ratio, $\sigma_{\mathrm{prod}}\times BR_{\mathrm{inv}}$, scaled to a centre-of-mass energy of 206\,\GeV for Higgs mass \Mh and some examples of the Higgs decay width \Gh. The discontinuities in the limits reflect the changes in the analysis used at this mass (see Figure \ref{f:pattern}).}
\end{figure}
\begin{figure}
  \centering
\includegraphics[width=1.\textwidth]{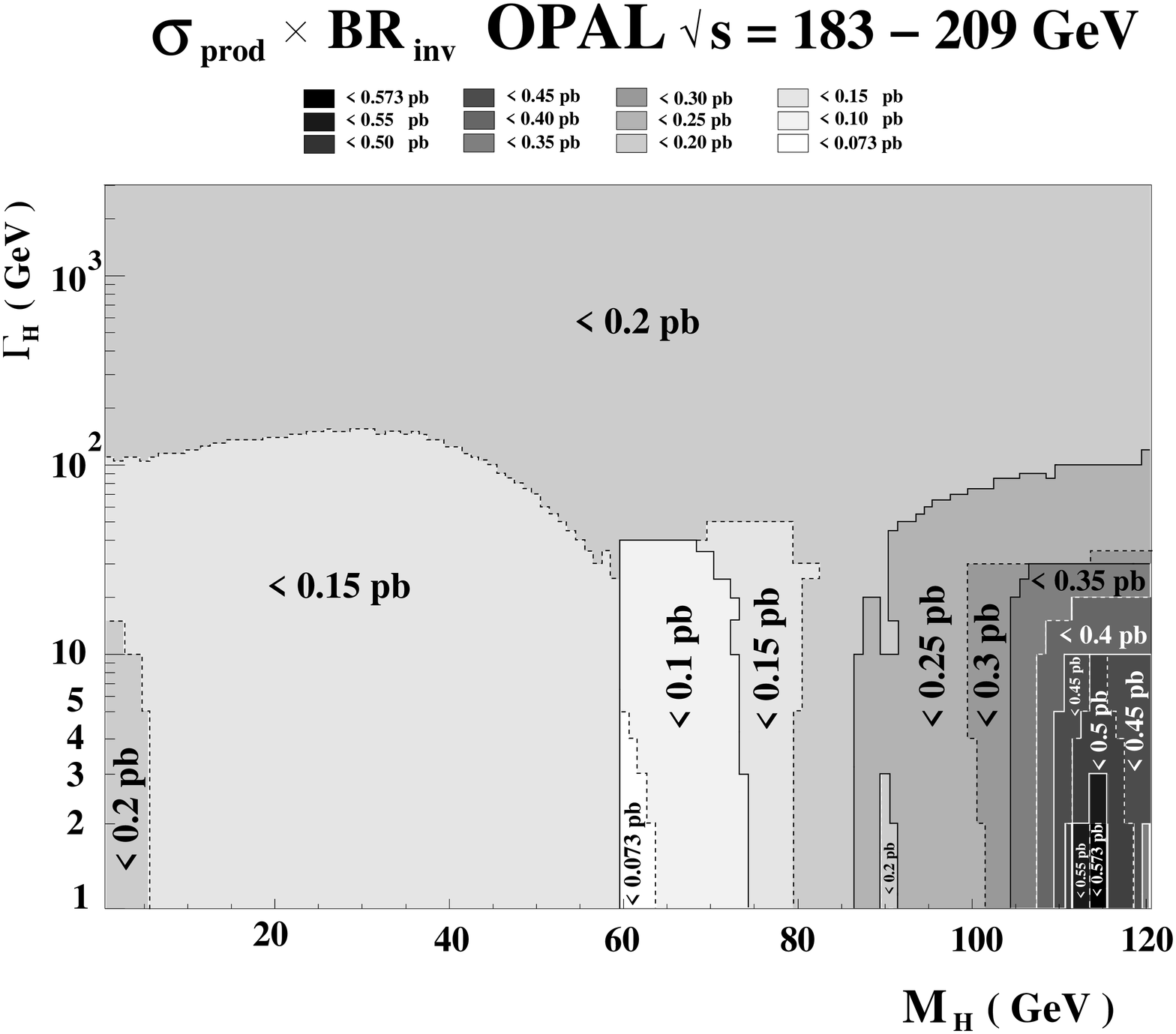}
 \caption{\label{f:final1} \sl The exclusion contours at 95\,\% CL on the model independent production cross-section times branching ratio, $\sigma_{\mathrm{prod}}\times BR_{\mathrm{inv}}$, for Higgs boson mass \Mh and the Higgs boson decay width \Gh up to 3\,\TeV (note a change in logarithmic scale below \Gh = 5\,\GeV for better visibility). Solid and dashed lines delimit areas of excluded upper limits. Cross-sections times branching ratio between 0.07\,pb and 0.57\,pb are excluded with the \OPAL data above $\sqrt{s} = 183$\,\GeV.}
\end{figure}

Examples of the projections of the observed upper cross-section limits together with the median expected upper limits and the corresponding one and two standard deviation bands on the expected limits are displayed in Figure~\ref{f:xsbands} for some choices of \Gh. Above a width of 300\,\GeV the exclusion plots look quite similar to the example displayed in Figure~\ref{f:xsbands}i) because the excluded limits do not change very much. The observed limits for \Gh $\gtrsim$ 60\,\GeV are well contained in the one standard deviation bands on the expectations and generally do not exceed two standard deviations except in Figure~\ref{f:xsbands} a) at \Mh = 114\,\GeV.
~The discontinuities in the graphs correspond to changes in the analyses. As one can observe, below \Gh $\lesssim 40$\,\GeV the analysis are changed more often. Therefore the chance is higher that in a few bins there are statistical fluctuations in the selected data, that lead to a deviation of more than one standard deviation around the median. Also the data selected are highly correlated, as one can see for example in the upward fluctuation around \Mh = 114 \,\GeV  visible in Figure~\ref{f:xsbands} a)-c). All results for the observed upper limits on \mysec~are summarised in a contour plot (Figure~ \ref{f:final1}) in the scanned (\Mh,\Gh)-plane. Above \Gh = 200\,\GeV the observed upper limits are in the range of 0.15\,pb to 0.18\,pb for all \Mh and vary very little. For such large \Gh the recoil mass distribution of the Higgs tends to be more and more uniformly stretched out over the mass range explored. There is not much difference in the selection of signal events for a Higgs boson with e.g. a width of 400\,\GeV or 600\,\GeV in the considered range of Higgs masses. This prevents any specific discriminating kinematical properties from being assigned to the expected signal as signal masses of a broad kinematical range are selected with roughly equal probability. Therefore only one likelihood analysis is used in this part of the search area, selecting the same subset of data and background. Since the upper limit on the model-independent cross-section refers to a production cross-section at a centre-of-mass energy of 206\,\GeV, it must become independent of (\Mh,\Gh) for an extremely large \Gh. In this case the shape of the Higgs signal would just be a box, weighted with the production cross-section from 1\,\GeV to the kinematic limit of about 115\,\GeV. The data are then compared to an approximately constant signal expectation. Hence the upper limit on the cross-section is approximately independent of the (\Mh,\Gh) hypothesis at a value of roughly 0.16\,pb. For resonances with a decay width smaller than 200\,\GeV there are regions where the limits are below 0.15\,pb or even 0.1\,pb for \Mh between 60 and 74\,\GeV. In this mass range the number of data events selected is smaller than expected. Above \Mh of 85\,\GeV the upper limits become larger than 0.2\,pb and rise considerably for small widths below 40\,\GeV (see Figure~\ref{f:xsbands} a-e). This is due to the fact that the Higgs mass approaches the kinematic limit and the likelihoods which rely on kinematical variables like $p^{\mathrm{T}}_{\mathrm{MIS}}$ lose discrimination power. A maximal value of 0.57\,pb is observed for \Mh of 114\,\GeV and \Gh of 1\,\GeV corresponding to a circa two-$\sigma$ excess in the data.\par  
It should be kept in mind that no optimisation of the search has been performed for \Gh below 5\,\GeV. In the region of heavy Higgs boson mass $\gtrsim$~105\,\GeV and small width a search using recoil mass spectra would be more sensitive. Therefore this region is more sensitively covered by searches that have been performed by the \LEP experiments documented in~\cite{unknown:2001xz}.
\subsection{Interpretation of the result in the stealthy Higgs scenario}
\label{s:stealthy}
\begin{figure}
  \centering
  \includegraphics[width=1.0\textwidth]{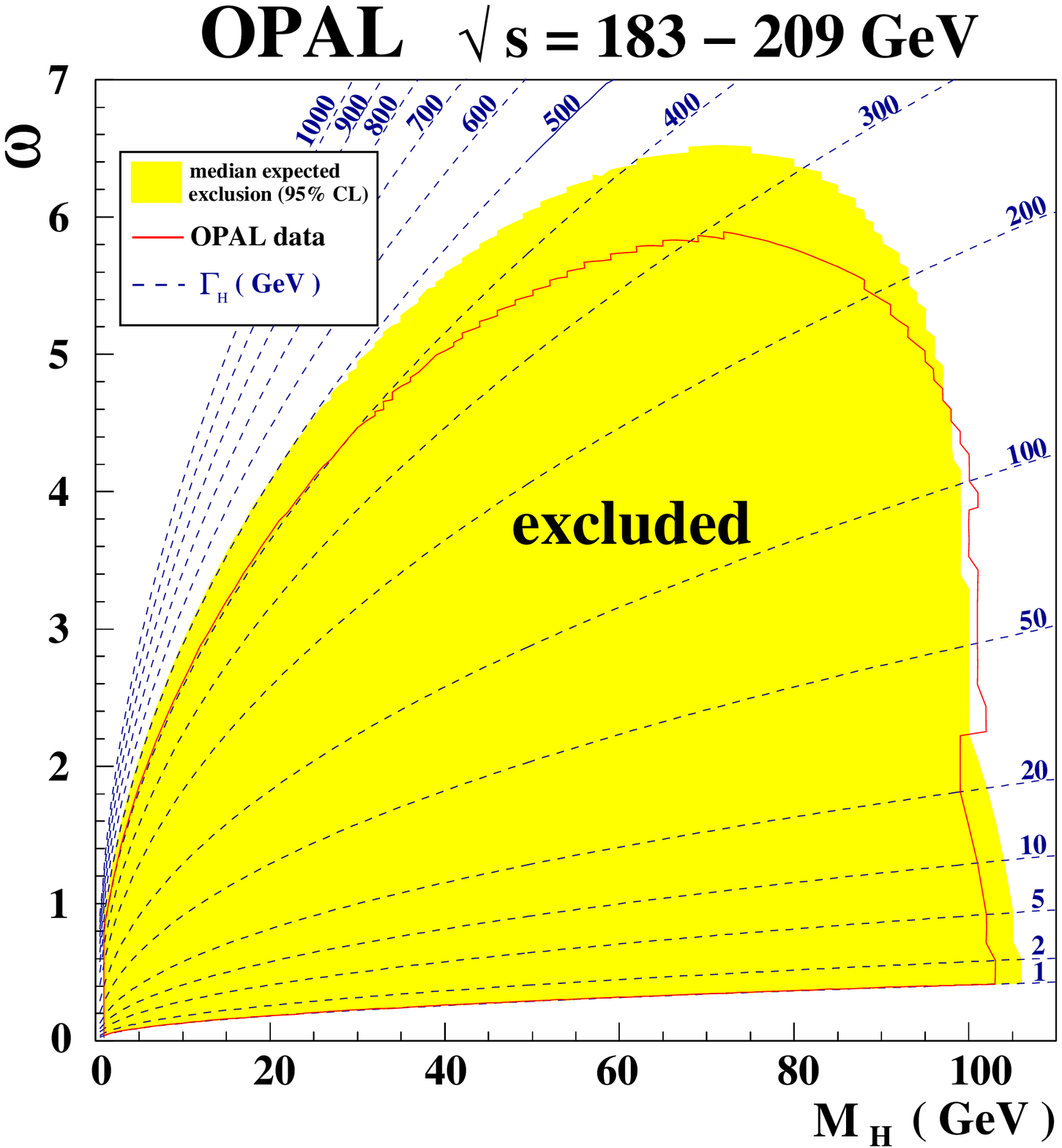}
  \caption{\label{f:final2} \sl The exclusion contours at 95\,\% CL on the Higgs mass \Mh and the Higgs-phion coupling $\omega$ of the stealthy Higgs model. The values of $\omega$ are related to the decay width \Gh via $\Gamma_{\mathrm{H}}(\Mh) = \Gamma_{\mathrm{SM}}(\Mh) + \frac{\omega^2 v^2}{32\, \pi\, \Mh}$, in the case of massless phions (see Equation~\ref{eq:higgs_width}). Contours of fixed \Gh are also shown in the plot as dashed lines.}
\end{figure}

Interpreting the width \Gh of a Higgs boson according to Equation~\ref{eq:higgs_width}, and setting $m_{\mathrm{phion}}$ to zero, it was possible to set limits on $\omega$ in the stealthy Higgs scenario.
A range from $\omega = 0.04 $ to $\omega = 24.45$ was probed.
The excluded regions are shown in Figure~\ref{f:final2} at 95\,\% confidence level (CL) in the $\omega$-$M_{\mathrm{H}}$ parameter space. To illustrate the Higgs boson width according to Equation~\ref{eq:higgs_width}, contours of fixed \Gh corresponding to a given mass $M_{\mathrm{H}}$ and coupling $\omega$ are added to the plot.
~The maximum excluded invisible width is about \Gh = 400\,\GeV for Higgs boson masses $\lesssim$ 35\,\GeV, decreasing slowly to \Gh = 115\,\GeV for \Mh = 100\,\GeV. The minimal exclusion of $\omega = 0.04$ is observed at \Mh = 1\,\GeV and the maximal exclusion is $\omega = \omax$ for \Mh = 73\,\GeV.
~For $\omega $ between 0.04 and 0.59 a Higgs mass from 1 to 103\,\GeV could be excluded. The maximal excluded Higgs mass was 103\,\GeV for width between 1 and 3\,\GeV, compared with the expected exclusion of 106\,\GeV.\par
 The results presented in this study extend the previous decay-mode independent searches for new scalar bosons with the \OPAL detector \cite{Abbiendi:2002qp} to regions of larger couplings and higher Higgs boson masses. In \cite{Abbiendi:2002qp} an interpretation within the stealthy Higgs model yielded a maximal excluded coupling $\omega$ for masses around 30\,\GeV, where $\omega$ was excluded up to $\omega = 2.7$. That study excluded Higgs boson masses up to \Mh = 81\,\GeV. It should be pointed out that the decay-mode independent searches also studied Higgs widths between 0.1 and 1\,\GeV and therefore cover the gap between searches within scenarios assuming a narrow decay width of the invisibly decaying Higgs boson \cite{unknown:2001xz} and the search presented in this paper up to \Mh = 81\,\GeV.

\section{Conclusions}

A dedicated search was performed in the channel $\mathrm{e}^+\mathrm{e}^- \to \mathrm{H}\mathrm{Z}$ with $ \mathrm{Z}\to \qq$ and the non-\SM decay $\mathrm{H} \to E_{\mathrm{MIS}}$ final state allowing for invisible decay widths of the Higgs boson from 1\,\GeV up to 3\,\TeV. The data taken by the \OPAL detector at \LEP above the W pair threshold were analysed. No indication for a signal was found and upper limits were set on \mysec. The maximal upper limit is 0.57\,pb at \Mh = 114\,\GeV and \Gh = 1\,\GeV. Over the scanned region of the (\Mh,\Gh)-plane upper limits are generally of the order of 0.15\,pb, especially for large values of \Gh $ \gtrsim$ 400\,\GeV or Higgs boson masses $\lesssim$ 85\,\GeV. \par 
The limits were interpreted in the stealthy Higgs scenario assuming the presence of a large number of massless singlet states. Limits were calculated on the coupling $\omega$ to a hidden scalar sector of the Higgs boson with a given mass \Mh. A large part of the parameter plane kinematically accessible with \LEP\,2 was excluded extending a previous exclusion published in \cite{Abbiendi:2002qp}. Values for $\omega$~between 0.04 (\Mh = 1\,\GeV)~and~\omax~(\Mh = 73\,\GeV) were excluded, and for certain values of $\omega$ Higgs boson masses are excluded up to \Mh = 103\,\GeV. The possible non-detection of a light Higgs boson at the \LEP searches due to non-\SM invisible Higgs boson decays is therefore restricted to the case of extremely large decay widths $\gtrsim$ 400\,\GeV.  
\section*{Acknowledgements}
\noindent We particularly wish to thank the SL Division for the efficient operation
of the LEP accelerator at all energies
 and for their close cooperation with
our experimental group.  In addition to the support staff at our own
institutions we are pleased to acknowledge the  \\
Department of Energy, USA, \\
National Science Foundation, USA, \\
Particle Physics and Astronomy Research Council, UK, \\
Natural Sciences and Engineering Research Council, Canada, \\
Israel Science Foundation, administered by the Israel
Academy of Science and Humanities, \\
Benoziyo Center for High Energy Physics,\\
Japanese Ministry of Education, Culture, Sports, Science and
Technology (MEXT) and a grant under the MEXT International
Science Research Program,\\
Japanese Society for the Promotion of Science (JSPS),\\
German Israeli Bi-national Science Foundation (GIF), \\
Bundesministerium f\"ur Bildung und Forschung, Germany, \\
National Research Council of Canada, \\
Hungarian Foundation for Scientific Research, OTKA T-038240, 
and T-042864,\\
The NWO/NATO Fund for Scientific Research, the Netherlands.\\

\clearpage

%

%
%

\clearpage

\begin{table}
  \small
  \begin{center}
  \begin{tabular}{|cc|c|c|c|} \hline
 \multicolumn{2}{|c|}{\hspace*{2mm}\textbf{binned} \boldmath$\sqrt{s}$\unboldmath \hspace*{6mm}  \textbf{nominal} \boldmath$\sqrt{s}$\unboldmath \textbf{ (GeV)}} & \textbf{ year}  & \textbf{int. luminosity (pb$^{-1}$)}& \textbf{accid. veto (\%)}\\
    \hline
~$>$~180--186  &   183          &  1997        &                40.0     &                  3.37 \\
~$>$~186--193  &   189          &  1998        &               199.8     &                  2.24 \\
~$>$~193--198  &   196          &  1999        &                70.4     &                  2.53 \\
~$>$~198--203  &   200          &  1999        &               112.0     &                  2.96 \\ 
~$>$~203--209  &   206          &  2000        &               206.9     &                  2.22 \\  
  \hline 
  \end{tabular}
  \caption{\label{t:lumicor}
 \sl Breakdown of the analysed integrated data luminosities according to the centre-of-mass energies. The data was binned in five nominal centre-of-mass energies. The last column states the reduction of the signal efficiencies and expected background rates due to accidental triggering of the forward energy veto in the preselection, which is not modelled in the \MC.}
  \end{center}
\end{table}


 \begin{table}
\small
 \begin{center}
 \begin{tabular}{|l|c|c|c|c|c|} \hline
 \textbf{cut}& \hspace*{7mm}\boldmath $ \gamma \gamma $ \unboldmath \hspace*{5mm}& \hspace*{3mm}\textbf{qq( \boldmath$ \gamma $\unboldmath)} \hspace*{2mm} & \textbf{4-fermion}&\hspace*{3mm} {\bf total SM} \hspace*{2mm} & \hspace*{4mm}\textbf{data}\hspace*{4mm}\\ \hline

 (1)-(5)       &48795 & 15639 &  4880 & 69314 &    74178\\ \hline
 (6)           &  148 & 10359 &  1394 & 11901 &    11779\\ \hline
 (7)           &   62 &  9128 &  1336 & 10526 &    10472\\ \hline
 (8)           &   44 &  4897 &  1167 &  6108 &     6264\\ \hline
 (9)           &   33 &  1061 &   964 &  2058 &     2116\\ \hline
 (10)          &   18 &   425 &   895 &  1338 &     1387\\ \hline
 (11)          &   18 &   423 &   879 &  1320 &     1368\\ \hline
 (12)          &    4 &    68 &   820 &   892 &      899\\ \hline
 (13)          &    4 &    60 &   441 &   505 &      498\\ \hline

 \end{tabular}
 \end{center}
\caption{\sl Expected number of \SM background 
 events after the preselection normalised to a data
luminosity of 629.1 $\mathrm{pb^{-1}}$. The total SM background after preselection is expected to be 505 $ \pm $ 5(stat) $ \pm $ 21(syst). The contributions of the different subclasses
are broken down in column two to four for the two-photon, two-fermion and four-fermion processes respectively.}\label{t:preselcuts}
 \end{table}


\begin{table}
  \small
  \begin{center}
     \begin{tabular}{|l|l|c|c|}
    \hline
\multicolumn{2}{|c|}{\textbf{analysis}}&\multicolumn{2}{|c|}{\textbf{background uncertainty}}  \\  
\multicolumn{2}{|c|}{label~~~ likelihood ~~~ reference mass range (GeV)} & \multicolumn{2}{|c|}{\textbf{ kinematic var.~ isol. lepton veto}} \\ \hline
\multicolumn{2}{|l|}{~A1~~~~~~~\hspace*{5mm} 1~~~~~~~~~ 1--120} &\hspace*{5mm}~~~~ 2.4\,\%~~~~~~ &2.4\,\%  \\
\multicolumn{2}{|l|}{~A2~~~~~~~\hspace*{5mm} 2~~~~~~~~~ 1--120} &\hspace*{5mm}~~~~ 1.6\,\%~~~~~~ &2.3\,\%  \\
\multicolumn{2}{|l|}{~A3~~~~~~~\hspace*{5mm} 1~~~~~~~~~ 50--80} &\hspace*{5mm}~~~~ 1.0\,\%~~~~~~ &2.5\,\%  \\
\multicolumn{2}{|l|}{~A4~~~~~~~\hspace*{5mm} 2~~~~~~~~~ 50--80} &\hspace*{5mm}~~~~ 1.6\,\%~~~~~~ &2.6\,\%  \\
\multicolumn{2}{|l|}{~A5~~~~~~~\hspace*{5mm} 2~~~~~~~~ 80--120} &\hspace*{5mm}~~~~ 1.1\,\%~~~~~~ &1.5\,\%  \\
\hline
\multicolumn{2}{|c|}{\textbf{choice for uncertainty} } & \hspace*{4mm}2.4\,\% & 2.4\,\% \\ \hline    
  \end{tabular}
  \caption{\label{t:sysres0}
\sl  Results of the study of systematic uncertainties of the expected background for the five kinds of analyses, labelled A1-A5, used in the search (see Figure \ref{f:pattern}) at a centre-of-mass energy of 206\,\GeV. Since the analysis labelled A1 covers the largest part of the search area, its uncertainty was chosen as representative uncertainty on the background due to the uncertainty in the kinematic variables and the isolated lepton veto at all centre-of-mass energies.}
  \end{center}
\end{table}

\begin{table}
  \small
  \begin{center}
     \begin{tabular}{|l|l|c|c|}
    \hline
\multicolumn{2}{|c|}{\textbf{signal hypothesis}}&\multicolumn{2}{|c|}{\hspace*{5mm}\textbf{efficiency uncertainty}}  \\  
\multicolumn{2}{|c|}{ \boldmath \Mh \unboldmath \textbf{(\GeV)}~~\boldmath \Gh \unboldmath \textbf{(\GeV)}} &\multicolumn{2}{|c|}{ \textbf{kinematic var.~ isol. lepton veto}} \\ \hline
\multicolumn{2}{|c|}{\textbf{~20}~~~~~~  \textbf{~~5}} &\hspace*{5mm}~~~~ 0.6\,\%~~~~~ & 0.6\,\%  \\
\multicolumn{2}{|c|}{\textbf{~20}~~~~~  \textbf{~20} } &\hspace*{5mm}~~~~ 0.4\,\%~~~~~ & 0.7\,\%  \\
\multicolumn{2}{|c|}{\textbf{~20}~~~~~  \textbf{~70}}  &\hspace*{5mm}~~~~ 0.3\,\%~~~~~ & 0.7\,\%  \\
\multicolumn{2}{|c|}{\textbf{~20}~~~~  \textbf{200}}   &\hspace*{5mm}~~~~ 0.1\,\%~~~~~ & 0.7\,\%  \\
\multicolumn{2}{|c|}{\textbf{~60}~~~~~~  \textbf{~~5}} &\hspace*{5mm}~~~~ 0.7\,\%~~~~~ & 0.8\,\%  \\
\multicolumn{2}{|c|}{\textbf{~60}~~~~~  \textbf{~20}}  &\hspace*{5mm}~~~~ 0.7\,\%~~~~~ & 0.8\,\%  \\
\multicolumn{2}{|c|}{\textbf{~60}~~~~~  \textbf{~70}}  &\hspace*{5mm}~~~~ 0.2\,\%~~~~~ & 0.8\,\%  \\
\multicolumn{2}{|c|}{\textbf{~60}~~~~  \textbf{200}}   &\hspace*{5mm}~~~~ 0.3\,\%~~~~~ & 0.7\,\%  \\
\multicolumn{2}{|c|}{\textbf{110}~~~~~~  \textbf{~~5}} &\hspace*{5mm}~~~~ 5.5\,\%~~~~~ & 0.7\,\%  \\
\multicolumn{2}{|c|}{\textbf{110}~~~~~  \textbf{~20}}  &\hspace*{5mm}~~~~ 2.9\,\%~~~~~ & 0.8\,\%  \\
\multicolumn{2}{|c|}{\textbf{110}~~~~~  \textbf{~70}}  &\hspace*{5mm}~~~~ 1.3\,\%~~~~~ & 0.8\,\%  \\
\multicolumn{2}{|c|}{\textbf{110}~~~~  \textbf{200}}   &\hspace*{5mm}~~~~ 0.1\,\%~~~~~ & 0.8\,\%  \\
\hline
\multicolumn{2}{|c|}{\textbf{all} \boldmath \Mh \unboldmath \textbf{and} \boldmath \Gh \unboldmath}            & \hspace*{5mm}1.9\,\% & 0.7\,\% \\ \hline                    
  \end{tabular}
  \caption{\label{t:sysres0a}
\sl  Results of the study of systematic uncertainties in twelve representative (\Mh,\Gh)-points at a centre-of-mass energy of 206\,\GeV. For each source the root-mean-square of the individual uncertainties in the twelve points was taken to get an (\Mh,\Gh) independent estimate of the uncertainty at all centre-of-mass energies.}
  \end{center}
\end{table}


\begin{table}
  \small
  \begin{center}
     \begin{tabular}{|l|c|c|}
    \hline

   \textbf{source}        &   \hspace*{5mm}   \textbf{background uncertainty}  \hspace*{3mm}  &   \hspace*{5mm}  \textbf{efficiency uncertainty}  \hspace*{3mm}  \\
    \hline

  kinematic variables         &             2.4\,\%     &                  1.9\,\% \\
  isolated lepton veto        &             2.4\,\%     &                  0.7\,\% \\
  limited MC statistics       &             1.0\,\%     &                  0.2\,\% \\
  prediction 2- and 4-f cross-sect.&        2.0\,\%     &                    -    \\
   total uncertainty           &            4.1\,\%     &                  2.0\,\% \\

  \hline 
  \end{tabular}
  \caption{\label{t:sysres}
  \sl Results of the study of systematic uncertainties of the background for the five analyses (see Table \ref{t:sysres0})  and of the signal efficiencies in twelve representative (\Mh,\Gh)-points (see Table \ref{t:sysres0a}) at a centre-of-mass energy of 206\,\GeV. The total uncertainty on background expectation and signal efficiency is applied at all centre-of-mass energies and for all (\Mh,\Gh) hypotheses.}
  \end{center}
\end{table}


\begin{table}
  \small
  \begin{center}

\begin{tabular}{|l|c|c|c|c|c|} \hline 
      \textbf{label~~~reference masses}   & \textbf{likelihood} & \textbf{2-fermion}&\textbf{4-fermion}& {\bf total SM} &\textbf{data} \\ \hline 

     A1\hspace*{15mm} 1--120 GeV  &type 1 &  11&  374 & 385 $\pm$ 4 $\pm$ 16 &     369 \\ \hline
     A2\hspace*{15mm} 1--120 GeV  &type 2 &  3 &  378 & 381 $\pm$ 4 $\pm$ 16 &     370 \\\hline
   
     A3\hspace*{15mm} 50--80 GeV & type 1 &  5 &  315 & 320 $\pm$ 3 $\pm$ 13 &     305 \\\hline
   
     A4\hspace*{15mm} 50--80 GeV & type 2 &  2 &  315 & 317 $\pm$ 3 $\pm$ 13 &     310 \\\hline 
      
     A5\hspace*{15mm} 80--120 GeV &type 2 &  8 &  247 & 255 $\pm$ 3 $\pm$ 11 &     253 \\\hline

     \end{tabular}
   \end{center}
   \caption{\sl The likelihood selection of events with a signal likelihood exceeding 0.2 according to the different search strategies. The individual contributions to the total \SM background of two-fermion and four-fermion background is broken down in the second and third column respectively. For the total \SM background the statistical and the systematic uncertainty is also given.}
   \label{t:lhoodsel}
\end{table}

\clearpage

\clearpage

\clearpage


\clearpage


\clearpage

%
\clearpage


\clearpage

\clearpage

\clearpage

\clearpage
%
%
\end{document}